\documentclass[prd,amsfonts,onecolumn,superscriptaddress,aps,nofootinbib,11pt]{revtex4}

\pdfoutput=1
\usepackage[utf8]{inputenc}
\usepackage{amsmath,amssymb,mathrsfs}
\usepackage{graphicx}
\usepackage{enumerate}
\usepackage{multirow}
\usepackage{float}
\usepackage[usenames,dvipsnames]{xcolor} 
\usepackage{hyperref}
\usepackage{slashed}
\usepackage{soul}
\hypersetup{colorlinks=true,
		breaklinks=true,
    linkcolor=Blue,          
    citecolor=Plum,        
    urlcolor=YellowOrange           
}

\begin{document}
\title{Probing ALP-Sterile Neutrino Couplings at the LHC}

\author{Alexandre Alves}
\email{aalves@unifesp.br}
\affiliation{Departamento de Física, Universidade Federal de São Paulo, Diadema, 09913-030, Brasil}

\author{A. G. Dias}
\email{alex.dias@ufabc.edu.br}
\affiliation{Centro de Ci\^encias Naturais e Humanas, Universidade Federal do ABC, Santo Andr\'e-SP, Brasil}

\author{D. D. Lopes}
\email{lopes.diego@ufabc.edu.br}
\affiliation{Centro de Ci\^encias Naturais e Humanas, Universidade Federal do ABC, Santo Andr\'e-SP, Brasil}
\affiliation{Institut für Theoretische Physik, Universität Heidelberg, Philosophenweg 16, 69120 Heidelberg, Germany}

\date{\today}

\begin{abstract}
 In this work, prospects to probe an overlooked facet of axion-like particles (ALPs) 
-- their potential couplings to sterile neutrinos --  are presented. We found that mono-photon searches have the potential 
to constrain ALP couplings to sterile neutrinos when a new heavy scalar boosts the ALP decay 
yields. Working within an effective field theory (EFT) approach, we scan the parameters space 
to establish the reach of the 13 TeV LHC to probe such couplings. We found regions 
of the parameters space evading several experimental constraints 
that can be probed at the LHC. Moreover, a complementary role between 
the LHC and various experiments that search for axions and ALPs 
can be anticipated for models where ALPs interact with sterile neutrinos. We also present the UV 
realization of a model having an axion-like particle, a heavy scalar and sterile neutrinos whose parameters are spanned by our EFT approach. The proposed model contains a type of seesaw mechanism for generating masses for the active neutrinos along with sterile neutrinos involving the high energy scale of the spontaneous breaking of the global symmetry associated to the ALP. Some benchmark points of this model can be discovered at the 13 TeV LHC with 300 fb$^{-1}$.

\end{abstract}
\maketitle
\section{Introduction}

The study of 
pseudo-Nambu-Goldstone bosons 
such as axions and axion-like particles (ALPs) has lead to many developments in 
high energy physics. Theoretically, these particles are expected to have two main features: 
they are light in comparison to the scale of spontaneous breaking of the global symmetry which 
originates them; and their couplings to the Standard Model (SM) particles are, at least, suppressed 
by the inverse of that same scale. A well known solution to the strong $CP$ problem is the 
Peccei-Quinn mechanism which is based on the implementation of an anomalous chiral global U(1)$_{PQ}$ 
symmetry \cite{Peccei:1977hh,Peccei:1977ur}, whose spontaneous breaking gives rise to the axion 
\cite{Weinberg:1977ma,Wilczek:1977pj}. The axion gets mass inversely proportional 
to the scale of spontaneous breaking of the U(1)$_{PQ}$ symmetry due the peculiar feature   
that this symmetry is also broken explicitly at the quantum level.  

The ALP has some interactions similar to the axion but, differently,  
its mass is not directly related to its couplings with the SM particles. In the ALP case, 
the associate spontaneously broken global U(1) symmetry may be generic, and also approximate 
due the existence of operators breaking it explicitly. Such operators give to the ALP a mass 
that is not necessarily related to the scale of the spontaneous broken of the U(1) symmetry.  
Even in this situation, the ALP mass can be considered naturally small by the argument that 
in the limit where they vanish, the symmetries are augmented \cite{tHooft:1979rat}.  
The search for axions and ALPs is currently among the main experimental efforts in particle physics. 
For the general aspects of axions and ALPs see    
\cite{Tanabashi:2018oca,Kim:2008hd,Jaeckel:2010ni,Ringwald:2012hr,Marsh:2015xka}, for example. 
In this work we will deal solely with ALPs. 

Many experiments and astrophysical observations furnish constraints on ALP masses 
and couplings to photons, gluons and electrons~\cite{Tanabashi:2018oca}. 
However, the couplings of ALPs to neutrinos and other dark particles have been overlooked 
in comparison to their couplings to more easily observable 
probes like photons and electrons. This is comprehensible though. In colliders, the search for 
a light scalar decaying invisibly is difficult in the gluon fusion production mode once the missing 
energy spectrum produced along with a visible jet, in a monojet-type signature, is expected to be soft. 
A good probe for the ALP($a$)-photon coupling is $Z\to \gamma+a\to 2(3)\gamma$~\cite{Jaeckel:2015jla,Alves:2016koo} but if $a$ decays invisibly, the monophoton signature would produce a not too hard photon that would be difficult to observe in colliders. In this work, we not only explore a mass region that is more promising for collider searches~\cite{Bauer:2017ris} but also consider a class of models where particles present harder spectra. 

Several studies have already considered the possible signals and prospects of ALPs 
in colliders for specific couplings with gauge bosons and fermions. 

For example, depending on the scenario considered, ALPs can be constrained through: 
$Z$ boson decay into two or three photons~\cite{Jaeckel:2015jla,Alves:2016koo} and other SM heavy 
resonances decay~\cite{Brivio:2017ije,Bauer:2018uxu}; monophoton, monojets, and
triphotons~\cite{Mimasu:2014nea}, decay into two photons and two leptons~\cite{Bauer:2017ris},  
non-resonant production~\cite{Gavela:2019cmq,Merlo:2019anv}, and photon pair production 
in ultra-peripheral 
collisions~\cite{Baldenegro:2019whq} signals at the LHC; production via Primakoff process 
in fixed target  experiments~\cite{Dobrich:2015jyk,Harland-Lang:2019zur,Dobrich:2019dxc} 
such as NA62 and SHiP~\cite{Hahn:1404985,Alekhin:2015byh}, in the FASER experiment~\cite{Feng:2018noy} 
and also in photon-beam experiments~\cite{Aloni:2019ruo}; flavor violation interactions 
via couplings to the $W^\pm$ bosons~\cite{Izaguirre:2016dfi,Gavela:2019wzg} 
and leptons~\cite{Bauer:2019gfk}; ALPs as mediators 
of interactions between the SM particles and dark matter~\cite{Dolan:2017osp}. For somewhat earlier collider probes of axion-like particles see Ref.~\cite{Jaeckel:2012yz}. 

Realistic models embodying ALPs predict at least one-loop couplings to all the Standard Model 
particles. At the tree level, ALPs are expected to participate in the scalar potential. 
Possible ultra-violet complete theories involving ALPs, by their turn, include new heavy fermions 
which might need new scalars 
if their masses are generated via a sort of spontaneous symmetry 
breaking mechanism. In those scenarios, a large scalar sector with the pseudoscalar ALPs 
and another new heavy scalar, at least, is plausible. Augmented fermionic and scalar sectors are very compelling sources of new physics which might be necessary in order to stabilize the SM potential and/or to trigger baryogenesis, for example.

Hidden valley, MSSM and NMSSM~\cite{Strassler:2006im,Strassler:2006ri,Miller:2003ay,Martin:1997ns} are examples of models where 
a heavy scalar is predicted to decay resonantly into lighter scalars. 
For example, in NMSSM, a heavy $CP$-even scalar can decay into two $CP$-odd ones giving very distinctive signatures~\cite{Ellwanger:2003jt}. 
In the context of axion-like particles, a heavy scalar might decay into two ALPs which, if not long-lived, decay inside the LHC detectors.  

The MiniBoone~\cite{Aguilar-Arevalo:2018gpe} and LSND~\cite{Aguilar:2001ty} experiments have reported an excess of appearance of electron neutrinos in a muon neutrino beam for over a decade which, in combination, amount to a $6\sigma$ statistical significance level. Explaining this observation would require a $3+N$ neutrino oscillation model thus evidencing the existence of sterile neutrinos. This might considered, so far, one of the most robust experimental evidence for physics beyond the SM paradigm. In combination, the class of models with new scalars and sterile neutrinos have been proposed to address several astrophysical observations related to dark matter and sterile neutrinos~\cite{LEE2014118,MERLE2015283,Schneider:2016uqi,POSPELOV200853,Kim:2008pp,Lindner:2010wr,Merle:2017jfn}, multi-component dark matter~\cite{Zurek:2008qg}, warm dark matter~\cite{Gelmini:2009xd}, self-interacting DM~\cite{Bringmann:2013vra}, models for electroweak phase transition with sterile neutrinos embedded~\cite{Fairbairn:2013uta}, stable Higgs sectors extended by an scalar DM portal~\cite{Khoze:2014xha}, inflationary models~\cite{Nurmi:2015ema}, and collider signatures~\cite{Shoemaker:2010fg}.

In view of these motivations, we want to explore the possibility of a new channel not previously studied, the resonant ALP pair production 
where one of the ALPs decays to sterile neutrinos and the other one to photons. By sterile neutrino, 
we mean a generic long-lived neutral fermion which originates from a singlet field under 
the Standard Model gauge group. 

We focus on models where a light ALP is allowed to decay to photons, sterile neutrinos, light charged leptons and hadrons. ALPs of order of GeV and heavier can be probed 
in gluon and photon fusion processes~\cite{Jaeckel:2012yz}.
Decays to electrons and muons have a very small impact in the branching ratios of the ALP to sterile neutrinos and photons for the ALP masses that are likely to be probed at the LHC, as we demonstrate in the next sections, and we will ignore them in our analysis.

If the scalar that decays 
into ALPs is heavy and the ALP is short lived, they get boosted in the laboratory frame and 
decay inside the LHC detectors leading to a pair of collimated photons mimicking a monophoton 
signal -- a photon jet. This channel has been intensely studied by the 
LHC collaborations in the search for dark particles~\cite{Aad:2014tda}, 
and also as signals of a high-mass resonance decaying into pairs 
of such effective photon jets via light scalar resonances~\cite{Aaboud:2018djx}. 

Our goal in this work is to explore the 13 TeV LHC reach to probe ALP-neutrino couplings in the monophoton channel
\begin{equation}
pp\to S\to aa\to \gamma\gamma+\not\!\! E_T\to \gamma_{jet} +\not\!\! E_T,
\end{equation}
where one heavy scalar, $S$, decays to two light ALPs, $aa$, which then decay to photons and sterile neutrinos that pass the detector leaving with no trace. From now on, we assume that the ALPs are boosted and lead to very colimated photons which hit a single detector cell mimicking a monophoton channel. This kind of signature with colimated photons has been explored in previous works like~\cite{Jaeckel:2015jla, Alves:2016koo}.

We will work within an effective field theory formulation of the interactions between the ALPs, the new scalar $S$, the sterile neutrino $N$ and the SM fields. However, we also provide an ultra-violet complete construction to exemplify a concrete model which motivates the choice of the EFT parameters. In this model, the heavy scalar and the ALP are the components of a same SM singlet complex field with self-interactions dictated by a scalar potential. It is reasonable to expect that the high energy scale in which the approximate U(1) symmetry related to the ALP is spontaneously broken participates in the neutrinos mass generation mechanism. Thus, the model is also built to furnish a mass generation mechanism that can lead to sterile neutrinos with masses at the MeV scale as well as active neutrinos with masses at the sub-eV scale.

Moreover, we are going to assume that the same new heavy states mediate both the $S$ and ALP 1-loop interactions to SM particles. After integrating out these heavy states, the new physics scale $\Lambda$ parametrizes all the new scalar interactions making our EFT model more predictive.

Our paper is organized as follows. In Section~\ref{sec-efcl} we present the effective Lagrangian on which 
our developments are based and discuss the main parameters entering in the decay widths. In the 
Section~\ref{sec:constr} we discuss the implementation of the several existing experimental constraints. Our analysis on the search for the ALP-sterile neutrino interactions with monophotons at the LHC are presented in 
Section~\ref{sec:monophoton}. A generic model for a complete theory at the ultra-violet including a seesaw mechanism to generate sterile neutrinos masses at the MeV scale as well as active neutrinos masses at the sub-eV scale is present in section~\ref{uvmodel}. Our conclusions are given in Section~\ref{conclu}. 

\section{EFT Lagrangian and partial decay widths}
\label{sec-efcl}
We parametrize the relevant interactions of the heavy scalar, $S$, the ALP, $a$, the sterile neutrino, $N$, 
and SM gauge bosons, electrons and muons, with the following effective Lagrangian
 
\begin{equation}
\begin{split}
\label{eq:Lagrangian}
\mathscr{L}_{int} &= \frac{c_{BB}}{\Lambda} S B^{\mu \nu} B_{\mu \nu} + \frac{c_{WW}}{\Lambda} S W_i^{\mu \nu} W_{\mu \nu}^i + \frac{c_{GG}}{\Lambda} S G_a^{\mu \nu} G_{\mu \nu}^a \\ 
&+ \frac{k_{BB}}{\Lambda} a B^{\mu \nu} \tilde{B}_{\mu \nu} + \frac{k_{WW}}{\Lambda} a W_i^{\mu \nu} \tilde{W}_{\mu \nu}^i + \frac{k_{GG}}{\Lambda} a G_b^{\mu \nu} \tilde{G}_{\mu \nu}^b \\ 
&+  f_S\, S\, a^2 -\lambda S^2 a^2 - \frac{c_{aN}}{\Lambda} \overline{N} \gamma^{\mu} \gamma^5 N \; \partial_{\mu} a - \frac{c_{ae}}{\Lambda} \overline{e} \gamma^{\mu}\gamma^5 {e} \; \partial_{\mu} a - \frac{c_{a\mu}}{\Lambda} \overline{\mu} \gamma^{\mu} \gamma^5 \mu \; \partial_{\mu} a. 
\end{split}
\end{equation}
We assume that all the scalar couplings are mediated by the same new physics scale $\Lambda$ as discussed in the Introduction. The parameters $c_{BB}, c_{WW}, c_{GG}$, $k_{BB}, k_{WW}, k_{GG}$, $c_{aN}$, $c_{ae}$ and $c_{a\mu}$
are all dimensionless whereas $f_S$ has dimension of mass. The coupling to gauge bosons preserve the SM group symmetry. The field strength tensors of the $U(1)_Y$, $SU(2)_L$ and $SU(3)_C$ gauge bosons are, respectively, $B_{\mu\nu}$, $W_{\mu\nu}$ and $G^a_{\mu\nu}$. The dual field strength of the $U(1)_Y$ gauge boson field is defined as
$\tilde{B}_{\mu\nu}=\frac{1}{2}\epsilon_{\alpha\beta\mu\nu}B^{\alpha\beta}$, with the totally anti-symmetric tensor being such that $\epsilon_{0123}=1$, and analogously to the other fields. In the Section~\ref{uvmodel} we give an example of a concrete ultra-violet complete model 
containing heavy fields which can lead to the low energy effective Lagrangian like the Eq. (\ref{eq:Lagrangian}). Despite we have included the quadrilinear term $S^2a^2$ it does not play any relevant role in our analyses, once it gives a subdominant contribution for the $a$ production through $S$ decay. An example of a simple potential, where $S$ and $a$ are the components of a complex SM singlet field, leading to a mass term to the ALP field and the $S-a$ interactions is  given in Section~\ref{uvmodel}. 

Using the equation of motion for the neutrinos, we see that the interaction term with them 
is actually proportional to their masses. That is the reason why we do not expect 
that couplings to the SM neutrinos can be probed at all, only to a heavy sterile neutrino of mass $m_N$. An invisible ALP decay is therefore a clear signal of interactions with sterile neutrinos or some new dark state. By its turn, the $Saa$ coupling, $f_S$, has dimension of energy and can, in principle, be large.  

In spite of the simplifying assumption of a same new physics scale mediating both $S$ and ALP 
interactions, the couplings of these new states to these scalars are not necessarily the same. 
Both $S$ and $a$ couplings to the SM particles  
are supposedly given in terms of the structure and parameters of a complete high energy theory. 
For example, if $S$ and $a$ are originated from scalar singlets under the SM group 
they might have tree level couplings with the known fermions and the Higgs boson proportional 
to mixing angles. In this case, their couplings to the vector bosons, $\gamma$, $Z^0$, and $W^\pm$
can arise through one-loop order in perturbation theory involving new particles carrying 
SM quantum numbers. In special, the ALP couplings depend in general on the 
associated U(1) symmetry charges of the fermions and their properties under the SM 
group (see Section \ref{uvmodel}). If $S$ and $a$ are the real and the imaginary
components of the same complex scalar field, as the model presented in Section \ref{uvmodel}, 
the parameters $c_{BB}, c_{WW}, c_{GG}$ and $k_{BB}, k_{WW}, k_{GG}$ above can be related. 
Anyway, we stick to the general case where these parameters can be all different from 
each other. 

The partial widths of $S$ and $a$ are given right below 
\begin{eqnarray}
&& \Gamma (S \to a a) = \frac{f_S^2}{8 \pi m_{S}} \sqrt{1 - \frac{4 m_{a}^2}{m_{S}^2}},\nonumber \\
&& \Gamma (S \to \gamma \gamma) = \frac{(c_{BB} c_w^2 + c_{WW} s_w^2)^2}{4 \pi \Lambda^2} m_{S}^3,\nonumber \\
&&\Gamma (S \to g g) = \frac{2 c_{GG}^2}{\pi \Lambda^2} m_{S}^3,\nonumber \\
&&\Gamma (S \to Z Z) = \frac{(c_{WW} c_w^2 + c_{BB} s_w^2)^2 (m_{S}^3 - 4 m_{S} m_{Z}^2 + 
\frac{6 m_{Z}^4}{m_{S}})}{4 \pi \Lambda^2} \sqrt{1 - \frac{4 m_{Z}^2}{m_{S}^2}},\nonumber \\
&&\Gamma (S \to \gamma Z) = 
\frac{c_w^2  s_w^2 (c_{BB} - c_{WW})^2 (m^2_{S} - m^2_{Z})^3}{2 \pi \Lambda^2 m_{S}^3},\nonumber \\
&&\Gamma (S \to W^{-} W^{+}) = \frac{c_{WW}^2 (m_{S}^3 - 4 m_{S} m_{W}^2 + 
\frac{6 m_{W}^4}{m_{S}})}{2 \pi \Lambda^2} \sqrt{1 - \frac{4 m_{W}^2}{m_{S}^2}},\nonumber \\
&& \Gamma (a \to \gamma \gamma)= \frac{(k_{BB} c_w^2 + k_{WW} s_w^2)^2}{4 \pi \Lambda^2} m_{a}^3,\nonumber \\
&& \Gamma (a \to N \overline{N}) = \frac{c_{a N}^2}{2 \pi \Lambda^2} m_{a} m_{N}^2 
\sqrt{1 - \frac{4 m_{N}^2}{m_{a}^2}},\nonumber \\
&& \Gamma (a \to \ell^+\ell^-) = \frac{c_{a \ell}^2}{2 \pi \Lambda^2} m_{a} m_{\ell}^2 
\sqrt{1 - \frac{4 m_{\ell}^2}{m_{a}^2}},\nonumber \\
&& \Gamma (a \to g g) = \frac{2 k_{GG}^2}{\pi \Lambda^2} m_{a}^3.
\label{widtha}
\end{eqnarray} 

In the formulae of Eqs.(\ref{widtha}) above, $m_S$ and $m_a$ denote the  heavy scalar and 
the ALP masses, respectively; $m_Z$ and $m_W$ the masses of the $Z$ and $W$ bosons, respectively; $\ell= e,\mu$;
and $s_w^2\equiv{\rm sin}^2\theta_W\approx 0.231$ is the sine squared of the electroweak mixing angle. 
In the ALP mass range from $~0.4$ GeV up to 2 GeV, the effective ALP-gluon coupling cannot describe 
the coupling to hadrons correctly as it does not take the mass thresholds into account. 
For this mass range we take the main hadronic modes individually in our calculations 
as showed in Ref.~\cite{Aloni:2018vki}. Yet, as we are going to see, after the neutral pion channels 
is open, the monophoton signature vanishes and the prospects for the LHC observation die out 
if $m_a\gtrsim 0.4$ GeV effectively.   
Nevertheless, it is possible to choose a fermion content in a UV-complete realization of the 
model in which the couplings to gluons are suppressed raising the branching ratio of the other 
channels (see Section~\ref{uvmodel}). 

\begin{figure}[ht!]
\includegraphics[scale=0.45]{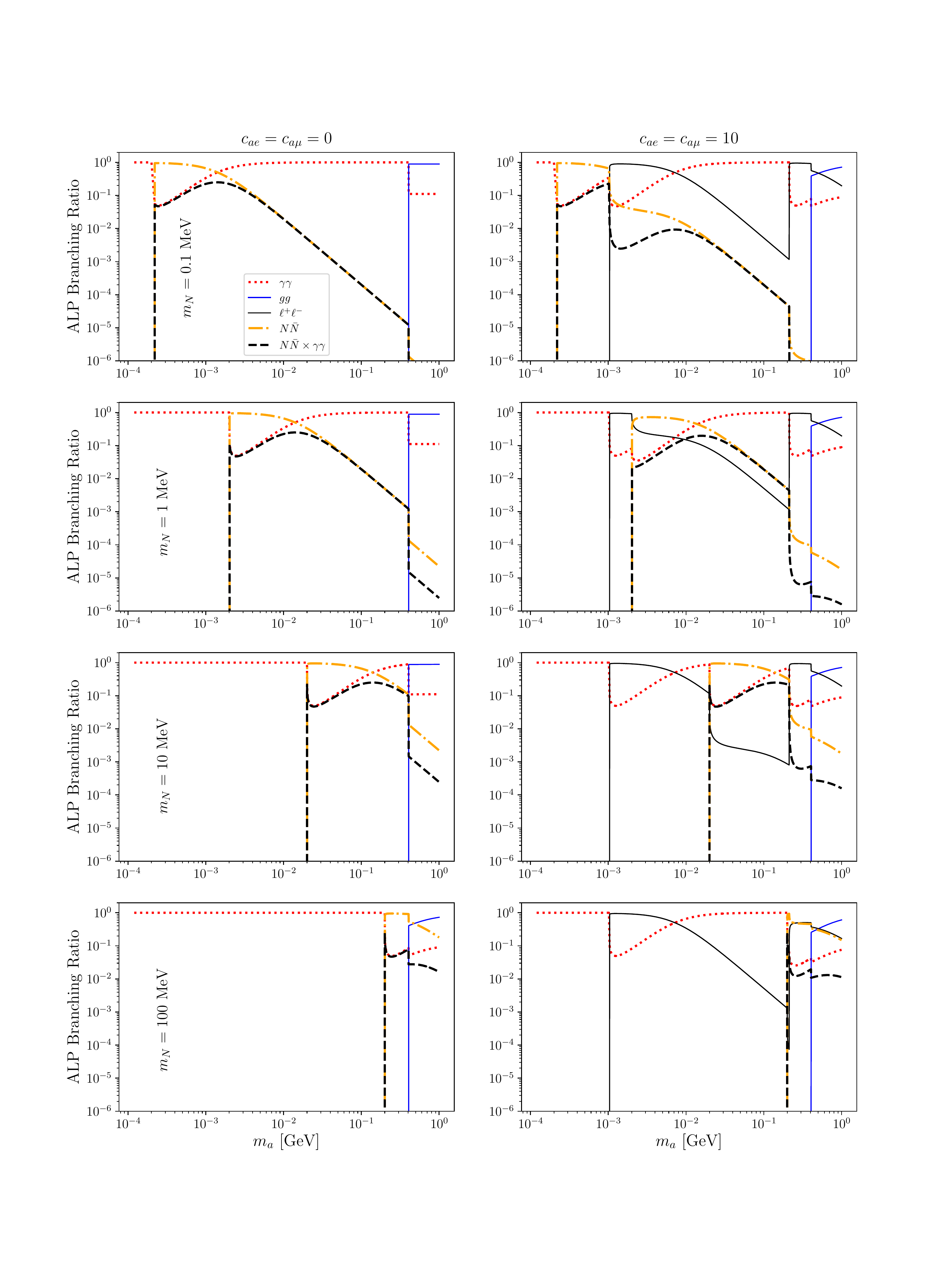}
\hspace*{-1cm}
\caption{The branching ratios of the ALP in the various channels allowed by the interactions of Eq.~(\ref{eq:Lagrangian}) as a function of the ALP mass $m_a$. The left(right) panels have ALP-leptons couplings set to 0(10). The sterile neutrino mass is fixed in 0.1, 1, 10 and 100 MeV from first to the last row. The dashed black lines denoted by $N\bar{N}\times\gamma\gamma$ represent the product $BR(a\to\gamma\gamma)\times BR(a\to NN)$.}
\label{fig:alp_branching}
\end{figure}

Three main factors dictate the number of monophoton signal events that can be produced in $pp$ 
collisions at the LHC. The resonant production cross section of the ALPs, 
$\sigma(pp\to S\to aa)$, the branching ratio $BR(S\to aa)$, 
and the product $BR(a\to\gamma\gamma)\times BR(a\to N\bar{N})$.

In Fig.~(\ref{fig:alp_branching}), we display the branching ratios of the ALP for masses 
up to 1 GeV in eight different scenarios. In all of them, we fix $m_S = 1000$ GeV, 
$f_S = 100$ GeV, and $k_{BB} = k_{WW}= k_{GG}=1$. The sterile 
neutrino mass is fixed to 0.1, 1, 10 and 100 MeV at the first, second, third and fourth row, 
respectively, and ALP-leptons coupling is set to 0 in the plots of the first column, and at 10 
in the plots of the second column. In all cases we keep  $c_{aN}/\Lambda=0.001$ 
GeV$^{-1}$ fixed as well.

\begin{figure}[t!]
\includegraphics[scale=0.56]{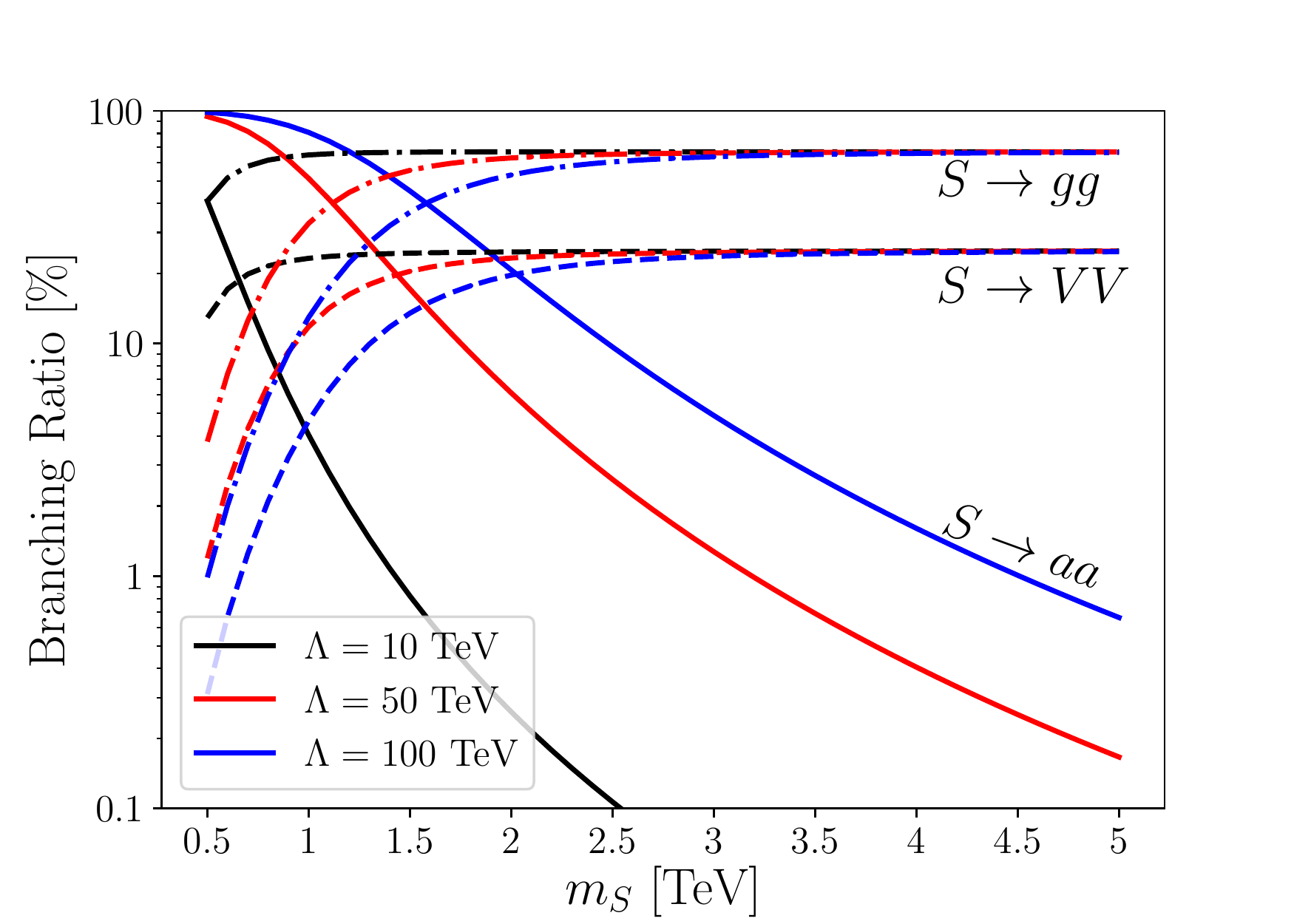}
\caption{The branching ratios of the scalar $S$, as a function of its mass, in the various channels allowed by the interactions of Eq.~(\ref{eq:Lagrangian}). We fixed  $m_a = 0.1$ GeV, $\Lambda = 10$ TeV, $f_S = 100$ GeV, and $c_{BB} = c_{WW} = c_{GG} = 1$ to compute these branching ratios. The $S\to VV$ channel represents the combination of the $ZZ$ and $WW$ branching ratios.} 
\label{fig:branchingS}
\end{figure}

Showing scenarios for different ALP-leptons couplings anticipates the two frameworks that we are working with -- the EFT approach, where $c_{a\ell}$ is free, and the UV complete model of Section~\ref{uvmodel} where this coupling is suppressed. We basically see from Fig.~\eqref{fig:alp_branching} that switching off the leptons couplings leaves more room for $\gamma\gamma$ and $NN$ decays naturally. In special, we are interested in the product $BR(a\to\gamma\gamma)\times BR(a\to NN)$, depicted as dashed black lines. If the sterile neutrino mass is larger than 0.1 MeV, we see that including leptons decays have a minor impact actually. This is because the ALP-sterile neutrino mass is proportional to its mass.

As $m_N$, the sterile neutrino mass, increases, the region of ALP masses where decays to $\gamma\gamma+NN$ shrinks and as soon as $m_N$ approaches the hadrons threshold, the hadrons channel dominates and decays to photons and neutrinos become rare. Finally, notice that for $m_N<100$ MeV, there exist ALP mass regions where  $BR(a\to\gamma\gamma)\times BR(a\to NN)$ can reach its maximum of $\sim 0.25$. These spots will be the most promising ones to search for monophoton events at the LHC.

\begin{figure}[t!]
\includegraphics[scale=0.6]{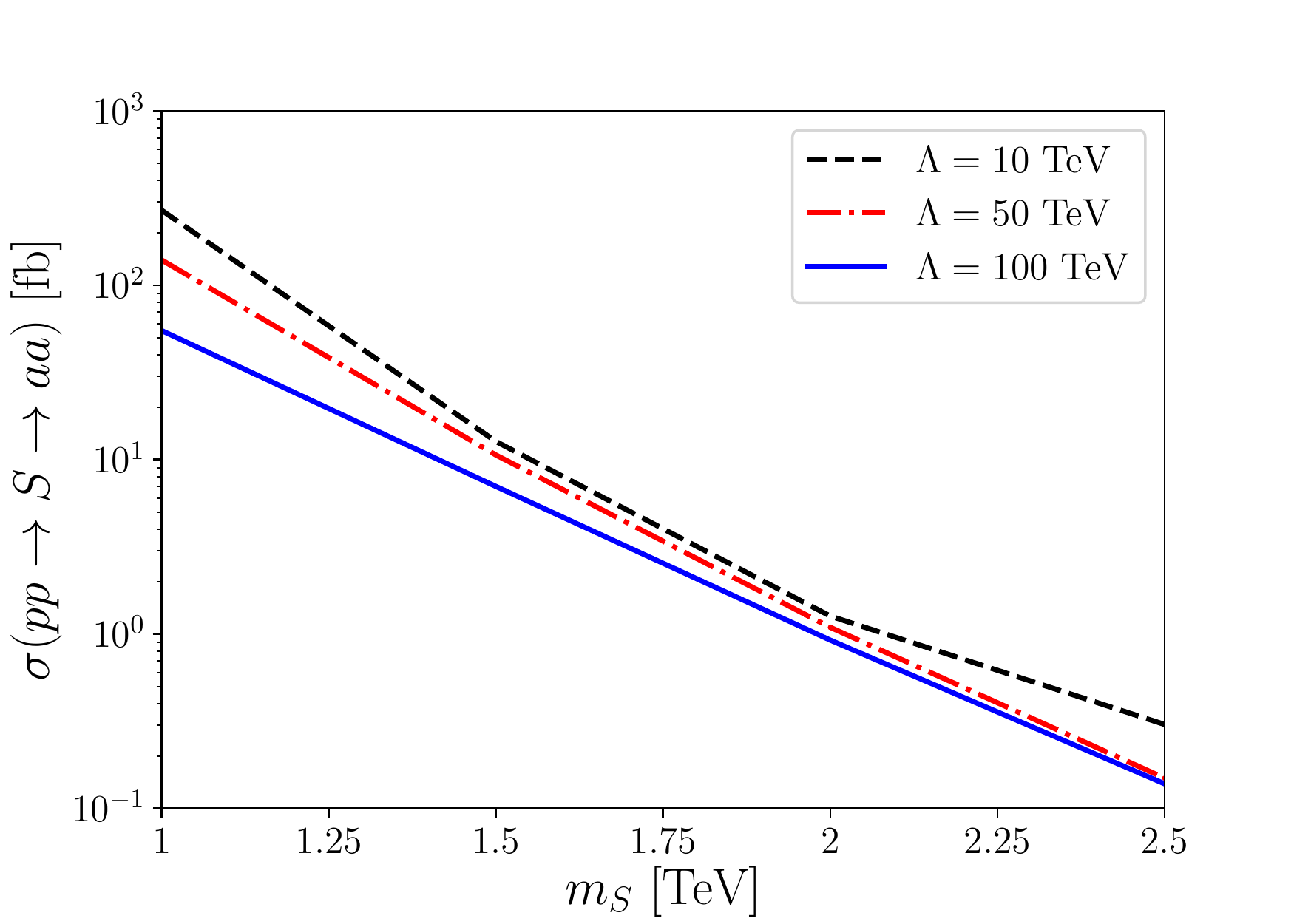} 
\caption{The resonant production cross section of two ALPs as a function of the $S$ mass for three fixed values of $\Lambda$. We fixed  $m_a = 0.1$ GeV,  $f_S = 100$ GeV, and $c_{BB} = c_{WW} = c_{GG} = 1$.} 
\label{fig:xsec}
\end{figure}

The dominant branching ratios of the scalar $S$, as a function of its mass, are shown in Figure~\eqref{fig:branchingS}. The solid, dashed, and dot-dashed lines show the branching ratios of $S$ to $aa$, $VV=ZZ+WW$, and $gg$, respectively. In upper lines (in black) we set $\Lambda=10$ TeV, middle lines(in red), $\Lambda=50$ TeV, and in lower lines(in blue), we fix $\Lambda=100$ TeV. The ALP mode dominates for $S$ masses up to $\sim 1.5$ TeV if $\Lambda=100$ TeV but, only up to $\sim 500$ GeV when $\Lambda=10$ TeV. For $m_S=1$ TeV fixed, $BR(S\to aa)$ drops from nearly 100\% for $\Lambda=100$ TeV to a few percent when $\Lambda=10$ TeV.  This dependence upon the scale $\Lambda$ helps to soften a bit the drop in the production cross section of ALP pairs shown in Figure~\eqref{fig:xsec}, especially for large $S$ masses. For an 1 TeV $S$ scalar, the production cross section ranges from tens to hundreds of femtobarns when $\Lambda$ varies from 10 to 100 TeV. The variation decreases for $m_S=2$ TeV, and the cross sections barely depend on $\Lambda$.

Next, let us start discussing the search for signals of the model at the LHC by first evaluating possible constraints to the model.

\section{Experimental Constraints} 
\label{sec:constr}

The interactions of Eq.~\eqref{eq:Lagrangian} can be probed in a number of ways and several experimental constraints apply. In special, resonant production and decay of the $S$ scalar into dijets, diphotons, $ZZ$, $WW$, $Z\gamma$ and $aa\to \gamma\gamma+\not\!\! E_T$ are subject to collider constraints. We also apply constraints on the lifetime of the ALPS and the induced $Z$ boson decays to photons.

We discuss the details of each one of these constraints right below. All the following experimental constraints were obtained from the 13 TeV LHC data, except for the last one of $Z$ boson decays to photons that come from the 8 TeV LHC run and Tevatron data.

\begin{enumerate}
\item \underline{Diphoton searches} -- The diphoton final state provides a clean experimental signature with excellent invariant-mass resolution and moderate backgrounds. 

According to~\citep{Aaboud:2017yyg}, searches for new phenomena in high-mass diphoton final states with the ATLAS experiment at the LHC gives an upper limit on 
\begin{equation}
\sigma(g g \to S)\times \left[Br(S \to \gamma \gamma)+BR(S\to aa)BR^2(a\to\gamma\gamma)\right]<0.5\; \hbox{fb} 
\label{eq:diphoton}
\end{equation}
at 95 \% confidence level.

\item \underline{Dijet searches} -- 
The search for resonances decaying into pairs of jets by the CMS \citep{Sirunyan:2018xlo}, with an integrated luminosity of $36 \ \hbox{fb}^{-1}$, places the limit on the resonance production cross section times the branching ratio $Br(S \to j j)$ times the acceptance $A$ given by
\begin{equation}
\sigma(g g \to S) Br(S \to j j)\times \epsilon_{acc} < 1.66\; \hbox{pb}, \end{equation}
at $95$\% confidence level. Here, $\epsilon_{acc}$ is the acceptance for narrow resonances, with the kinematic requirements $| \Delta \eta_{jj} | < 1.3$ for the dijet system and $| \eta_j | < 2.5$ for each of the identified jets, where $\eta_j$ is the rapidity of a jet.

\item \underline{$ZZ$ pairs} -- 

A search for a new scalar resonance decaying to a pair of $Z$ bosons by the CMS~\citep{Sirunyan:2018qlb} with an integrated luminosity of $35.9 \ fb^{-1}$, where the $Z$ boson pair decays are reconstructed using the $4l$, $2l2q$, and $2l 2 \nu$ final states, imposes the following upper limit on the resonance production cross section times the branching ratio $Br(S \to Z Z)$
\begin{equation}
\sigma(g g \to S) Br(S \to Z Z) < 10^{-3}\; \hbox{pb},
\end{equation}
at $95$\% confidence level.

\item \underline{$WW$ pairs} -- Searches for neutral heavy resonances performed in the $W W \to e \nu \mu \nu$ decay channel either directly or via leptonic tau decays with additional neutrinos, with an integrated luminosity of $36.1 \ fb^{-1}$, were made by the ATLAS experiment at the LHC in~\citep{Aaboud:2017gsl}.  According to these searches we must have the upper limit
\begin{equation}
\sigma(g g \to S) Br(S \to W W) < 0.5\; \hbox{pb},
\end{equation}
at $95 \%$ confidence level.

The search in the $e \nu \mu \nu$ decay mode is complementary to searches performed in other decay modes. In particular, the sensitivity to low mass
resonances is higher in the fully leptonic final state than in final states that include jets due to background from jet production.

\item \underline{$Z\gamma$ searches} --
The $Z(\to \ell^+\ell^-) \gamma$ final state can be reconstructed with high efficiency and good invariant mass resolution. 

Based on $Z\gamma$ searches of a heavy resonance by ATLAS~\citep{Aaboud:2017uhw}, with $36.1$ fb$^{-1}$, we set an upper limit on the resonance production cross section times the branching ratio $Br(S \to Z\gamma)$ given by
\begin{equation}
\sigma(g g \to S) Br(S \to Z\gamma) < 0.5\; \hbox{fb},
\end{equation}
at $95 \%$ confidence level.

\item \underline{Monophoton searches} -- Models that predict a photon with high momentum and large $\not\!\! E_T$ in $pp$ collisions have been constrained by the LHC. The results of a search of such events in the ATLAS study of Ref.~\citep{Aaboud:2017dor} with an integrated luminosity of $36.1$ fb$^{-1}$ gives us an upper limit on the total cross section for the process
\begin{equation}
\sigma(\gamma+\not\!\! E_T)\times \epsilon_{acc} < 2.3\; \hbox{fb},
\label{eq:monoph}
\end{equation}
at $95 \%$ confidence level. Here, $\epsilon_{acc}$ is the acceptance for monophoton final states with the kinematic requirements $E_T^{\gamma} > 150$ GeV, $|\eta| < 1.37$ or $1.52 < |\eta| < 2.37$ and $\Delta \phi(\gamma, \not\!\! E_T) > 0.4$. Also it is required events without jets or just one jet with $p_T > 30 GeV$, $|\eta| < 4.5$ and $\Delta \phi(jet, \not\!\! E_T) > 0.4$, no charged leptons in the final state and $\not\!\! E_T > 300$ GeV.

\item \underline{ALP lifetime} -- We require that the ALP decays inside the electromagnetic calorimeter which, at ATLAS and CMS, is situated approximately at 1 meter from the interaction point. 
The distance that the boosted ALP  produced in the decay of the heavy scalar $S$  travels from the interaction point before decaying is given approximately by \cite{Jaeckel:2015jla, Alves:2016koo} 
\begin{eqnarray}
l_{decay} \approx \frac{m_{S}}{m_a \Gamma_a} \times 10^{-16}\; {\rm m}. 
\label{ldecay}
\end{eqnarray}

The total width of the ALP can be computed from
\begin{eqnarray}
\Gamma_a &=& \Gamma(a \to \gamma \gamma) +  \Gamma(a \to N \overline{N})\theta(m_a - 2 m_N)\nonumber\\ 
&+& \Gamma(a \to e^{-} e^{+})\theta(m_a - 2 m_e) + \Gamma(a \to \mu^{-} \mu^{+})\theta(m_a - 2 m_\mu)\nonumber\\
&+& \sum_i \Gamma(a \to h_i))\theta(m_a-m_{h_i}),
\end{eqnarray}
where $h_i$ denotes a hadronic final state of total mass $m_{h_i}$. Substituting this equation in Eq.~\eqref{ldecay} leads to the constraint
\begin{eqnarray}
\label{lifetime}
\frac{m_a}{m_S} \Gamma_a > 10^{-16}\; \hbox{GeV}.
\end{eqnarray}

\item \underline{Photonic decays of the $Z$ boson} -- Important constraints from searches for $Z$ boson decays into photons are given by the CDF Collaboration~\cite{Aaltonen:2013mfa}: $Br(Z \to \gamma \gamma) < 1.45 \times 10^{-5}$ and the ATLAS Collaboration~\cite{Aad:2015bua}: $Br(Z\to \gamma\gamma\gamma) < 2.2\times 10^{-6}$.

The azimuthal opening angle between the two photons coming from the ALP in the decay $Z \to a\gamma \to \gamma+\gamma\gamma$ is given by 
\begin{eqnarray}
\Delta \phi \sim \frac{4 m_a}{m_Z}.
\end{eqnarray}

Taking the resolution of the CDF detector to be 120 mrad \citep{Latino:2000um}, we see that the two photons from the ALP will be colimated if $m_a<2.7$ GeV, and the CDF constraints apply~\cite{Alves:2016koo}. The resolution of the LHC detectors is $\Delta\phi = 20$ mrad. The search for $3\gamma$ decays of the $Z$ boson performed by the ATLAS Collaboration~\cite{Aad:2015bua} thus apply if the photons from the ALP decay are not collimated, that is it, for $m_a>0.46$ GeV. 

In both cases, the relevant branching ratio of this $Z$ decay channel is 
\begin{eqnarray}
Br(Z \to \gamma \gamma \gamma) &=& Br(Z \to a \gamma) \ Br(a \to \gamma \gamma) \\ \nonumber
&=& \frac{m_Z^3}{6 \pi \Lambda^2 \Gamma_Z} \left(1 - \frac{m_a^2}{m_Z^2} \right)^3 \frac{s_w^2 c_w^2 (k_{BB} - k_{WW})^2}{1 + \frac{8 k_{GG}^2}{(k_{BB} c_w^2 + k_{WW} s_w^2)^2}}.
\end{eqnarray}

Neglecting the ALP mass in comparison to the $Z$ boson mass we now have another constraint on the parameters in the EFT Lagrangian
\begin{eqnarray}
\frac{(k_{BB} - k_{WW})^2}{1 + \frac{8 k_{GG}}{(k_{BB} c_{w}^2 + k_{WW} s_{w}^2)^2}} \le \frac{6 \pi^2 \Lambda^2 \Gamma_Z}{m_{Z}^3 s_{w}^2 c_{w}^2} \times Br_{Z}^{exp},
\end{eqnarray}
where $Br_Z^{exp}$ is the CDF limit if $m_a<2.7$ GeV, and the ATLAS limit if $m_a>0.46$ GeV. As we are going to see, the CDF limit is, effectively, the bound to be respected.
\end{enumerate}

 Now, we are going to investigate the prospects of the LHC to observe monophoton events from the production of ALPs taking into account all the constraints just described.

\section{Search for ALP-sterile neutrino interactions with monophotons at the LHC}
\label{sec:monophoton}

\subsection{Sampling of parameters and ALP lifetime}
We are interested in the regime where one ALP from $S\to aa$ decays to two colimated photons and the other one to neutrinos, a monophoton signal therefore, as we discussed earlier. 

The EFT Lagrangian of Eq.~(\ref{eq:Lagrangian}) encompasses 11 independent parameters plus the $S$, $N$ and ALP masses, totaling 14 parameters. 
Instead of choosing a few representative benchmark points, we performed a thorough exploration of the parameters space looking for insights of the interplay among those parameters to find the most promising spots for a collider search. In order to take into account all the constraints simultaneously, we generated $2 \times 10^6$ random sets of parameters and tested them against all the constraints described in Section~(\ref{sec:constr}). The range of the parameters drawn from log-uniform distributions are the following
\begin{eqnarray}
 c_{BB} &\in & [0.001, 1],\; c_{WW} \in [0.001,  1],\; c_{GG} \in [0.001, 1], \nonumber \\ 
 k_{BB} &\in & [0.001, 1],\; k_{WW} \in [0.001, 1],\; k_{GG} \in [0.001, 1] \nonumber \\
 c_{aN} &\in & [0.1, 200],\; c_{ae}=c_{a\mu} \in [0.001, 1],\; f_S \in [500, 1500]\; \hbox{GeV}, \nonumber \\
 m_a &\in& [0.001, 2]\; \hbox{GeV},\; m_N \in [10^{-9}, 1]\; \hbox{GeV},\; m_S \in [0.5,2]\;\hbox{TeV},\; \Lambda \in [1, 100]\; \hbox{TeV}\; .
 \label{eq:ranges}
\end{eqnarray}

The range of the $c$ and $k$ coefficients are chosen with the assumption that they should lie in the perturbative regime of a concrete model. 
Also, we require $\frac{c_{aN}}{\Lambda}m_N < 1$ in order to keep ourselves in the perturbative regime as well. 

The ALP mass and its coupling to photons are very constrained by a variety of experiments. In Ref.~\cite{Bauer:2017ris}, a mass range of a few MeV is established as the one which is more promising for collider searches. The ALP-neutrino and ALP-leptons couplings, $c_{aN}$, $c_{ae}$, and $c_{a\mu}$, respectively, considered in Eq.~\eqref{eq:ranges}, might lie in natural ranges from the point of view of an UV complete theory, as exemplified in Section~\ref{uvmodel}. The parameters $\Lambda$, $f_S$, $m_S$ and $c_{aN}$ also look natural from the point of view of the model constructed in Section \ref{uvmodel}.

The requirement that the ALPs must decay inside the detector, given by Eqs.~\eqref{ldecay} and \eqref{lifetime}, strongly constrains the ALP mass. Actually, only 30\% of all points generated pass this constraint. After imposing this constraint though, a very small fraction of ALP pairs is expected to decay both outside the first meter of the detector and we can safely assume that all the ALP pairs selected from $\ell_{decay}<1$ m will indeed decay in this region. On the other hand, a very small fraction of ALP pairs is expected to either both decay outside the detector or one decays inside and the other outside, as we show in Appendix A.

\begin{figure}[t!]
\includegraphics[scale=0.6]{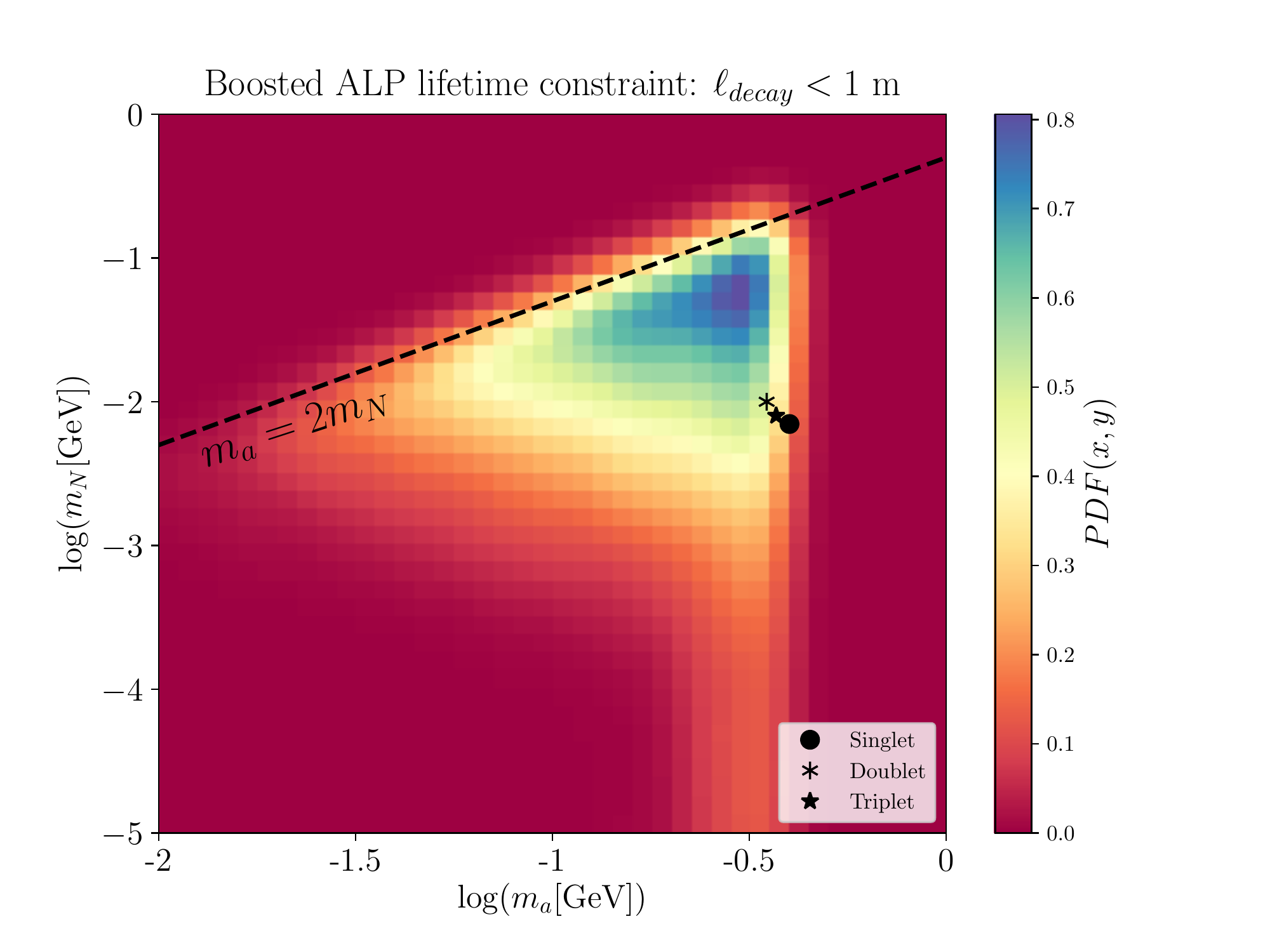}
\caption{Points in the plane of masses $m_a$ {\it versus} $m_N$ which satisfy Eqs.~\eqref{ldecay} and \eqref{lifetime}. The points where randomly generated for the parameters in the same intervals of Eq.~\eqref{eq:ranges}. The dot, asterisk and star represent examples of a singlet, doublet and triplet representation of the UV complete model presented in Section~\ref{uvmodel}, respectively.}
\label{plot-lifetime}
\end{figure}

In the subsequent heatmaps, we display the two-dimensional probability density function (PDF) of the points estimated as the ratio of the number of points contained in a given pixel and the total number of parameters points that pass all the constraints. In Fig.~\eqref{plot-lifetime}, in particular,
we see that the bulk of viable points that pass all the constraints lie above $m_a\sim 100$ MeV and $m_N\sim 1$ MeV. Notice that the neutrinos decay channel make it easier to respect this requirement once the bulk of the points lie in the region where $m_a>2m_N$. In this plot and the subsequent heatmaps, we also show representative points of the UV complete model to be constructed in Section~\ref{uvmodel}. We refer the reader to that section for discussions about these points.

Concerning the neutrino masses, considering only one type of neutrino, the  branching ratio of the ALP into neutrinos turns out to be proportional to $(m_N/m_a)^2$.

If $m_N << m_a$, $Br(a \to N \overline{N})$ gets diminished, making the cross section too small to give any detectable signal. 
The cosmological upper limit on the sum of the SM neutrinos masses~\citep{Tanabashi:2018oca}, $\sum_{\nu} m_{\nu} < 0.72$ eV, by its turn, imposes an upper limit on the SM neutrino masses which make their branching ratios very small. This is an immediate consequence of the ALP-fermion interaction that scales with the fermion mass. For this reason, a missing energy signal at colliders can only come from ALP decays to a new heavier neutrino or dark fermion. 

There are laboratory and cosmological constraints over the sterile neutrinos mixing 
with the active ones for the mass scales below GeV which we are considering here, as can be seen in Ref.~\cite{Adhikari:2016bei} and references therein.  We discuss these constraints in Section~\ref{uvmodel} where an ultraviolet completed model that can account for the effective interactions of the ALP and mass generation for the neutrinos is presented.  

\subsection{Details of the signal and backgrounds simulations}
In order to estimate the reach of the LHC to constrain signals of ALPs in the monphoton channel, 
we simulated events for the signal $p p \to \ S \to \ a a \to \ \gamma \gamma +N  \overline{N}$, 
and its main SM backgrounds: (1) the irreducible $Z(\to \ \nu \overline{\nu}) + \gamma$, 
and (2) the reducible $W(\to \ l \nu) + \gamma$ with a missing electron or muon coming from 
the $W$ boson, and (3) the reducible  $W(\to e \nu_e)$ where the electron is misidentified 
as a photon, following the ATLAS study of Ref.~\cite{Sirunyan:2017ewk}. 
The simulations were performed at Leading Order using \texttt{MadGraph5}~\citep{Alwall:2014hca} and the cross sections of the main backgrounds and the signals were rescaled by higher order QCD K-factors as discussed ahead. The QCD corrections have a larger impact in the gluon fusion production of the signal compared to the electroweak production of the backgrounds. Hard jet emission are expected to have a mild impact in the kinematic distributions used to select the signal and background events but they hardly affect the sensitivity prospects of the LHC for the monophoton searches. We simulated parton hadronization and showering with \texttt{Pythia8}~\citep{Sjostrand:2007gs} and detector effects with \texttt{Delphes3}~\citep{deFavereau:2013fsa}. A photon is considered to be isolated if particles with $p_T>0.5$ GeV inside a cone of $\Delta R=0.5$ centered in the photon 3-momentum do not deposit a total fraction of their transverse momentum above 0.12 of the photon $p_T$.

We fixed $m_a = 0.1$ GeV, $m_N = 0.04$ GeV, $\Lambda = 10$ TeV, $f_S = 100$ GeV, $k_{GG} = 0$ and $c_{BB} = c_{WW} = c_{GG} = k_{BB} = k_{WW} = c_{aN} = 1$ as our benchmark parameters for the simulation at the  $13$ TeV LHC, generating $10^5$ events for the signals and each one of the backgrounds. We checked that this benchmark point is not excluded by any constraints but we also point out that this benchmark point is used just as an starting point for our calculations as described right below.
The mass of the ALP does not play any role in the kinematics of the events in the range considered in this work, but the heavy scalar mass does. A heavier scalar passes the cut requirements more easily, but its  production cross section decreases with the mass. 

In order to estimate cut efficiencies for signals, we generated events for masses from $m_S=500$ up to $m_S=2000$ GeV in steps of 250 GeV and the efficiency of the intermediary masses are then interpolated. In the benchmark point, $BR(a\to\gamma\gamma)\times BR(a\to N\bar{N})=0.13$, almost reaching the maximum of 0.25, while $BR(S\to aa)=0.07$. The production cross section of the $S$ scalar of 1 TeV is 6.8 pb, and the cross section after the decay chain $S\to aa\to \gamma\gamma+N\bar{N}$ turns out to be around 70 fb.

In the subsequent analysis, we compute the number of signal events $N_S$ for an integrated luminosity $L$ as follows
\begin{eqnarray}
 N_{S_0} &=& L\times \sigma_0(pp\to S)\times BR_0(S\to aa)\times BR_0(a\to\gamma\gamma)BR_0(a\to N\bar{N})\times \epsilon^{cut}_0\times\epsilon^{eff}_0 \nonumber\\
 N_S &=& N_{S_0}\times \frac{\sigma(pp\to S)}{\sigma_0(pp\to S)}\times \frac{BR(S\to aa)}{BR_0(S\to aa)}\times \frac{BR(a\to\gamma\gamma)BR(a\to N\bar{N})}{BR_0(a\to\gamma\gamma)BR_0(a\to N\bar{N})}\times \frac{\epsilon^{cut}\times\epsilon^{eff}}{\epsilon^{cut}_0\times\epsilon^{eff}_0},
\end{eqnarray}
where the variables denoted with a zero subscript are computed with the benchmark parameters and the other variables are computed for any other choice of parameters. The factors $\epsilon^{cut}$ denotes the cut efficiency as a function of the $S$ scalar, while $\epsilon^{eff}$ embodies other efficiency factors like the photon detection efficiency evaluated by \texttt{Delphes3}. 
The sensitivity of the LHC to the signal will depend on the parameters of the model through the branching ratios and the production cross section of the $S$ scalar which can be rescaled from the their benchmark values. The masses of the particles impact the cut efficiency and, as we discussed above, is affected only by $m_S$. 

The simulated events were generated requiring to pass the following basic cuts
\begin{eqnarray}
\label{basic-cuts}
p_{T}^{\gamma} > 165 \ \hbox{GeV}, \ |\eta_{\gamma}| < 2.5, 
\end{eqnarray}
where $p_{T}^{\gamma}$ and $\eta_{\gamma}$ denote the transverse momentum and pseudorapidity of the hardest photon of the event, respectively. In the case of the $W\to e+\nu_e$ background, we required $p_T^e > 165$ GeV and $|\eta_e| < 2.5$. In the subsequent analysis, the electron from this background is treated as a photon as discussed next.

\begin{table}[t!]
\centering
\begin{tabular}{ c|c|c|c } 
\hline
 Process &  Cuts Eq.~\eqref{basic-cuts2} & $\epsilon_{cut}\times\epsilon_{eff}$($\not\not\!\! E_T > 170\; \hbox{GeV}$) & $\epsilon_{cut}\times\epsilon_{eff}$($\not\not\!\! E_T > 500\; \hbox{GeV}$) \\
\hline
\hline
Signal, $m_S=1$ TeV \\
benchmark point & 48.9 & 0.7 & 0.16\\ 
\hline
$Z + \gamma$ &  16.7 & 0.7 & 0.029\\ 
\hline
$W + \gamma$  & 4.4 & 0.25 & 0.0046\\ 
\hline
Total\\ leading+subleading & 29.9 & 0.69 & 0.030 \\
\hline 
\hline
\end{tabular}
\caption{The cross sections after cuts of Eq.~\eqref{basic-cuts2}, in fb, for the benchmark signal, the leading backgrounds $Z+\gamma$ and $W+\gamma$, and the total backgrounds including the estimate of subleading contributions. In the rightmost columns we show the overall cut efficiencies for $\not\not\!\! E_T > 170\; \hbox{GeV}$ and $\not\not\!\! E_T > 500\; \hbox{GeV}$ and the other cuts of Eq.~\eqref{basic-cuts2} held fixed. K-factors were applied (see text).}
  \label{tab1}
\end{table}

 Besides the main contributions of $Z\gamma$ and $W\gamma$, the backgrounds might receive various subleading contributions. The most important comes from $W\to e+\nu_e$ where the hard electron is misidentified as a photon in the detector. The experimental study of Ref.~\cite{Sirunyan:2017ewk} depicts all the subleading contributions, many of them are hard to reproduce, for example, spikes in the electromagnetic calorimeter or effects from the beam halo. We choose to rescale our number of backgrounds by the factor $\epsilon_{corr}=1+\frac{\hbox{subleading}}{\hbox{leading}}$, where "subleading" is the total number of subleading backgrounds, and "leading" the number of $Z\gamma$ and $W\gamma$ events, respectively. The subleading contributions amount to around 30\% of the total backgrounds where the electron misidentifcation contribution from $W$ decays account for around half of the subleading contributions for the basic cuts of Eq.~\eqref{basic-cuts}. 
 
 The additional cuts shown right below, still following Ref.~\citep{Sirunyan:2017ewk}, are imposed to clean up background events further 
\begin{eqnarray}
\label{basic-cuts2}
& & \not\not\!\! E_T > 170\; \hbox{GeV}, \ p_{T}^{\gamma} > 175\; \hbox{GeV}, \ |\eta_\gamma|<1.44,\nonumber \\ 
& &\Delta \phi (\vec{\not\not\!\! p}_T, \vec{p}_T^{j}) < 0.5,\; \Delta \phi (\vec{\not\not\!\! p}_T, \vec{p}_T^{\gamma}) > 2.0,
\end{eqnarray}
 where $\Delta \phi (\vec{\not\not\!\! p}_T, \vec{p}_T)$ is the angle between the missing momentum vector $\vec{\not\not\!\! p}_T$ and the momentum vector $\vec{p}_T^\gamma$ of the photon and $\vec{p}_T^j$ of the leading jet in the transverse plane. Apart from the missing $E_T$, all the other cuts are identical to those from  Ref.~\cite{Sirunyan:2017ewk}. The hard $\not\!\! E_T$ and photon transverse momentum cuts favor our resonant ALP pair production and decay to a monophoton state. Moreover, such a hard photon has a high trigger efficiency. The $\Delta \phi (\vec{\not\not\!\! p}_T, \vec{p}_T^{j}) < 0.5$ requirement rejects events were the missing energy is produced by mismeasured jets. Finally, $\Delta \phi (\vec{\not\not\!\! p}_T, \vec{p}_T^{\gamma}) > 2.0$ selects events were the missing particles and the photon are produced back-to-back. This later selection cut effectively rejects events with additional hard jets, that is the reason why we expect our LO simulations would not be affected by extra hard jets.

The cut efficiencies and the cross section of production of the signal and backgrounds before and after cuts of Eq.~\eqref{basic-cuts2} are shown in the Table~\ref{tab1} for $\not\not\!\! E_T > 170\; \hbox{GeV}$ and $\not\not\!\! E_T > 500\; \hbox{GeV}$. The backgrounds are efficiently eliminated by hard missing transverse momentum cuts while the signals maintain a good cut efficiency especially if the $S$ scalar is heavy. 

The backgrounds normalizations after the cuts of Eq.~\eqref{basic-cuts2} were taken from the ATLAS simulation of Ref.~\cite{Sirunyan:2017ewk}. Our simulations for the $Z+\gamma$ and $W+\gamma$ backgrounds were used to calculate the correction factors for the cut efficiency in terms of harder $\not\not\!\! E_T$ thresholds keeping all the other cuts of Eq.~\eqref{basic-cuts2}. We checked that our normalizations after cuts, with the NLO QCD K-factors of 1.46~\cite{Alwall:2014hca} and 2.41~\cite{Alwall:2014hca} for the $Z+\gamma$ and $W+\gamma$, respectively, are 0.7 and 2.7 times the ATLAS normalizations, respectively. The $W\to e+\nu_e$ background is severely affected by hard missing transverse energy cuts, nevertheless we assume a conservative approach concerning the subleading backgrounds and just rescale the main backgrounds by the factor $\epsilon_{corr}$ discussed in the previous paragraph once it is hard to evaluate the impact of cuts on the other subleading backgrounds.

Beside the main backgrounds, the signals were rescaled by a conservative K-factor of 2~\cite{Catani:2003zt}, irrespective of the
heavy scalar mass.

\subsection{Discussion of the results}

\begin{figure}[t!]
\includegraphics[scale=0.5]{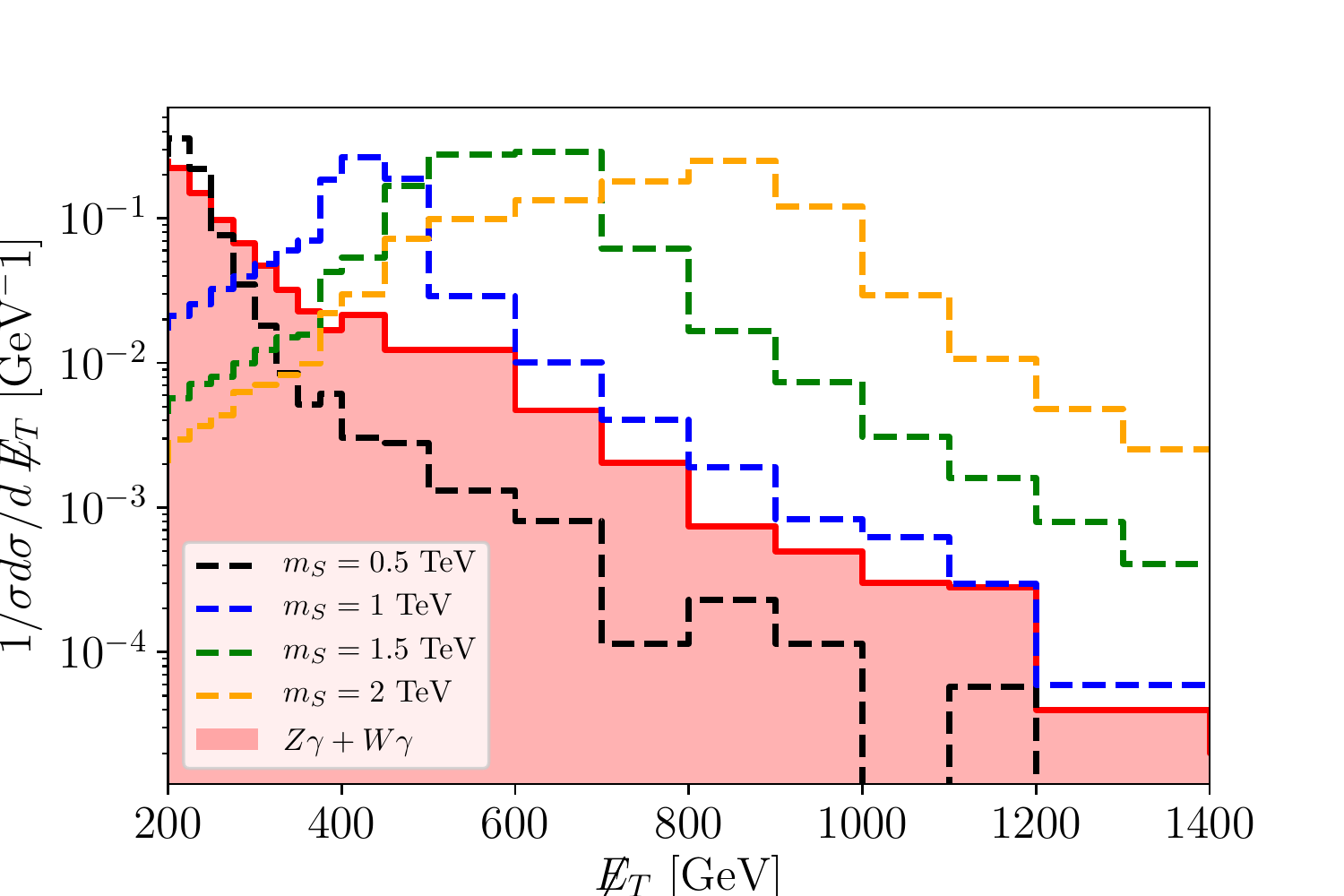}
\includegraphics[scale=0.5]{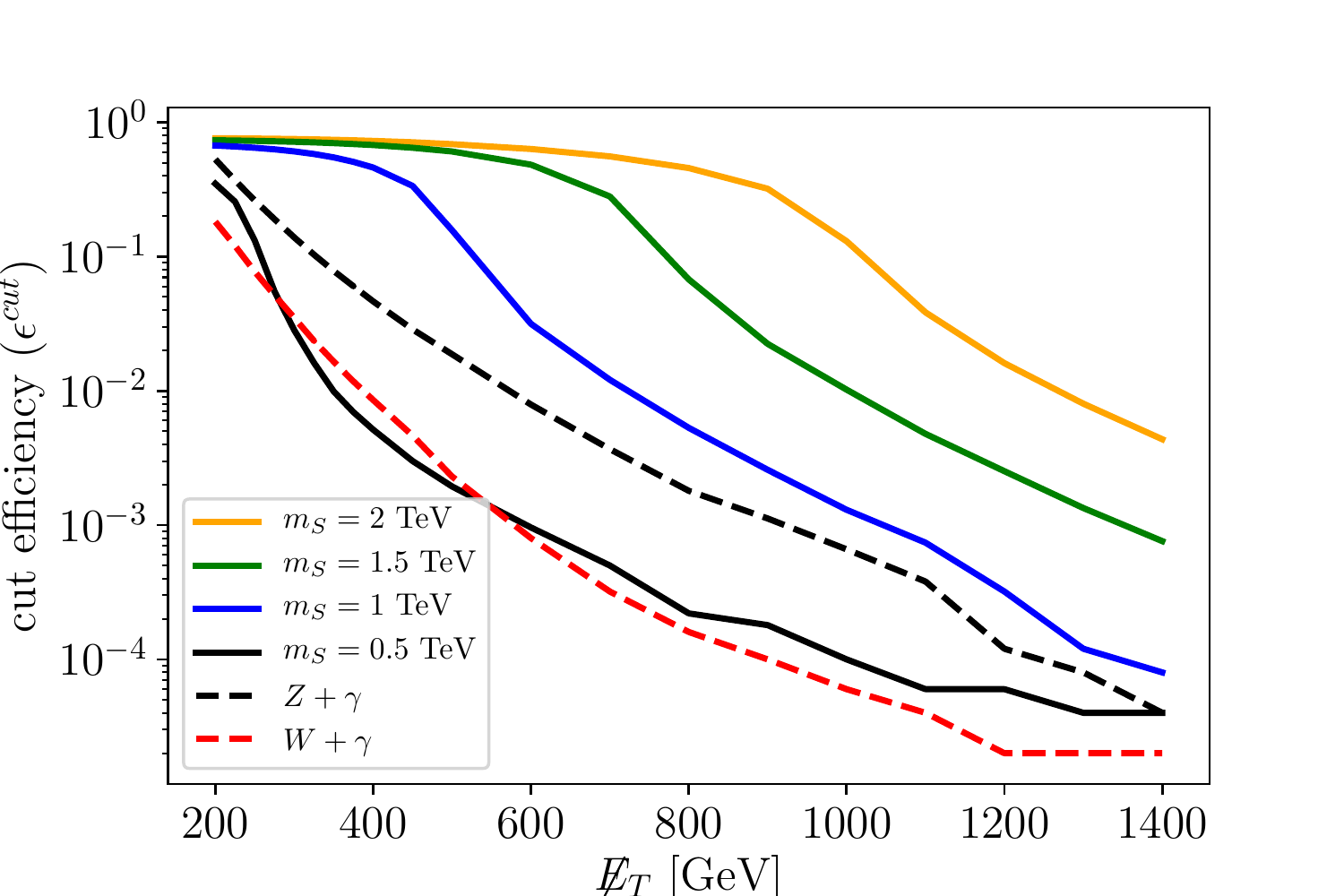}
\caption{The normalized $\not\not\!\!\! E_T$ distribution for signals and the main backgrounds are displayed in the left panel. In the right panel, we show the cut efficiency in terms of the $\not\not\!\! E_T$ threshold.}
\label{fig:cut_eff}
\end{figure}

The missing transverse energy is the most effective variable to eliminate these backgrounds. We scanned the $\not\not\!\!\! E_T$ cut from 200 up to 1500 GeV in order to investigate the best cut. The $\not\not\!\!\! E_T$ distribution of some representative signals and the main backgrounds and their cut efficiencies are shown in Fig.~\eqref{fig:cut_eff}. We observe that the peak of the signal distributions scale roughly with the half the mass of the $S$ scalar but the backgrounds are concentrated around $\not\not\!\!\! E_T\sim 200$ GeV.

For each point of the parameters space that passed all the constraints, we calculate the signal significance, $N_\sigma$, with a metric that interpolates the gaussian and poissonian regimes at the same time it allows for including systematic effects in the background normalization without overestimating the significance~\cite{LiMa,2008NIMPA.595..480C} fixing a 10\% systematic uncertainty in the total background normalization which seems to be realistic in view of the estimates from Ref.~\cite{Sirunyan:2017ewk}.

The missing $E_T$ cut was chosen in order to maximize the number of points of the parameter space with $N_\sigma\geq 2$, that is it, the maximum number of points that can be probed at the LHC at 95\% CL. We chose to impose 
\begin{equation}
\not\not\!\!\! E_T > 1\; \hbox{TeV}
\end{equation}
on the missing transverse energy of the events to get a more reliable estimate of the remaining backgrounds in the tail of the $\not\not\!\! E_T$ distribution. With this cut, around 26\% of the parameters space that fulfills the constraint requirements have $N_\sigma\geq 2$.

The heat maps shown next display the density of points in selected bidimensional slices of the parameters space with the following extra requirement: $N_\sigma\geq 2$, and eventually with $N_\sigma<2$ or $N_\sigma\geq 5$. 

From these plots we want to understand which regions of the parameters space are likely to be probed at the LHC. In colder regions, the chance that a point of the parameters space can be probed at LHC is higher than in hotter regions. Drawing contour lines where $N_\sigma=2$ would define the regions that could be probed by the LHC searches for monophoton signals. We found more convenient to show the density of points that can be probed though. It is important to notice that encountering a point outside these $2\sigma$ regions means that it is not possible to reach this level of statistical significance with 3000 fb$^{-1}$ but if a given point is found inside a certain 2D-slice of the parameters space it does not guarantee a signal, it is necessary to check if it falls inside all the other probable 2D-slices and then compute its statistical significance in this search channel. Looking at just one particular bidimensional region might be misleading. Yet, once its impossible to display the whole parameters space, we hope that showing 2D-slices of the most important parameters is a good guide for model building and further searches using the EFT approach.

Once we set $c_{ae}=c_{a\mu}$, 78 possible bidimensional slices of the parameters space can be displayed, in principle, but just a subset of those parameters are actually important to our purposes. Concerning the parameters related to the $S$ scalar, $c_{BB}$ and $c_{WW}$ are less important than $m_S$ and $c_{GG}$ which control the production cross section and one of the main decay modes of the scalar $S$, and $f_S$ that controls its branching ratio to ALP pairs. Therefore, we do not consider the $c_{BB}$ and $c_{WW}$ dimensions. In the case of the parameters related to the ALP, we consider all of them but $k_{BB}$ and $k_{WW}$,  which are combined in the new parameter 
\begin{equation}
 k_{\gamma\gamma} = \frac{4(c_w^2k_{BB}+s_w^2k_{WW})}{\Lambda}\; ,
 \label{eq:gaa}
\end{equation}
which is effectively involved in $BR(a\to\gamma\gamma)$ and that can probed directly in many experiments devised to detect axions and ALPs, once we are interested mainly in points with large ALP branching ratios to photons and neutrinos, thus those points where the decays to hadrons are closed. We also left the $c_{ae}$ and $c_{a\mu}$ aside in this analysis once their behavior is obvious -- increasing their values after the leptons channel is open increases the branching ratio to lepton and decrease the number of signal events into $\gamma_{jet}+\not\!\! E_T$. This way, we choose the following 6 independent parameters to be investigated in more detail: the masses $m_a$ and $m_N$, the couplings $k_{\gamma\gamma},\; c_{aN},\; f_S $, and the new physics scale $\Lambda$.

\begin{figure}[t!]
\includegraphics[scale=0.4]{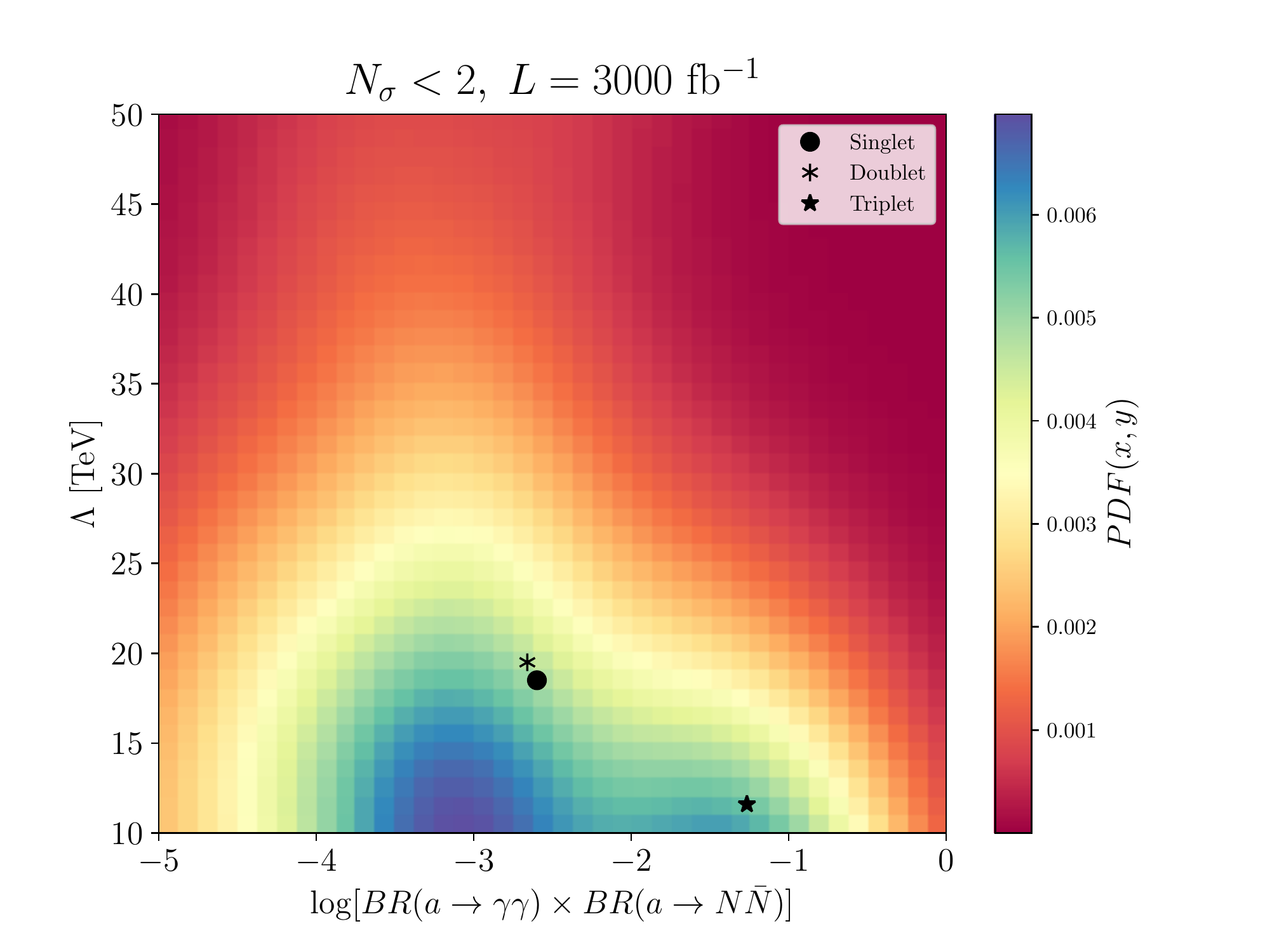}
\includegraphics[scale=0.4]{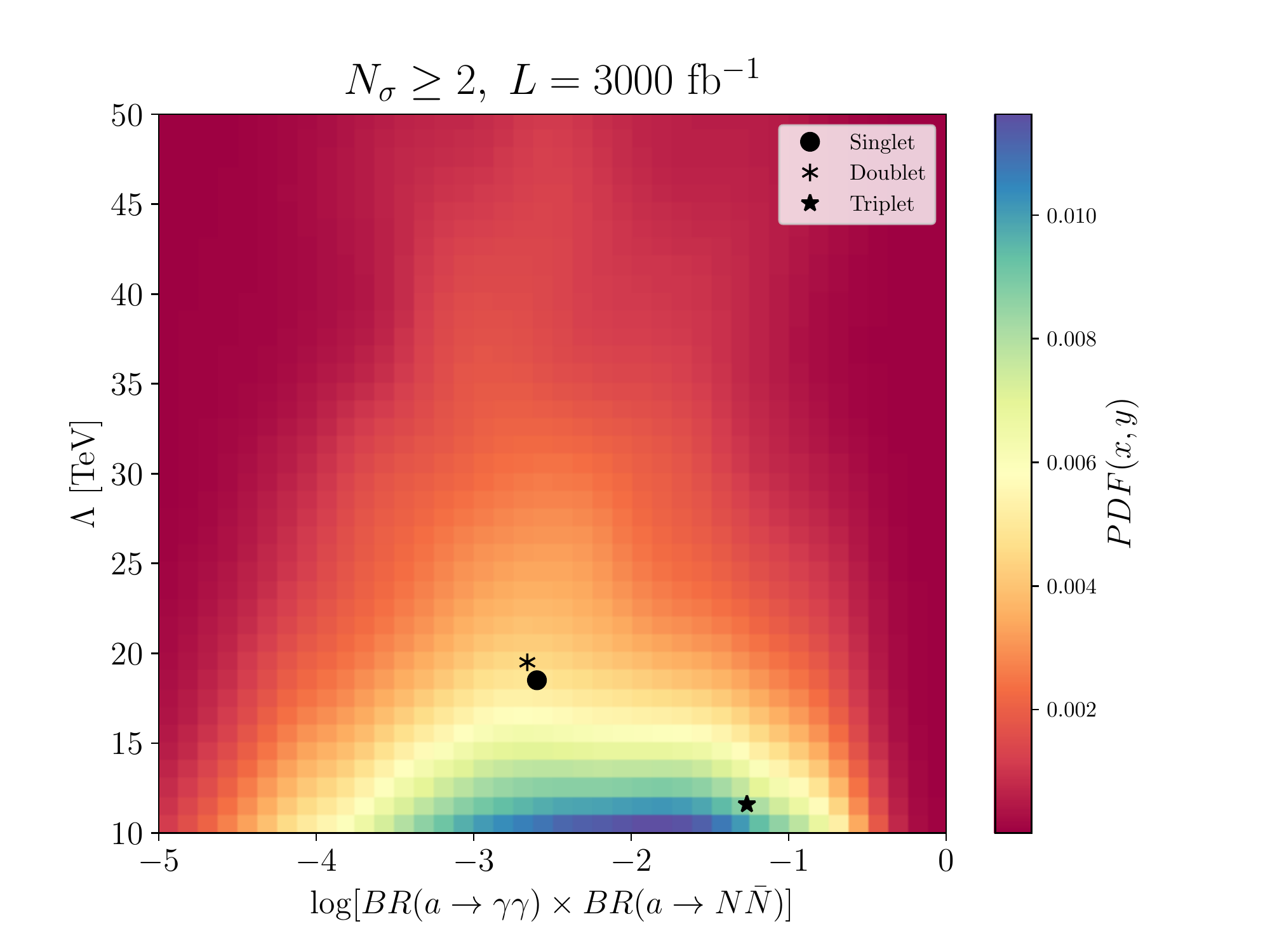}\\
\includegraphics[scale=0.4]{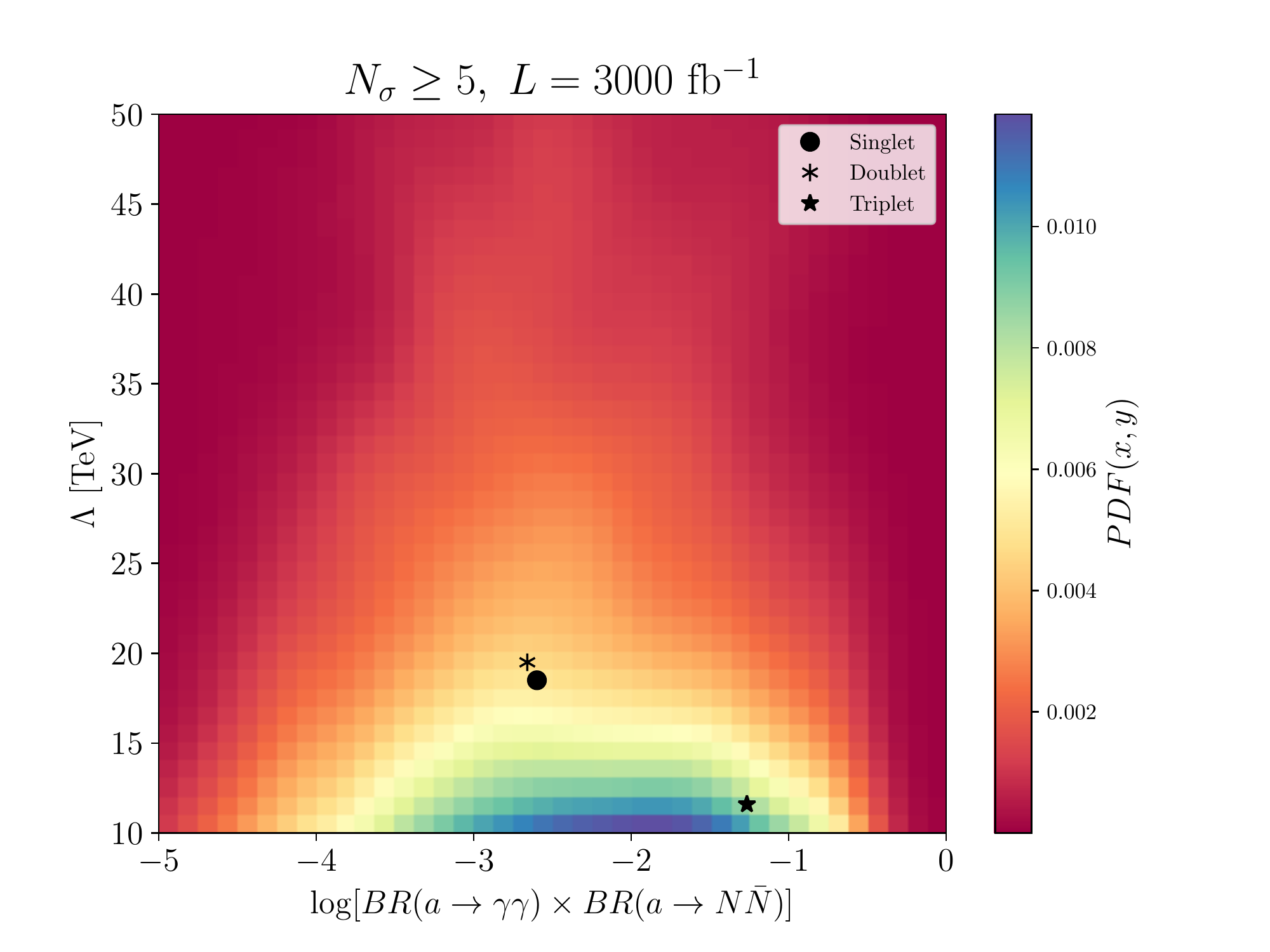}
\includegraphics[scale=0.4]{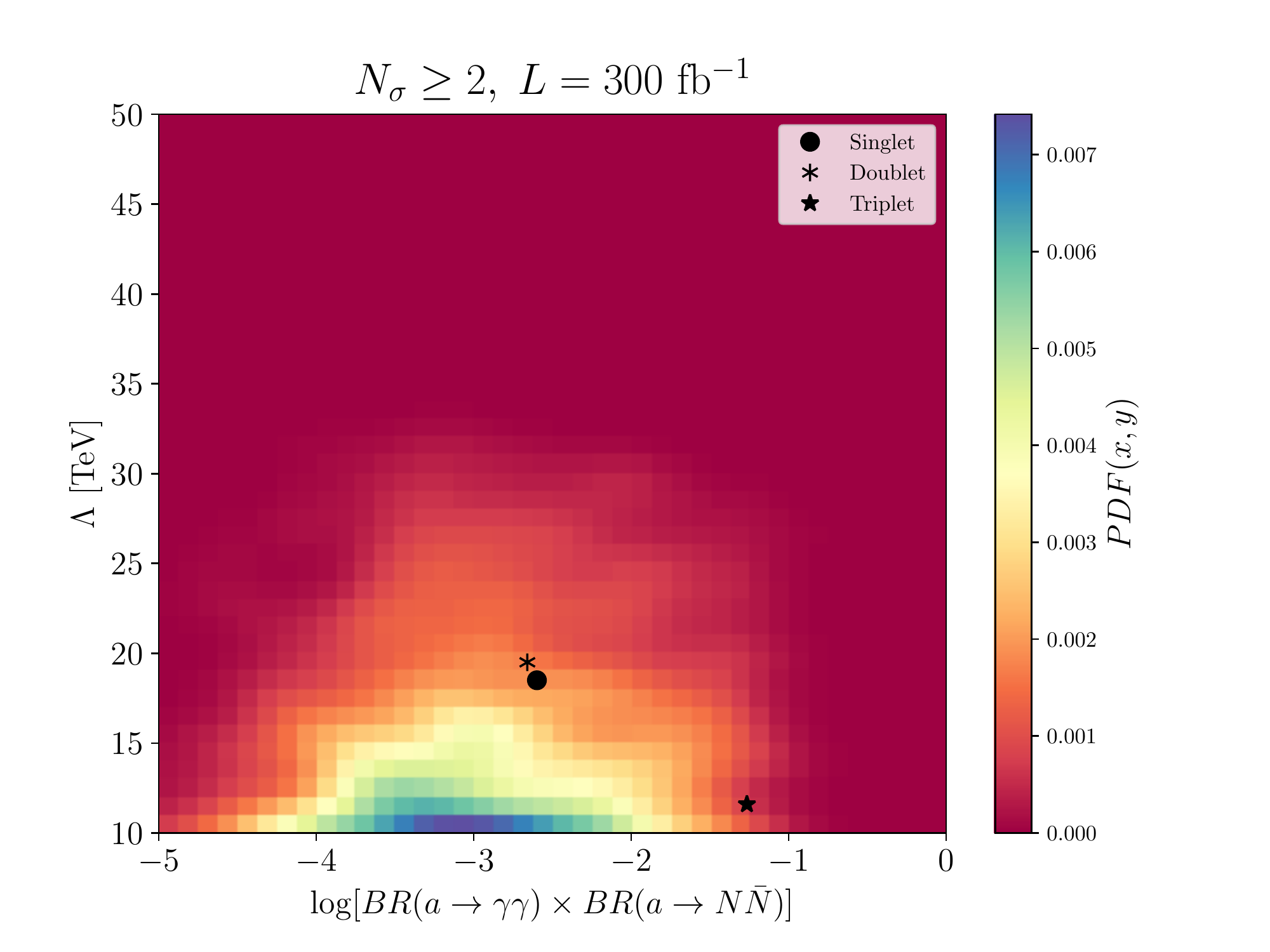}
\caption{Heat maps in the plane $BR(a\to\gamma\gamma)BR(a\to N\bar{N})$ {\it versus} $\Lambda$. The points are more concentrated in the colder regions. We display the cases where $L=3000$ fb$^{-1}$ and $N_\sigma<2(\geq 2)[\geq 5]$ in the upper left(upper right)[lower left] panel, and the case $L=300$ fb$^{-1}$ and $N_\sigma\geq 2$ in the lower right panel. The dot, asterisk and star represent examples of a singlet, doublet and triplet representation of the UV complete model presented in Section~\ref{uvmodel}, respectively.}
\label{fig:brprod-Lam}
\end{figure}
\begin{figure}[t!]
\includegraphics[scale=0.6]{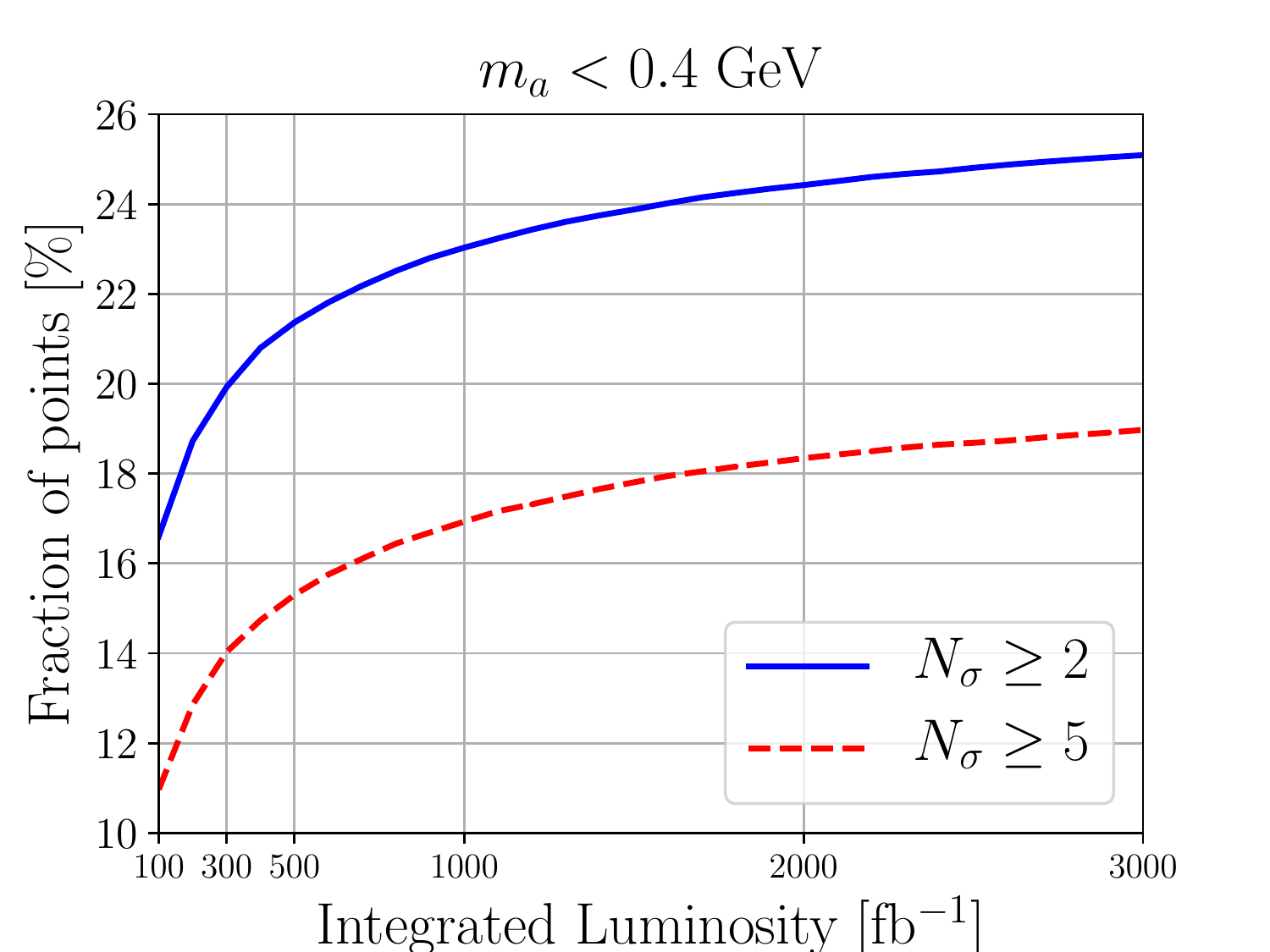}
\caption{The fraction of sampled points of the parameters space that have $N_\sigma\geq 2$ (solid line) and $N_\sigma\geq 5$ (dashed line) at the 13 TeV LHC in the monophoton channel as a function of the integrated luminosity.}
\label{fig:probe}
\end{figure}

Out of the 15 bidimensional slices that can be formed with the six parameters chosen, we display those whose interplay we found more important to be understood. 
Before discussing these plots, however, we show in Fig.~\eqref{fig:brprod-Lam} the points distributed in the $BR(a\to\gamma\gamma)\times BR(a\to N\bar{N})$ {\it versus} $\Lambda$ space. These are two key variables to understand which models can be tested in colliders. In the upper panels, we show the distribution of points with $N_\sigma<2$ at left, that is it, those points that cannot be probed at $\sim 95$\% CL at the LHC, and, at right, the points for which the LHC will present sensitivity with $N_\sigma\geq 2$. We can identify two clearly distinct regions. Points with $N_\sigma<2$ have typically  $BR(a\to\gamma\gamma)\times BR(a\to N\bar{N})<10^{-3}$, while those with $N_\sigma\geq 2$ have $BR(a\to\gamma\gamma)\times BR(a\to N\bar{N})>10^{-3}$ and are more concentrated in the $\Lambda<20$ TeV region. 
In the lower panels we show the prospects for an $N_\sigma\geq 5$ observation, for 3000 fb$^{-1}$ and $N_\sigma\geq 2$ observation after 300 fb$^{-1}$ at the left and right panel, respectively. The densities are qualitatively the same of the $N_\sigma\geq 2$ case but, of course, the number of points with $N_\sigma\geq 5$ is smaller than $N_\sigma\geq 2$. Overall, 26\% of the points have $N_\sigma\geq 2$ for 3000 fb$^{-1}$, while 14\% have $N_\sigma\geq 5$ for the same luminosity. 
We display the reach of the LHC to probe the EFT model for various luminosities in Fig.~\eqref{fig:probe} for ALP masses below the pions threshold. When the hadrons channel are open, the LHC looses sensitivity in the monopohoton channel and the monojet channel should more promising.

\begin{figure}[t!]
\includegraphics[scale=0.4]{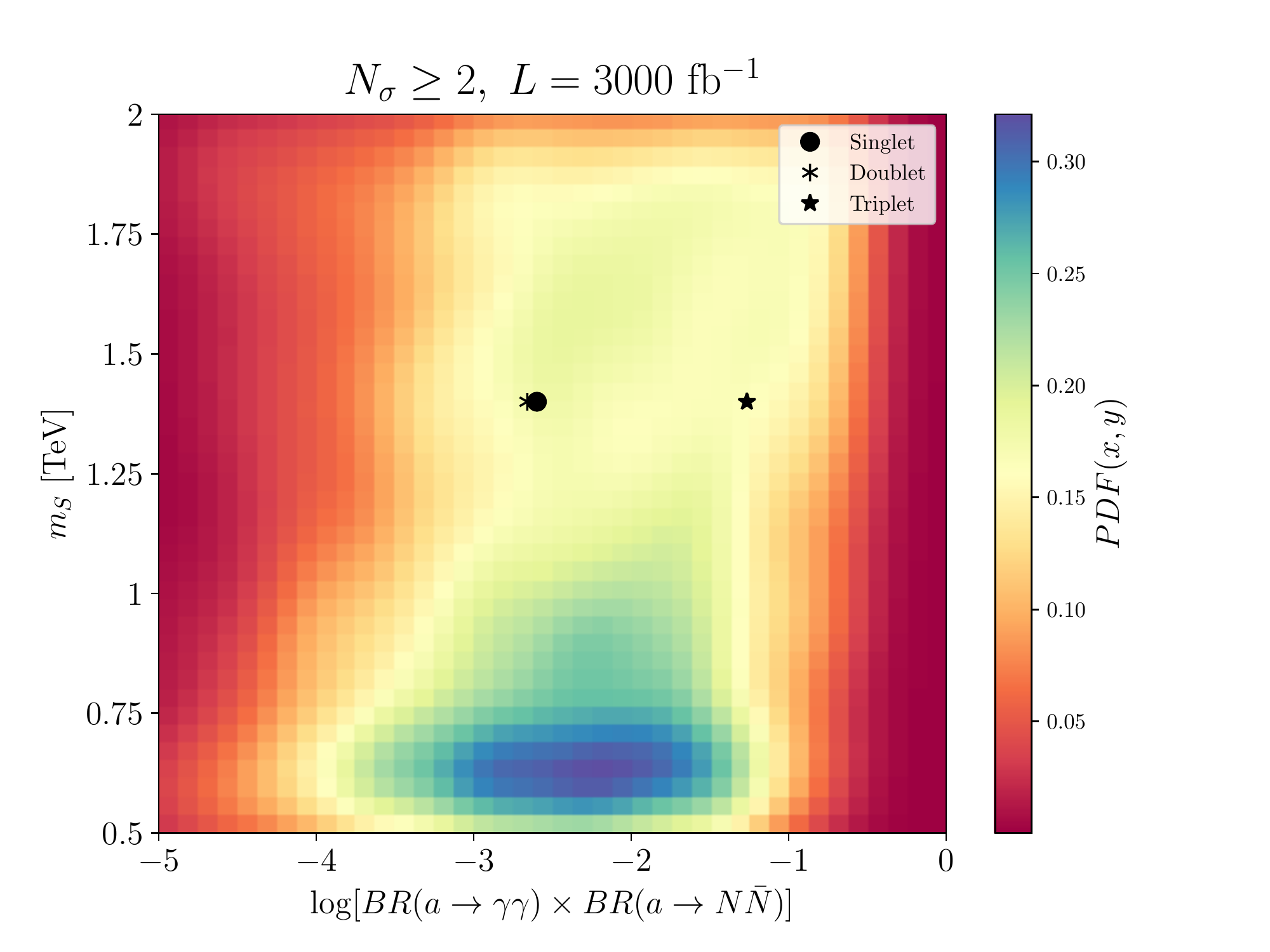}
\includegraphics[scale=0.4]{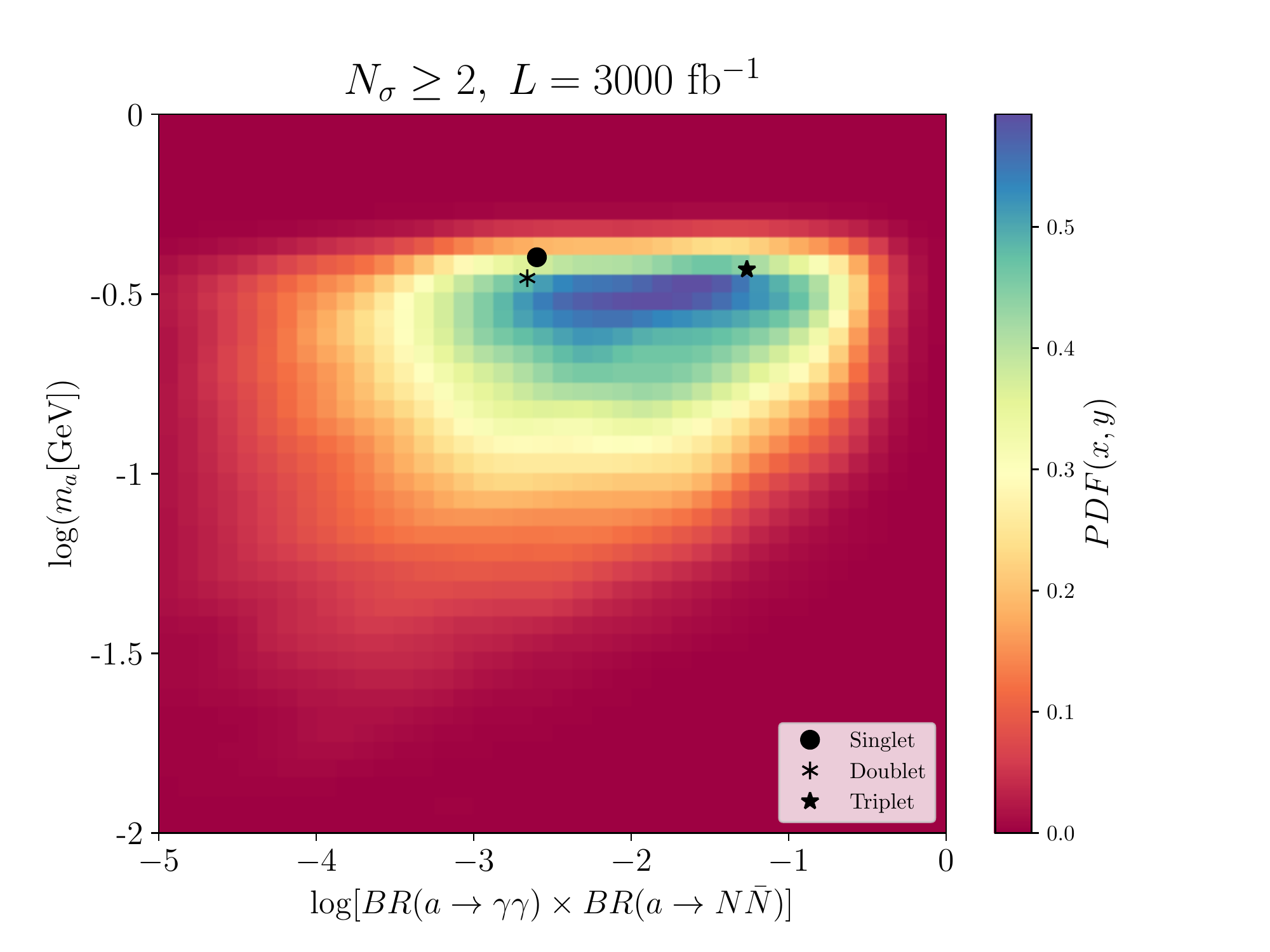}\\
\includegraphics[scale=0.4]{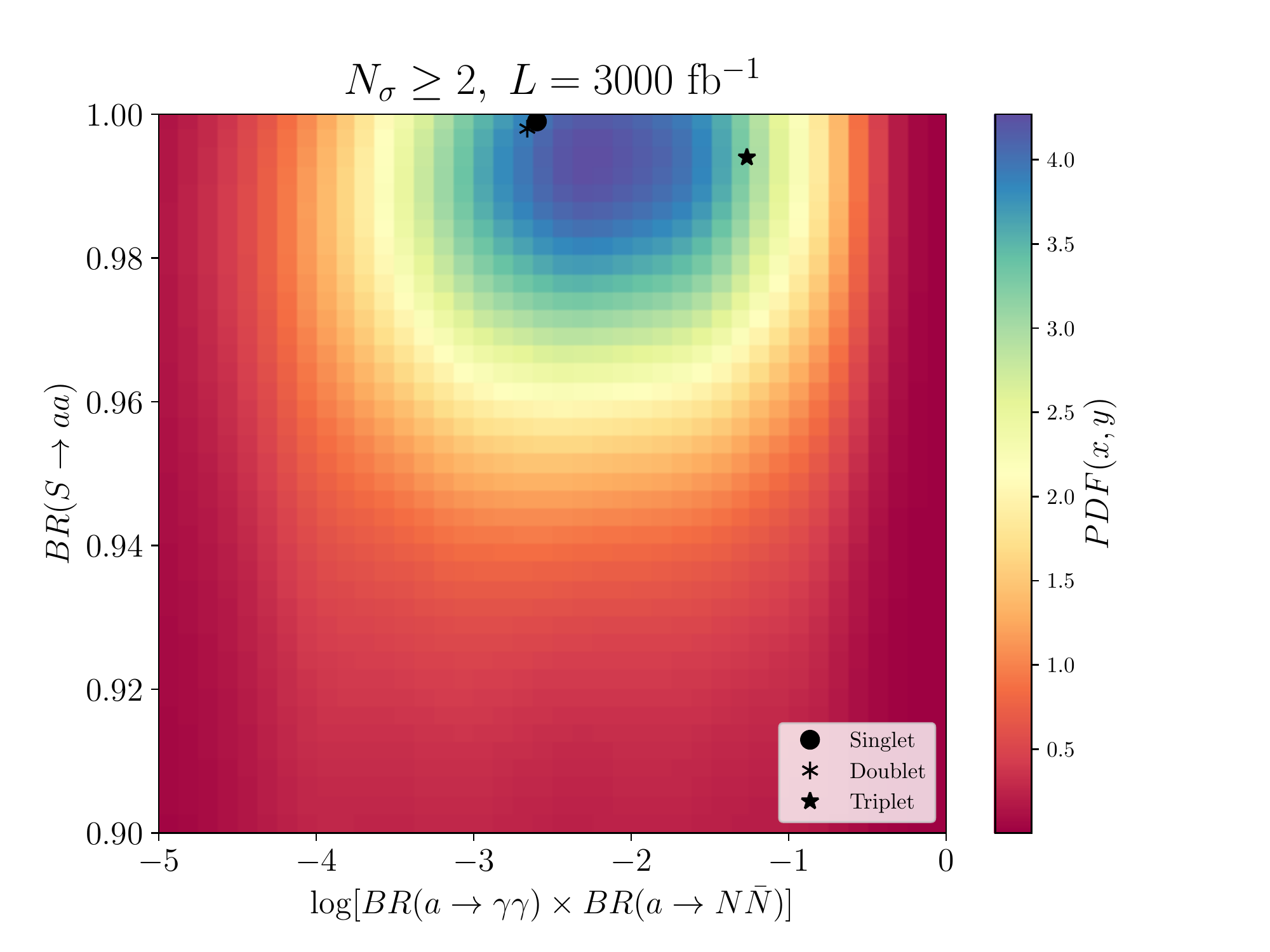}
\includegraphics[scale=0.4]{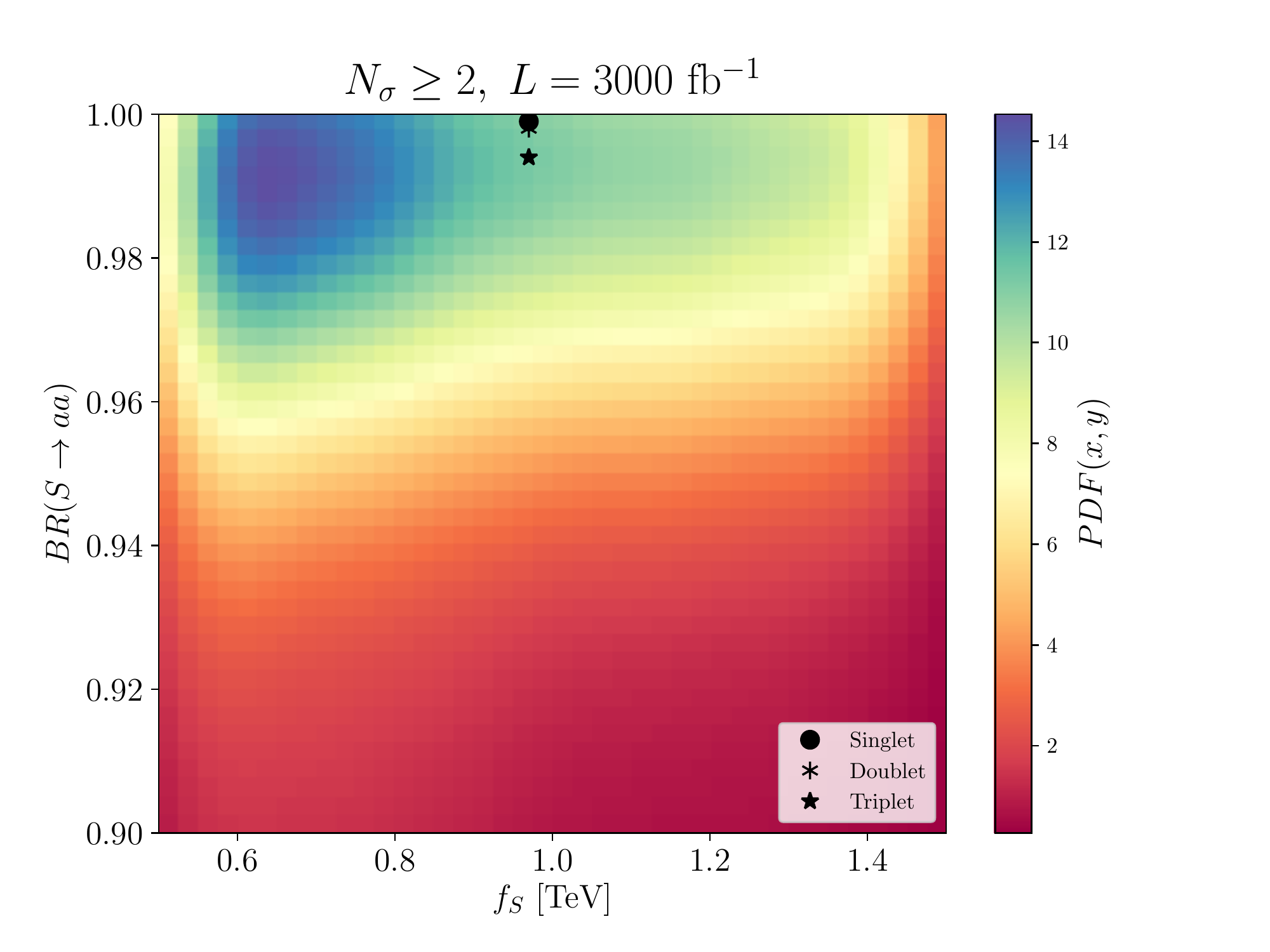}
\caption{
Heat maps in the plane $BR(a\to\gamma\gamma)\times BR(a\to N\bar{N})$ {\it versus} $m_S$ and $m_a$ are shown at the upper left and upper right panels, respectively. In the lower panels we display $BR(S\to aa)$ {\it versus} $BR(a\to\gamma\gamma)\times BR(a\to N\bar{N})$ and $f_S$ at the left and right, respectively. In all cases we show the regions that can be probed at $2\sigma$ and $L=3000$ fb$^{-1}$. The dot, asterisk and star represent examples of a singlet, doublet and triplet representation of the UV complete model presented in Section~\ref{uvmodel}, respectively.}
\label{fig:brs-params}
\end{figure}

Now let us show the correlation between the branching ratios of $S\to aa$, $a\to\gamma\gamma$ and $a\to N\bar{N}$, and some key parameters of the EFT formulation. First of all, in the upper panels of Fig.~\eqref{fig:brs-params} we display the product  $BR(a\to\gamma\gamma)\times BR(a\to N\bar{N})$ and the $S$ mass at left, and the ALP mass at right. The mass of the $S$ scalar is concentrated in the $\sim 500$--$800$ GeV range, the ALP mass above $\sim 100$ MeV and the product of branching ratios not far from 1\%. Note that above $\sim 400$ MeV, the signal in monopohotons fades away.
The preferred value of the product of the branching ratios  of the ALP into photons and sterile neutrinos, by its turn, can be explained in view of the ALP lifetime constraint depicted in Fig.~\eqref{plot-lifetime}. The larger the ALP mass, the easier is avoiding the lifetime constraint but it cannot be smaller than two neutrino masses nor large enough to open decays into pions  and muons in order that the monophoton channel becomes effective. Second, in the lower panels of Fig.~\eqref{fig:brs-params}, we show the correlation between the branching ratio of $S\to aa$  and the product $BR(a\to\gamma\gamma)\times BR(a\to N\bar{N})$, at left, and the parameters $f_S$ at right. This time, we learn that reaching the desired statistical significance is possible in regions where the $S$ decays almost all the time into ALP pairs and, naturally, that is possible mainly in the region with large trilinear $f_S$ couplings but not as large as possible at the expense of having a vanishingly small $c_{GG}$ once, in this case, there would be no $S$ production in gluon fusion.
 
\begin{figure}[t!]
\includegraphics[scale=0.4]{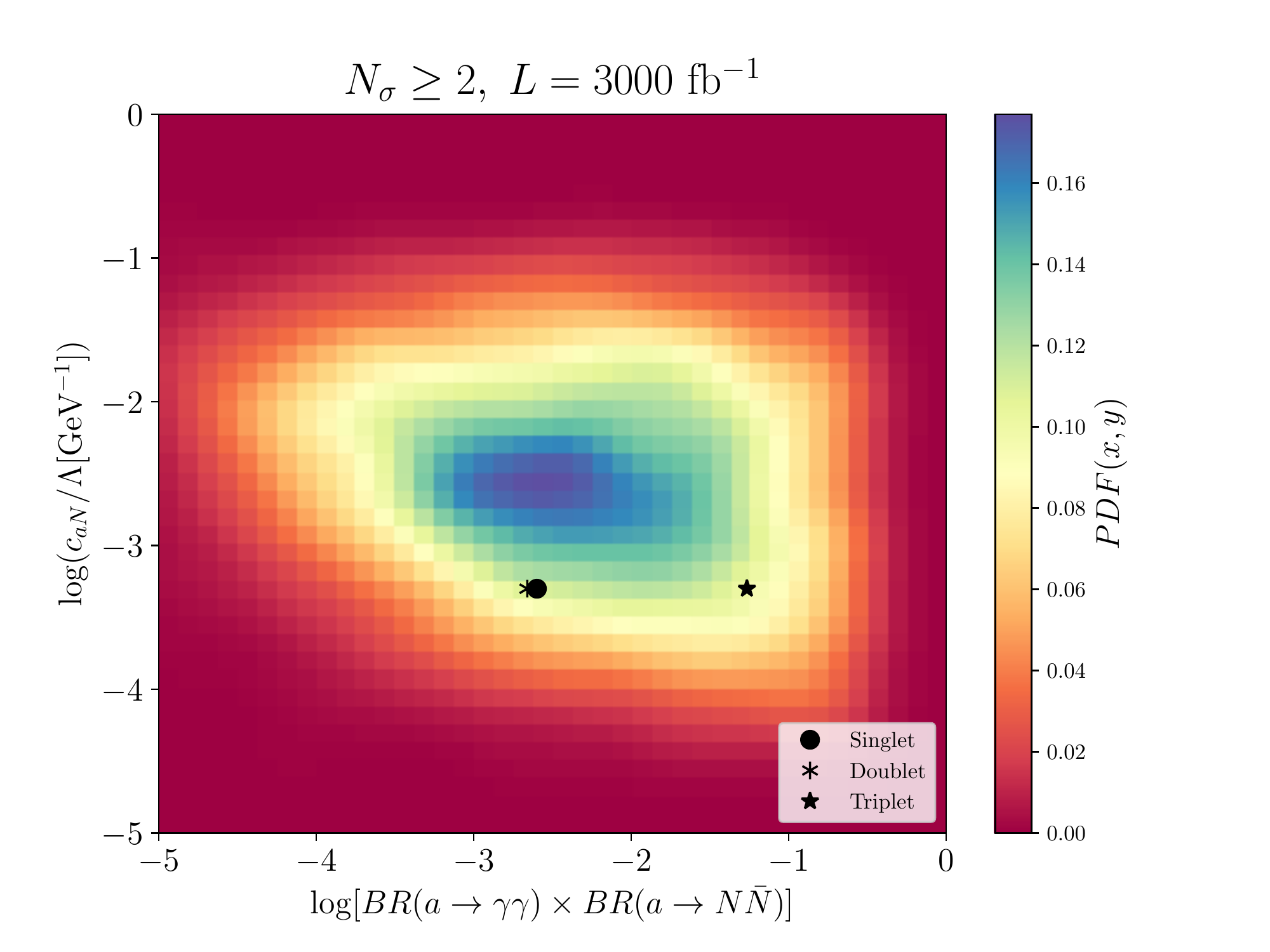}
\includegraphics[scale=0.4]{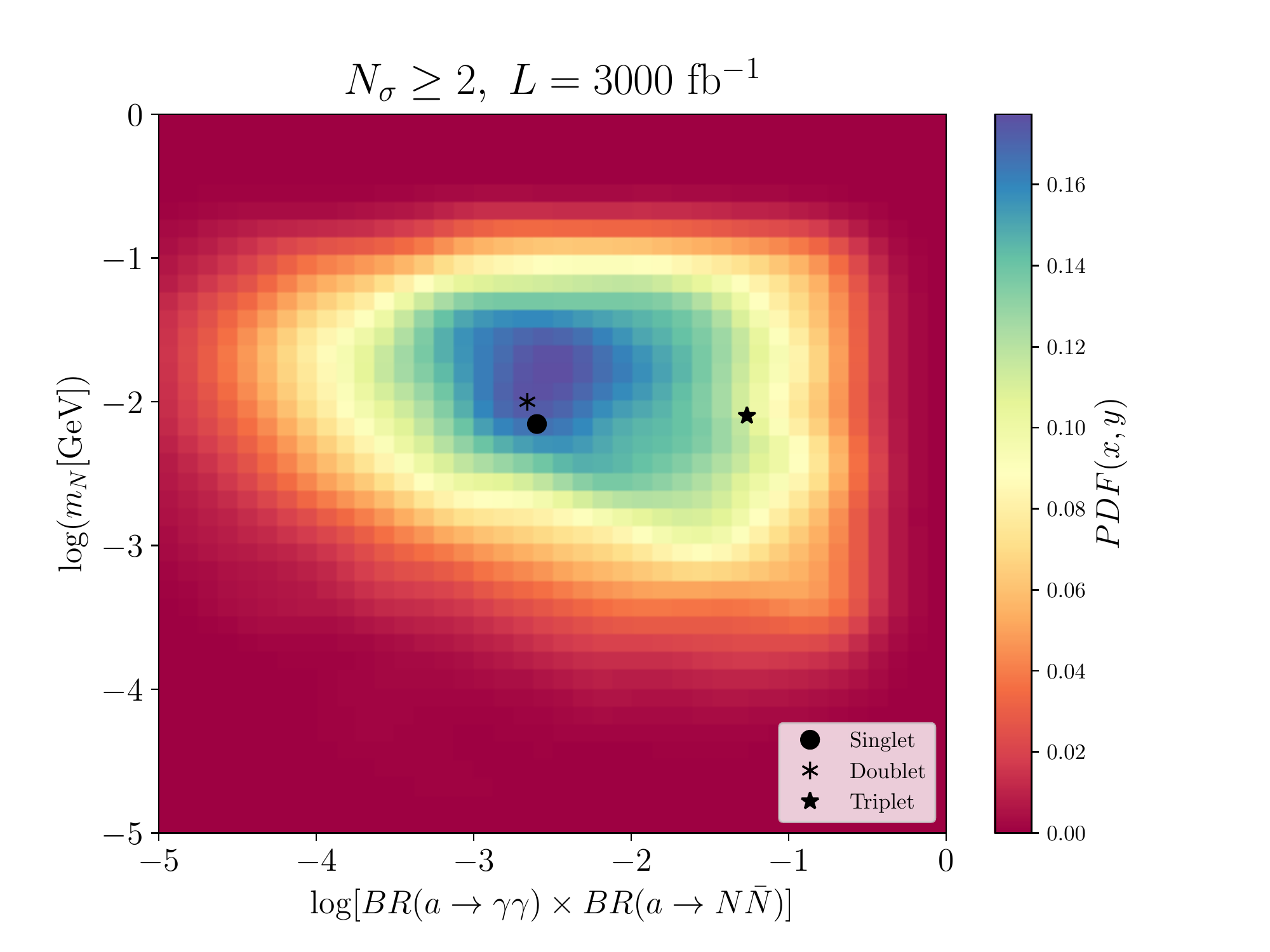}
\caption{Heat maps in the plane $BR(a\to\gamma\gamma)\times BR(a\to N\bar{N})$ {\it versus} $c_{aN}$ and $m_N$ are shown at the left and right panels, respectively. In all cases we show the regions that can be probed at $2\sigma$ and $L=3000$ fb$^{-1}$. The dot, asterisk and star represent examples of a singlet, doublet and triplet representation of the UV complete model presented in Section~\ref{uvmodel}, respectively.}
\label{fig:neu-params}
\end{figure}

Next, we investigate the role played by the parameters connected to the sterile neutrino, $c_{aN}$ and $m_N$. At the left panel of Fig.~\eqref{fig:neu-params}, we display the density of points with $N_\sigma\geq 2$ in the plane  $BR(a\to\gamma\gamma)\times BR(a\to N\bar{N})$ against $c_{aN}$, and against $m_N$ at the right panel, respectively. 
We have seen that the bulk of points that can probed at the LHC and pass all the constraints have $BR(a\to\gamma\gamma)\times BR(a\to N\bar{N})\sim 10^{-4}$--$10^{-1}$. Because $BR(a\to N\bar{N})\propto c_{aN}^2$, a large $c_{aN}$ is limited by the monophoton constraint of Eq.~\eqref{eq:monoph}. The neutrinos masses likely to be probed at the LHC in the monophoton channel is in the region $1$--$100$ MeV.

Finally, we discuss the effective coupling ALP-photon, $k_{\gamma\gamma}$, that can be probed by several experiments as discussed in the Introduction. The model considered here presents an extended spectrum compared to the axion and ALP models where these particles interact just with the photon or the SM gauge bosons, however, apart from collider searches, we do not expect that prospects from other types of experiments will get considerably impacted. We therefore assume that bounds and prospects from any experiment probing just the ALP-photon coupling applies to our case. For collider searches, on the other hand, we should expect that the future projections of reach and sensitivity get shrunk with the addition of new decay channels that compete with the photon channel.

\begin{figure}[t!]
\includegraphics[scale=0.4]{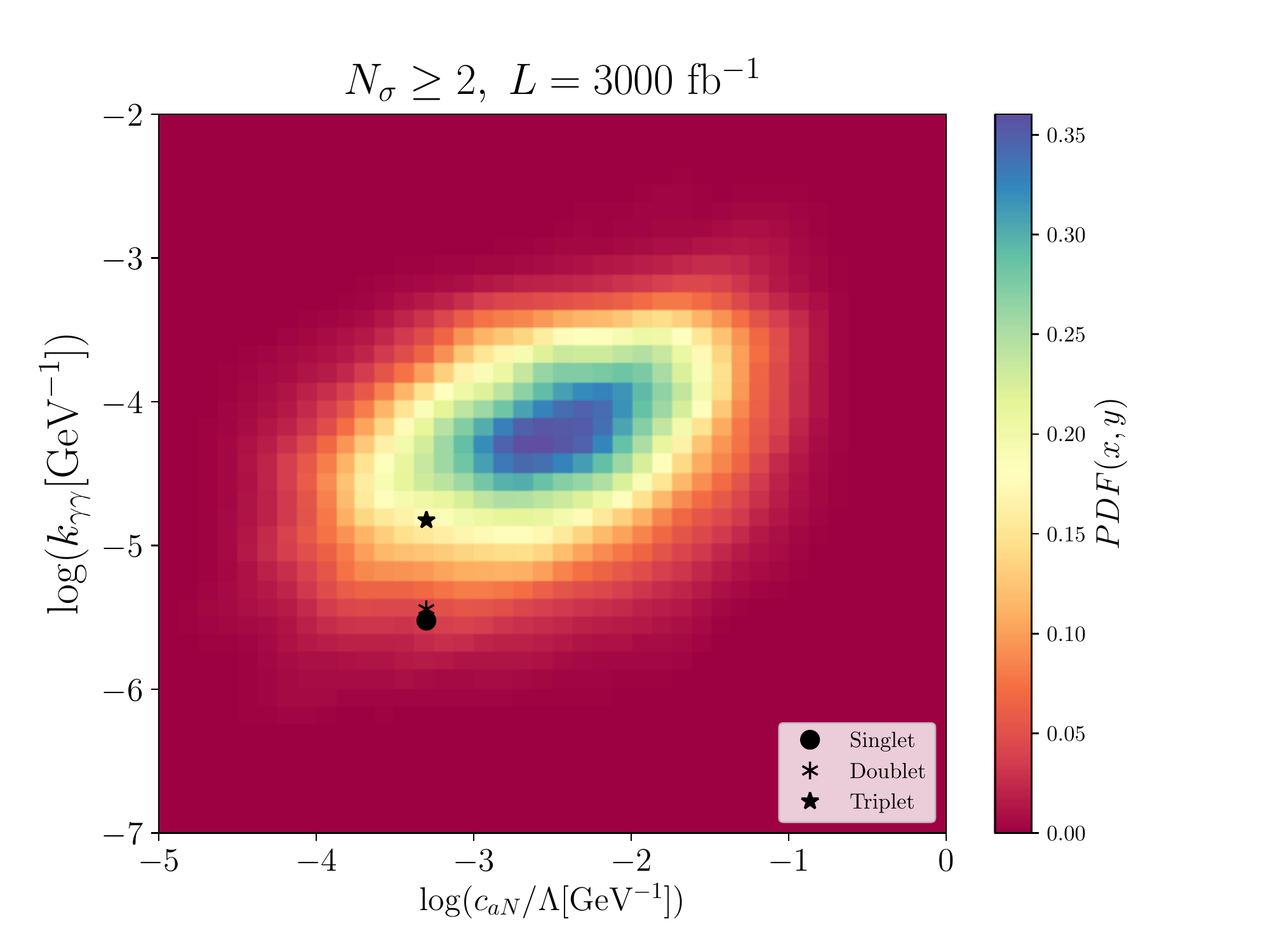}
\includegraphics[scale=0.4]{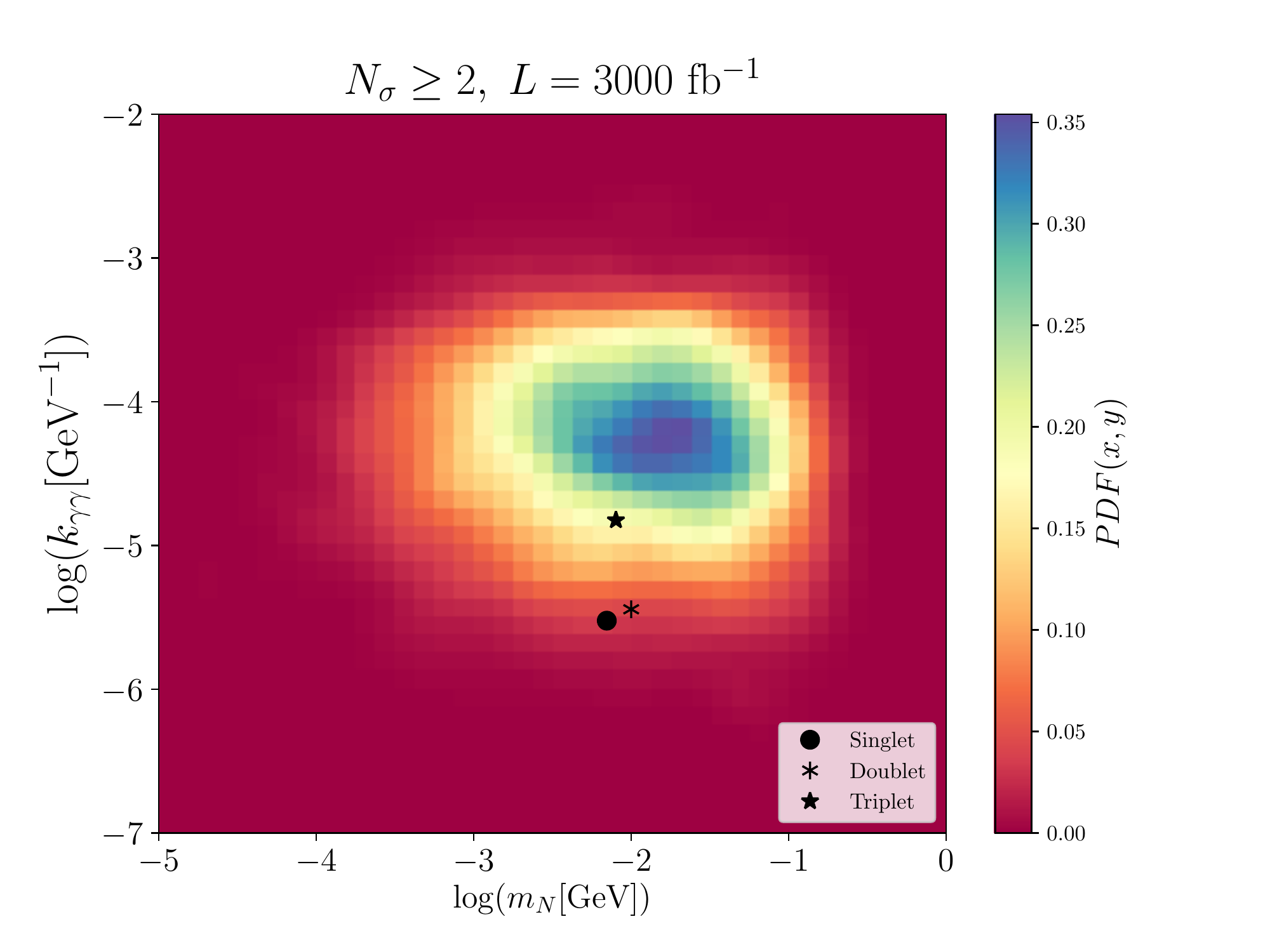}\\
\includegraphics[scale=0.4]{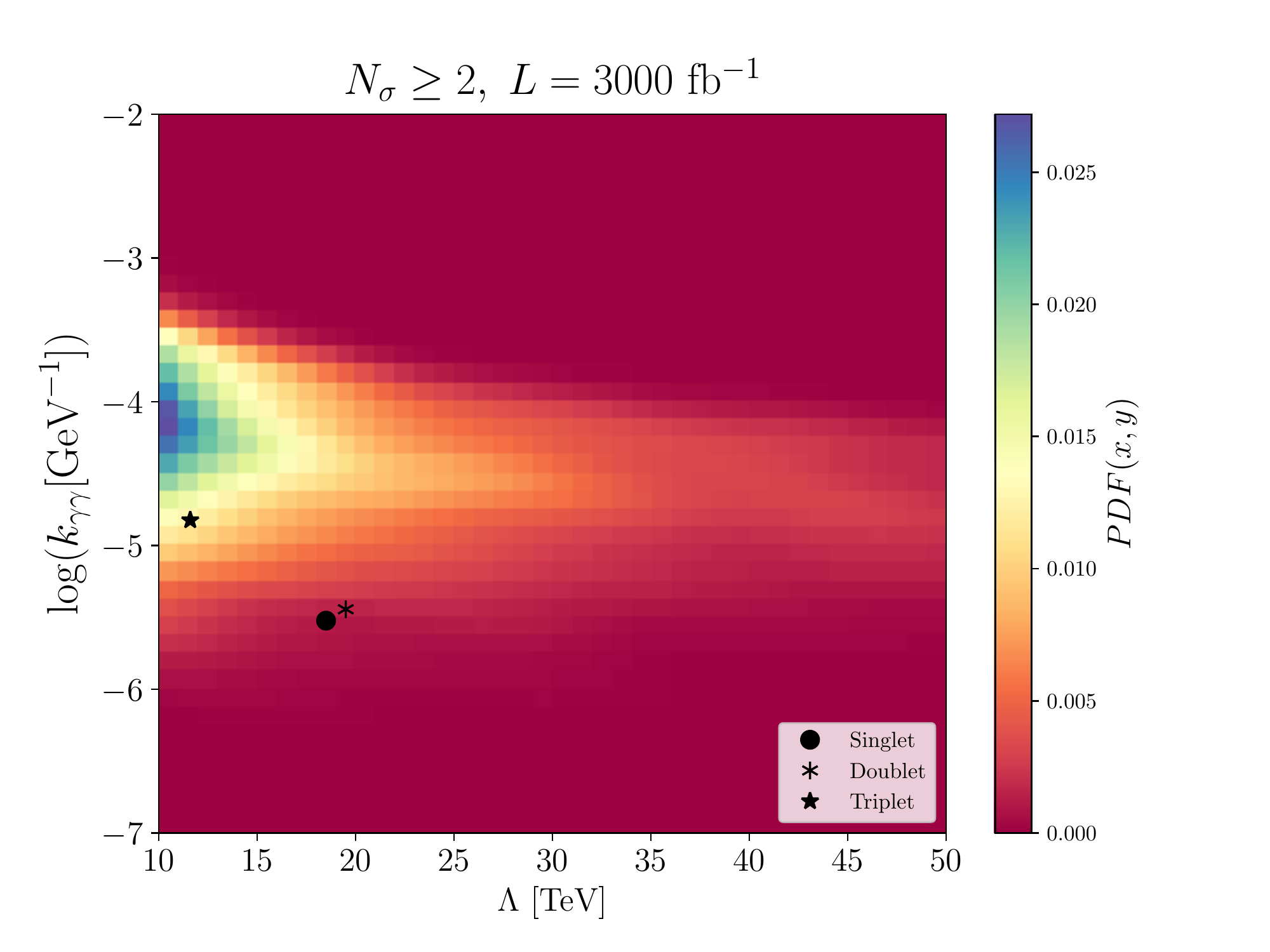}
\includegraphics[scale=0.4]{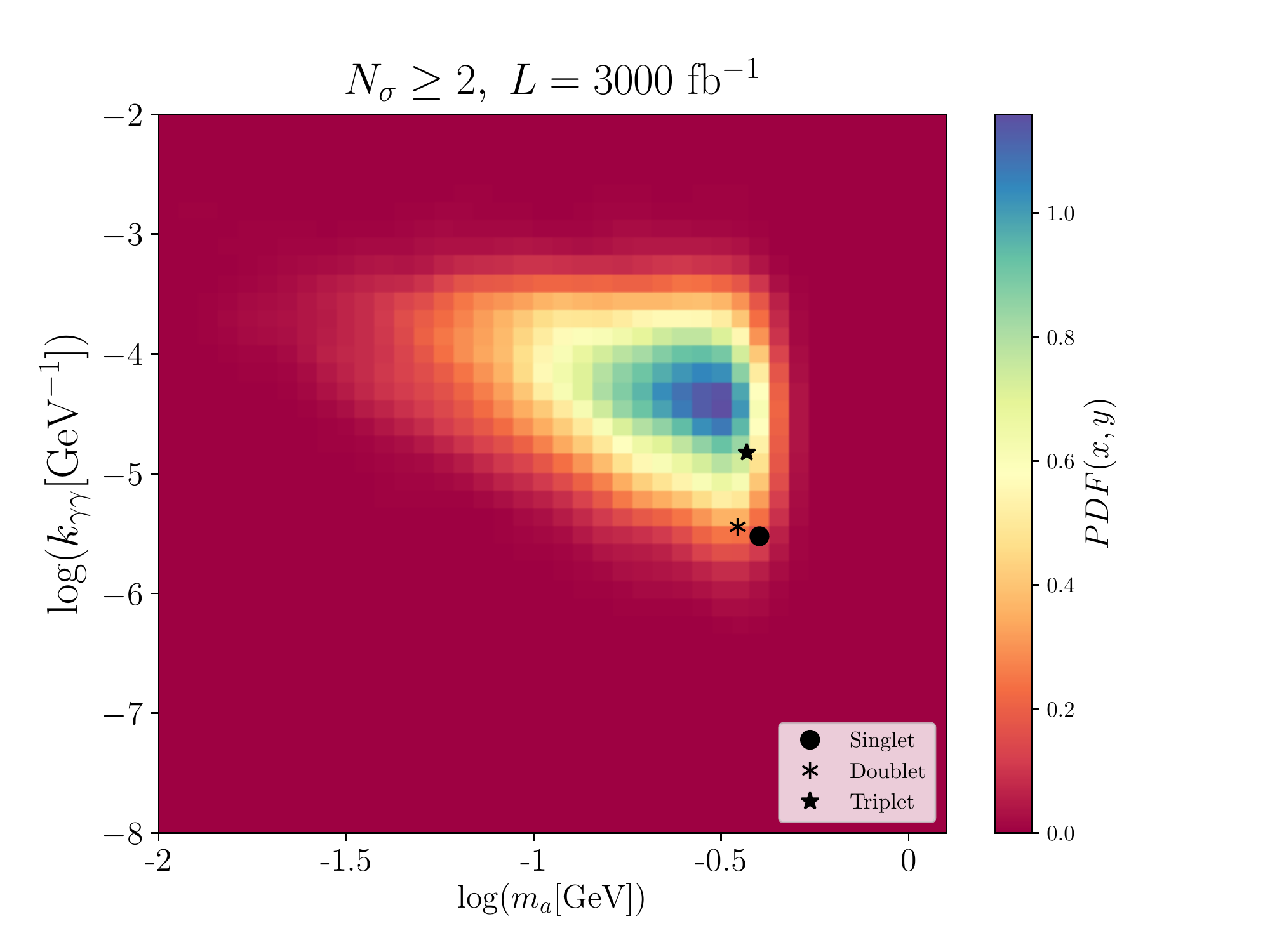}
\caption{Correlation between $k_{\gamma\gamma}$ and $c_{aN}$, $m_N$, $\Lambda$ and $m_a$ are shown at the upper left, upper right, lower left and lower right panel, respectively. For all plots, we impose $N_\sigma>2$ at the HL-LHC. The dot, asterisk and star represent examples of a singlet, doublet and triplet representation of the UV complete model presented in Section~\ref{uvmodel}, respectively.}
\label{fig:gaa-params}
\end{figure}

In the upper panels of Fig.~\eqref{fig:gaa-params}, the heat maps of $k_{\gamma\gamma}$ {\it versus} $c_{aN}$ and $m_N$ are shown at the left and right panels, respectively. ALP-photons couplings from $10^{-6}$ to $10^{-3}$ GeV$^{-1}$ will be probed by the LHC in monophoton searches. 

The lower panels show the $k_{\gamma\gamma}$ correlations against $\Lambda$ and the ALP mass at left and at right, respectively. Concerning the new physics scale $\Lambda$, we see that the majority of points have $\Lambda\lesssim 30$ TeV and, again,  $k_{\gamma\gamma}$ in the interval $10^{-6}$--$10^{-3}$ GeV$^{-1}$. Yet, we can observe points with very large $\Lambda$ when $k_{\gamma\gamma}\sim 10^{-5}$--$10^{-4}$ GeV$^{-1}$ and even smaller. When $\Lambda$ gets large, $BR(S\to aa)$ saturates to 100\% but it does not affect the branching ratio of the ALPs, then this tail towards large $\Lambda$ is the branching ratio of $S\to aa$ compensating for the drop in $BR(a\to \gamma\gamma)$.

The lower right panel Fig.~\eqref{fig:gaa-params} is of particular importance. These are two parameters that use to be directly constrained in many experiments hunting for axions and ALPs. In this particular EFT model with an sterile neutrino, the bulk of points that can be probed at the LHC at 95\% CL lies in a more or less triangular region as we can see in the plot. Roughly speaking, these points are concentrated in the region $m_a\times k_{\gamma\gamma}\in [0.1,0.4]\hbox{GeV}\times [10^{-6},10^{-3}]$ GeV$^{-1}$.

\begin{figure}[t!]
\includegraphics[scale=0.165]{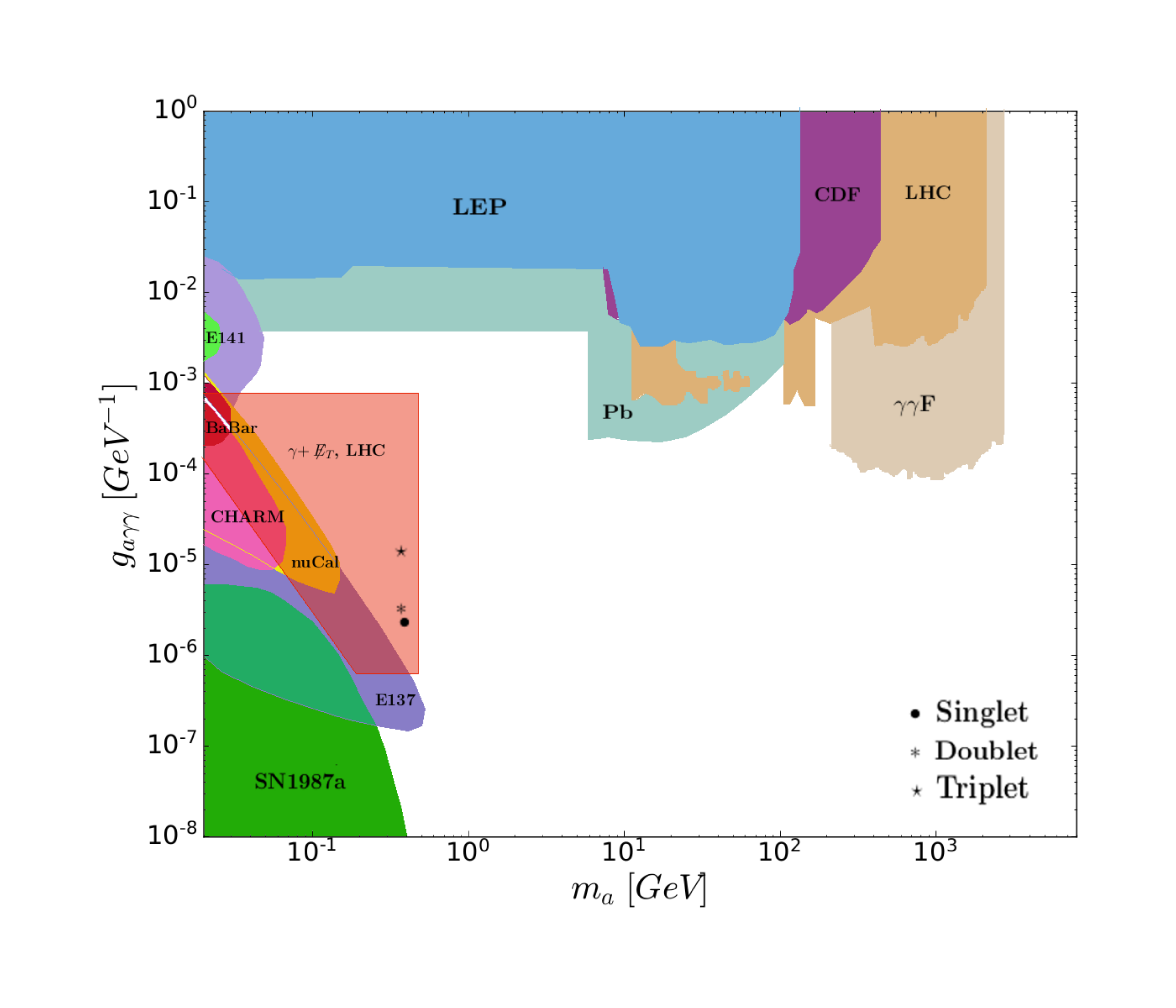}
\includegraphics[scale=0.165]{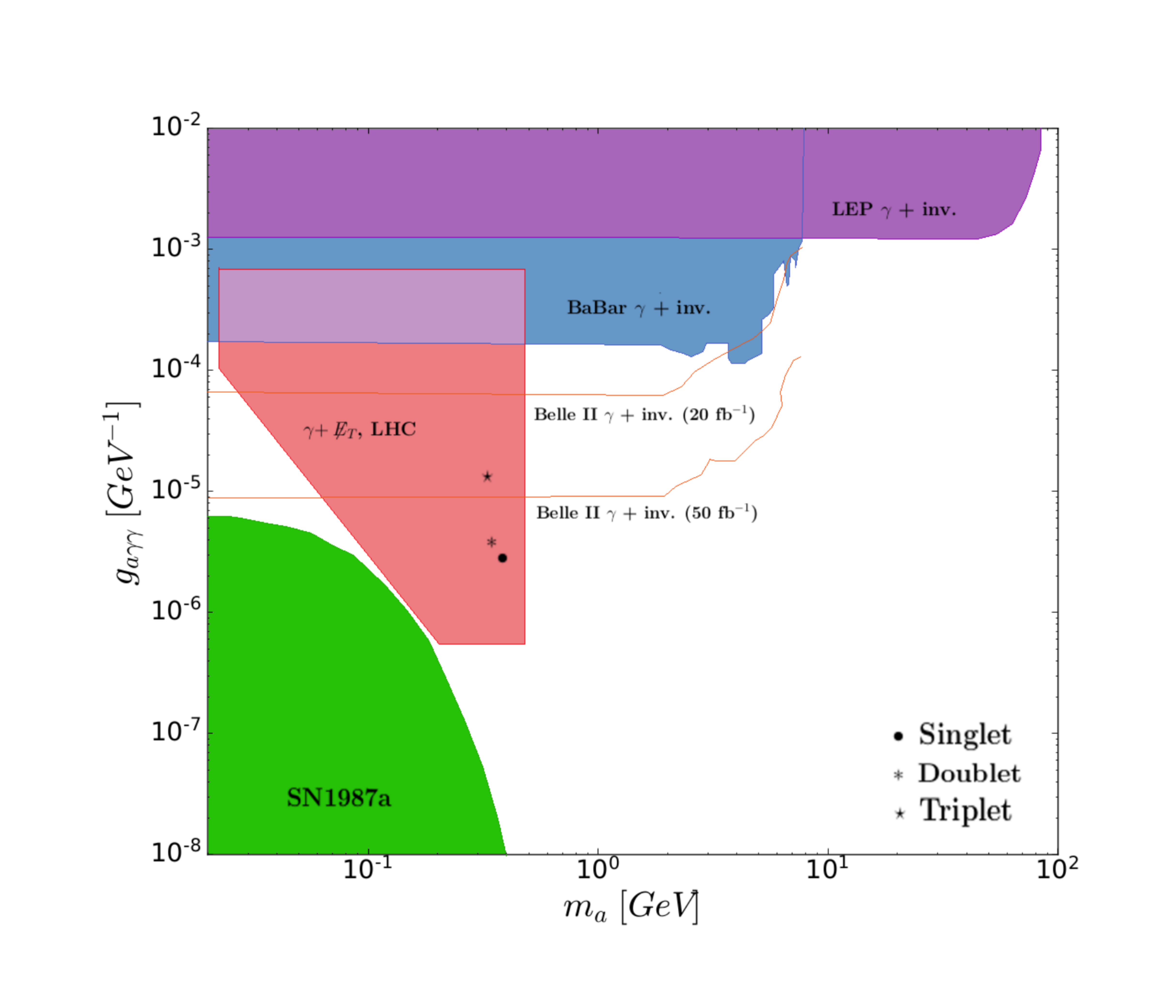}
\caption{ Comparison of the $\gamma+\not\not\!\! E_T$ search of this work, the red shaded area, at 95\% CL, against several experiments searching for ALPs coupling to photons \cite{Mimasu:2014nea,Jaeckel:2015jla,Bauer:2017ris,Brivio:2017ije,Aaltonen:2013mfa,Dolan:2017osp,Cadamuro:2011fd,Dobrich:2015jyk,Jaeckel:2017tud,Knapen:2017ebd}. The limits at the left panel are taken from Refs.~\cite{Dobrich:2019dxc} and \cite{Bauer:2018uxu}. The prospects of the collider experiments will shrink with the addition of the sterile neutrino decay channel but the LHC one obtained in this work. The green shaded area represents the constraints from supernova data. The points represent benchmarks of the model presented in Section~\ref{uvmodel}. Comparison of the LHC against LEP, Belle and BaBar prospects~\cite{Dolan:2017osp} to probe the ALP-photon coupling are shown at the  right panel where we also added the SN 1987 limit from Ref.~\cite{Dobrich:2019dxc}.}
\label{fig:money-plot1}
\end{figure}

\begin{figure}[t!]
\includegraphics[scale=0.23]{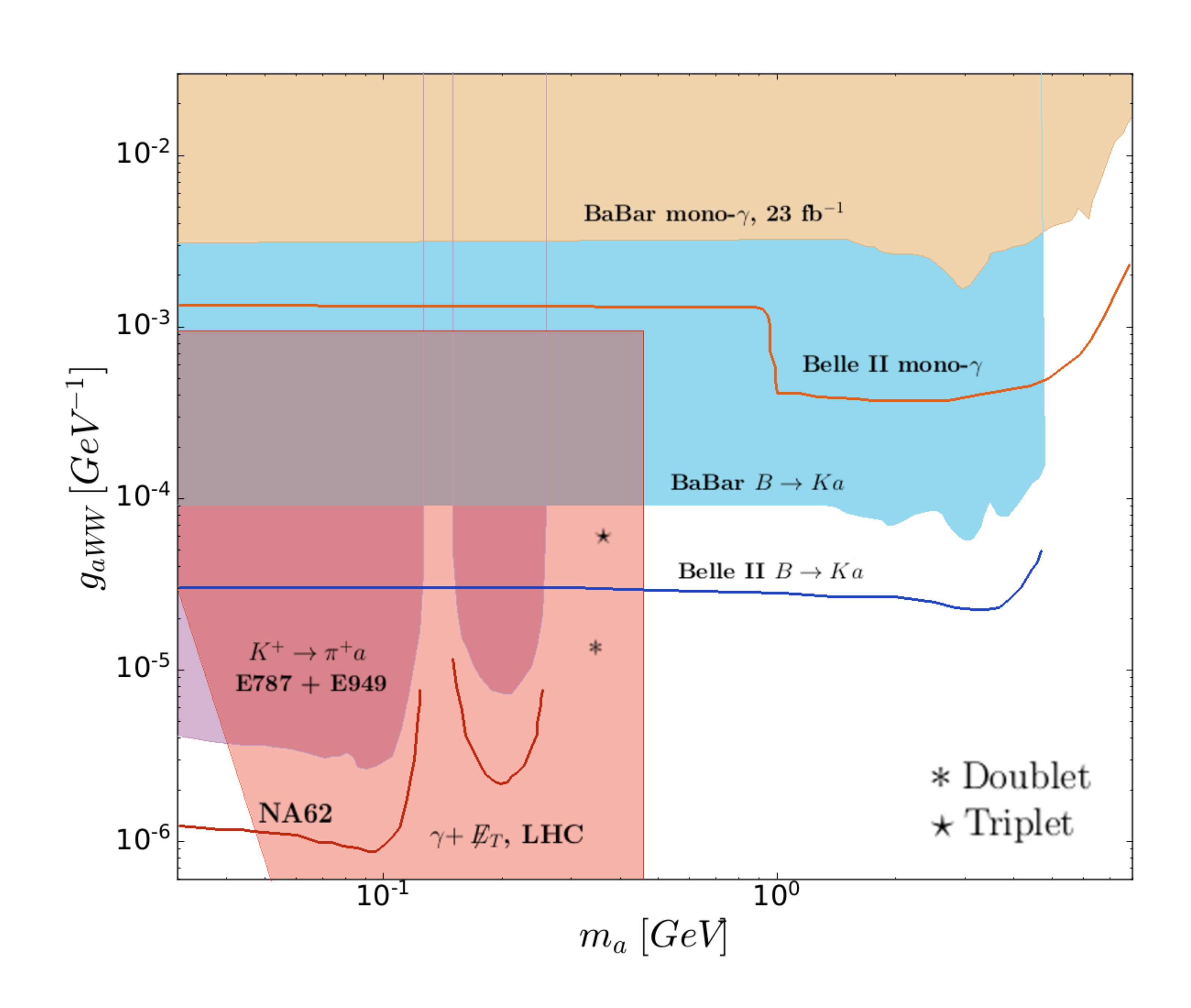}
\includegraphics[scale=0.38]{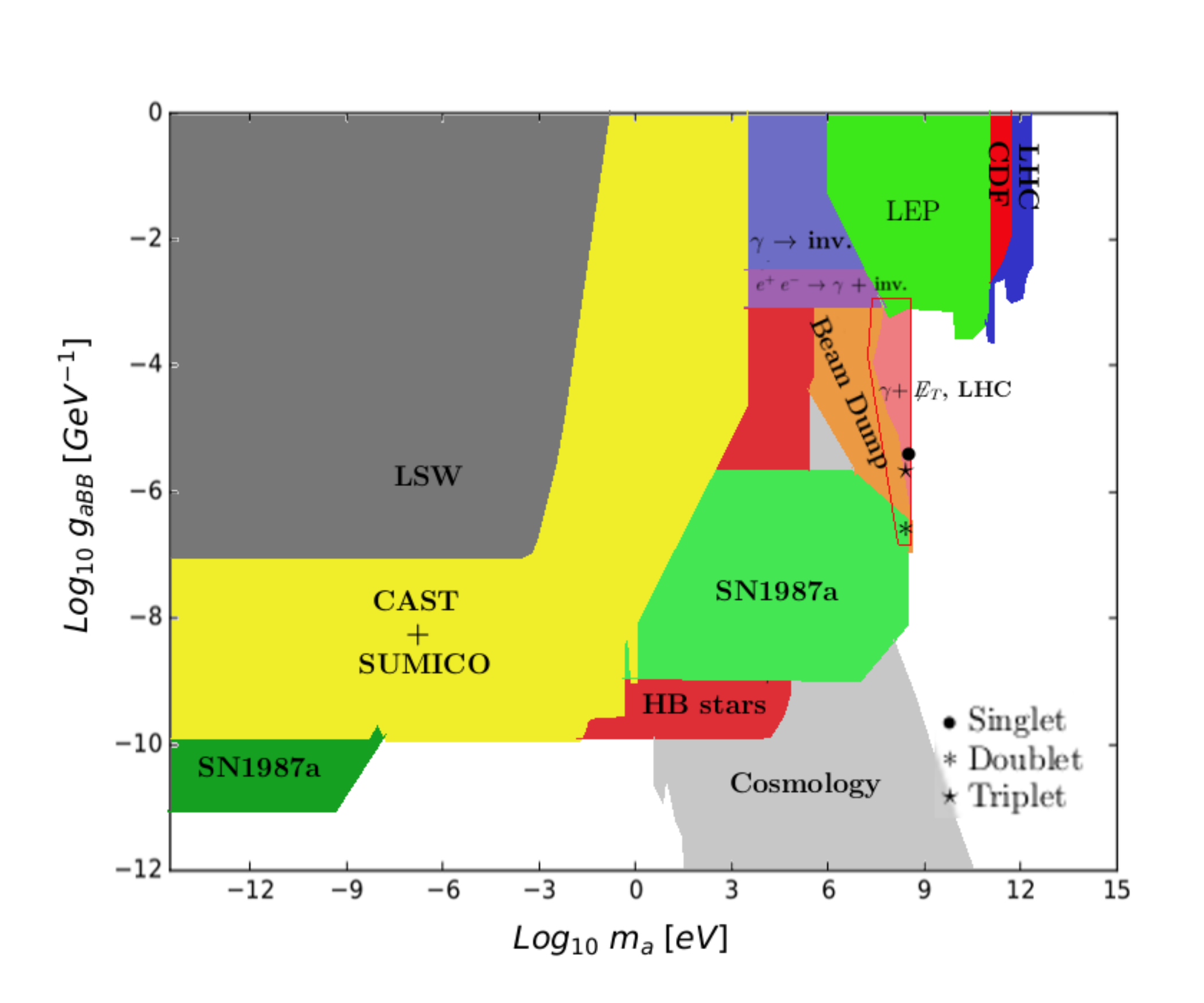}
\caption{ In the left panel \cite{Abe:2006qx,Aubert:2007hh,BABAR:2011aaa,Duh:2012ie,Artamonov:2005ru,Ceccucci:2014oza,Abouzaid:2008xm,Lees:2013kla,Artamonov:2009sz}, $g_{aW}$ is the ALP-coupling assuming $k_{BB}=0$ with the limits taken from~ Ref.~\cite{Izaguirre:2016dfi}. The points represent benchmarks of the model presented in Section~\ref{uvmodel}. The panel at right \cite{Jaeckel:2010ni,Alekhin:2015byh,Redondo:2008en,Cadamuro:2011fd,Hewett:2012ns,Jaeckel:2012yz,Mimasu:2014nea,Payez:2014xsa,Millea:2015qra} displays the $g_{aB}$ coupling of the ALP where $k_{WW}=0$ and is taken from Ref.~\cite{Jaeckel:2012yz}. In these plots, the regions that can be probed by the monopohoton channel, at 95\% CL, are contained in the light red shaded areas. The points still represent the benchmark points of the UV model presented in the next section.}
\label{fig:money-plot2}
\end{figure}

In order to compare the prospects of the HL-LHC to probe the ALP-sterile neutrino interaction with other experiments devised to investigate the ALP-photon coupling, we overlay our results, the red shaded areas, from the lower right plot of Fig.~\eqref{fig:gaa-params}  into the $m_a\times k_{\gamma\gamma}$ plane displaying several other prospects from other experiments in Figs.~\eqref{fig:money-plot1},  \eqref{fig:money-plot2}. In these plots, we adapted the definition of our $k_{\gamma\gamma}$ to match those of the works where we took the plots from when necessary. Reminding that this region corresponds to a confidence level of 95\% approximately. 

We expect that in the presence of a new decay channel, like into sterile neutrinos, the branching ratio into photons gets smaller. The prospects from other experiments shown in Figs.~\eqref{fig:money-plot1}, \eqref{fig:money-plot2} then are expected to shrink actually. Nevertheless, we think it is very interesting to compare the reach of the LHC with those experiments to demonstrate the usefulness of LHC searches for this model.

Interestingly, the HL-LHC sensitivity to monophoton events from $S\to aa\to \gamma_{jet}+\not\not\!\! E_T$ is complementary to the other experiments, covering a sizeable region of the $m_a\times k_{\gamma\gamma}$ space  not reached yet as we see in Fig.~\eqref{fig:money-plot1}. At the left panel of Fig.~\eqref{fig:money-plot1}, we show several constraints taken from Ref.~\cite{Dobrich:2019dxc,Bauer:2018uxu}, prospects from beam dump and the PrimEx recast~\cite{Aloni:2019ruo} experiments, regions previously covered by the LEP, Tevatron and LHC searches in photons channels, and supernova constraints (in green). In all these experiments, the rate at which ALPs decay to photons is an important parameter and suppressing it softens the current bounds and represent a limiting factor for the future projections. In the case where new decay modes are open to the ALP, just like sterile neutrinos or anything else, we should expect that the reach of these experiments get diminished. Searching for alternative ways to observe ALPs thus becomes important. 
Contrary to channels that rely on large decay rates into photons pairs, associate decay channels might probe regions where ALP-photon coupling is smaller in comparison. More importantly, we might be missing ALP signals by looking for signals that require large ALP couplings to photons or gluons. We also notice that some parts of the these parameters spaces which are expected to be probed at the LHC can also be searched for other experiments like Belle, BaBar and beam dump experiments providing further evidence for the model.

At the right panel Fig.~\eqref{fig:money-plot1}, we compare the reach of the HL-LHC with similar collider experiments in $\gamma+\not\not\!\! E_T$, at left, and also with other channels, at right. We show the current bounds of BaBar and LEP, and the projections of BaBar for very high integrated luminosities of 20 and 50 fb$^{-1}$~\cite{Dolan:2017osp} at the left panel along with the supernova constraint. Again, we see that $S\to aa\to \gamma_{jet}+\not\not\!\! E_T$ is likely to produce a signal in regions of the parameters space where models with no additional ALP decay modes and large ALP-photon couplings are expected to produce no signals. In particular, ALP-photon couplings down to $\sim 10^{-6}$ GeV$^{-1}$ can be probed if the ALP decays to sterile neutrinos. Notice that supernova constraints are at the verge of excluding ALP-photon couplings from this model.

At the left panel of Fig.~\eqref{fig:money-plot2}, we show the reach of the $\gamma_{jet}+\not\not \!\! E_T$ from the LHC in comparison to the Belle and BaBar reaches in dedicated channels, but now for an ALP with $k_{BB}=0$~\cite{Izaguirre:2016dfi}. Again, the LHC can probe smaller couplings to photons if the ALP decays to an sterile neutrino. The case where $k_{WW}=0$ is shown at the right panel of Fig.~\eqref{fig:money-plot2} with constraints taken from Ref.~\cite{Jaeckel:2012yz}. In this case, stronger bounds apply but it is still possible to probe a complementary region of the parameters space with monophotons. Again, a dark decay mode of the ALP will make all the previous collider bounds to shrink. Let us now present an explicit construction of a model for ALP-sterile neutrino couplings. 

\section{An ultra-violet complete theory for the ALP coupled to SM particles and sterile neutrinos} 
\label{uvmodel}

In this section, we present an UV complete model for ALP couplings to SM particles and sterile 
neutrinos that is effectively described by the EFT Lagrangian of Eq.~\eqref{eq:Lagrangian} 
in the low energy limit -- when the heavy new fields can be integrated out.

\subsection{ALP couplings to the SM gauge bosons and the heavy scalar}

Let us first consider a generic renormalizable model that extends the SM  
by the addition of a complex scalar singlet field $\Phi\sim(1,\,1,\,0)$ and  
a pair of left and right chiral fermionic fields 
$\Psi_{L,R}\sim(d_C^\Psi,\,d_L^\Psi,\,Y^\Psi/2)$, where the numbers in parentheses 
are the usual field transformation properties  
under the SM gauge groups SU(3)$_C$, SU(2)$_L$, 
and U(1)$_Y$, respectively. We assume an approximate 
global chiral U(1) symmetry 
\begin{equation}
\label{pqs}
        \Phi\rightarrow e^{i\alpha}\Phi,\hspace{0.5 cm}
        \Psi_L\rightarrow e^{i\frac{\alpha}{2}}\Psi_L,\hspace{0.5 cm}
        \Psi_R\rightarrow e^{-i\frac{\alpha}{2}}\Psi_R,
\end{equation}
with the remaining fields transforming trivially. Such a symmetry is taken to be 
broken explicitly in the scalar potential, $V(\Phi)$, and also spontaneously 
by the vacuum expectation value of the complex scalar singlet, $\langle\Phi\rangle\not=0$. 
The Lagrangian is 
\begin{equation}
\label{smext}
    \mathscr{L} = \mathscr{L}_{SM} +\partial_\mu\Phi^*\partial^\mu\Phi   +i\overline{\Psi}_L\slashed{D} \Psi_L+i\overline{\Psi}_R\slashed{D} \Psi_R 
    - y( \Phi\overline{\Psi}_L \Psi_R +h.c.) 
    - V(\Phi),
\end{equation}
in which $\mathscr{L}_{SM}$ represents the SM Lagrangian and the Yukawa 
coupling constant $y$ is taken to be real. The covariant derivative acting on 
$\Psi$ in Eq.~(\ref{smext}) is given by 
\begin{equation}
        D_\mu= \partial_\mu-ig_s T^\Psi_a G^a_\mu
        -ig\tau^\Psi_i W_\mu^i-ig^\prime\frac{Y^\Psi}{2}B_\mu,  
\end{equation}
where the generators of SU(3)$_C$ and SU(2)$_L$ in the $\Psi$ representation 
are normalized according to  Tr$[T^\Psi_aT^\Psi_b]=k^\Psi_C\delta_{ab}$ and 
Tr$[\tau^\Psi_i\tau^\Psi_j]=k^\Psi_L\delta_{ij}$, with the coefficients 
$k^\Psi_C\equiv k^\Psi_C(d_C^\Psi)$ and $k^\Psi_L\equiv k^\Psi_L(d_L^\Psi)$ 
depending on the representation dimension of the field. For the fermionic field 
in the fundamental representation of SU(3)$_C$ and SU(2)$_L$ we have 
$\Psi\sim(3,\,2,\,Y^\Psi/2)$, and it is defined that  
$k^\Psi_C(3)=k^\Psi_L(2)=\frac{1}{2}$.  

The scalar potential for the singlet field is 
\begin{equation}
    V(\Phi)=-\frac{1}{4}m^2(\Phi^2+\Phi^{*2})-\mu^2_\Phi\Phi^\dagger\Phi
    +\lambda_{1\Phi}(\Phi^\dagger\Phi)^2+\frac{\lambda_{2\Phi}}{2}(\Phi^4+\Phi^{*4}), 
    \label{Vphi}
\end{equation}
where for simplicity the parameters are taken as all real, with $m,\,\mu^2_\Phi>0$ and $\lambda_{1\Phi},\,\lambda_{2\Phi}>0$. The first and 
the last terms in Eq.~(\ref{Vphi}) are the only ones we consider to break explicitly 
the U(1) symmetry in Eq. (\ref{pqs}). Note that the Lagrangian in Eq. (\ref{smext}) with the potential 
in Eq.~(\ref{Vphi}) can be thought to result from the assumption of the invariance of the 
Lagrangian under the discrete $Z_4$ symmetry $\Phi\rightarrow -\Phi$, $\Psi_L\rightarrow i\Psi_L$ 
and $\Psi_R\rightarrow -i\Psi_R$.  We also work under the consideration that the couplings 
of $\Phi$ with the SM Higgs doublet, $H\sim(1,\,2,\,1/2)$, are negligible. Thus, interaction 
terms like $\lambda_{\Phi H}(H^\dagger H)\Phi^\dagger\Phi$ and 
$\lambda_{\Phi H}^\prime(H^\dagger H)\Phi^2$ can lead to different phenomenological scenario 
with processes like the 
heavy scalar $S$ decay into Higgs bosons, $S\rightarrow h\,h$, and  the 
Higgs boson decay into ALPs, $h\rightarrow aa$. Although interesting, these processes are 
outside the scope of this work.

The scalar singlet field, with its vacuum expectation value $\langle\Phi\rangle=v_\Phi/\sqrt2$, 
is decomposed as 
\begin{equation}
\label{philin}
    \Phi(x)=\frac{1}{\sqrt2}(v_\Phi+S(x)+i\,a(x)),  
\end{equation}
and we identify the heavy scalar and the ALP fields with $S(x)$ and $a(x)$, respectively. Thus, minimization of the potential leads to the following quadratic masses for these fields  
\begin{equation}
\label{msmax}
    \begin{split}
        m^2_S &=2(\lambda_{1\Phi}+\lambda_{2\Phi}) v_\Phi^2=2\lambda_{1\Phi} v_\Phi^2+\frac{1}{2}(m^2-m^2_a),\\
        m^2_a &=m^2-4\lambda_{2\Phi}v_\Phi^2=m^2-2(m^2_S-2\lambda_{1\Phi}v_\Phi^2).
    \end{split}
\end{equation}

We adopt the view that it is technically natural to assume 
$m^2,\,\lambda_{2\Phi}v_\Phi^2\ll\lambda_{1\Phi}v_\Phi^2$, 
which implies $m^2_a\ll m^2_S$, by the  
reason that the limit  $m^2\rightarrow 0$, $\lambda_{2\Phi}\rightarrow 0$ 
leads to an increasing of the symmetries of the model~\cite{tHooft:1979rat}. 
In the present case the symmetries are augmented by the U(1) chiral symmetry in 
Eq. (\ref{pqs}), which turns out to be exact at the classical level. 

Within the consideration that  the symmetry in Eq. (\ref{pqs}) is exact and    
if $\Psi_{L,R}$ transforms non-trivialy under SU(3)$_C$, i. e., $d_C^\Psi\neq1$, then  we would have in Eq. (\ref{smext}) a sort of KSVZ axion model~\cite{Kim:1979if,Shifman:1979if},  
which presents a solution to the strong CP problem through the Peccei-Quinn 
mechanism~\cite{Peccei:1977hh,Peccei:1977ur}. 

More precisely, the original KSVZ model is defined with the extra fermions as   
$\Psi_{L,R}\sim (3,\,1,\,Y^\Psi/2)$. Due to the anomalous feature of the U(1) symmetry $a(x)$ 
turns out to be the axion field, getting a mass 
$m_a= 5.70(7)\,{\rm meV}\left(\frac{10^9\,{\rm GeV}}{f_a}\right)$~\cite{diCortona:2015ldu,Borsanyi:2016ksw} (see \cite{Gorghetto:2018ocs} for additional higher loop and eletromagnetic corrections to this mass).
For the specific KSVZ type model in Eq. (\ref{smext}) we have $f_a=v_\Phi$. We see that our previous phenomenological analysis cannot contemplate an axion model of this type because it is not possible to have the axion mass in the interval  
$1\,{\rm MeV}\leq m_a\leq 0.26\,{\rm GeV}$ for $v_\Phi\geq 10^3$ GeV, which is the typical 
mass scale of $S$ according to Eq.~(\ref{msmax}). Thus, we use the model in Eq. (\ref{smext}) 
as prototype for the ALP interactions only, taking $m_a$ and the scale $v_\Phi$  
as independent free parameters in Eq.~(\ref{msmax}).

\begin{figure}[h!]
\includegraphics[scale=0.4]{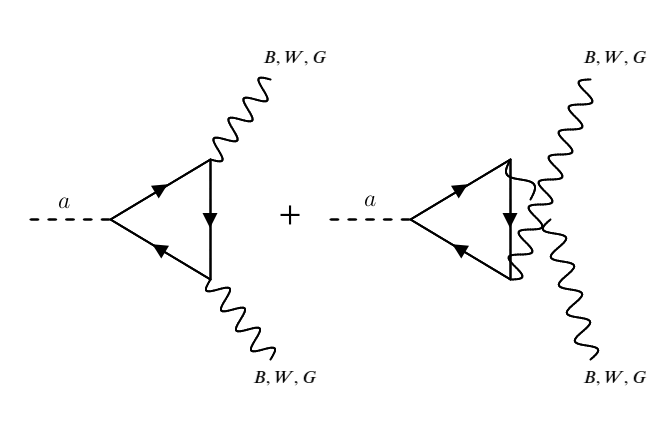}
\caption{Triangle diagrams contributing to the effective axion-like field to the vector bosons, $a\gamma\gamma$, $aGG$, $aW^+W^-$.}
\label{avv}
\end{figure}

With the vacuum expectation value  
$\langle\Phi\rangle$, the relevant terms from Eq.~(\ref{smext}) giving rise   
to the effective couplings of the ALP are in the Lagrangian 
\begin{equation}
    \label{axint}
    \begin{split}
    \mathscr{L}_{a\,\Psi\,V} =\frac{1}{2}\partial_\mu a\,\partial^\mu a - \frac{1}{2}m_a^2\, a^2 +i\overline{\Psi}\slashed{D} \Psi  - m_\Psi\overline{\Psi} \Psi-i\frac{m_\Psi}{v_\Phi} a\,\overline{\Psi}\gamma_5 \Psi,
    \end{split}
\end{equation}
where $\Psi=\Psi_L+\Psi_R$ and $m_\Psi=\frac{y}{\sqrt2}v_\Phi$. The couplings 
of $a(x)$ to the gauge fields are obtained through one loop triangle 
diagrams shown in Figure~\ref{avv}. The coupling of $a(x)$ 
to the abelian U(1)$_Y$ gauge field $B_\mu(x)$ can be straightforward computed compound 
an interaction vertex of $-i\frac{m_\Psi}{v_\Phi} a\overline{\Psi}\gamma_5 \Psi$ and 
two interaction vertices of $g^\prime\frac{Y^\Psi}{2}\overline{\Psi}\slashed{B} \Psi$ 
to form the one-loop triangle diagram. The result is 
\begin{equation}
    \label{abb}
    \mathscr{L}_{aBB} = \frac{\alpha_Y}{4\pi v_\Phi}\times[d_C^\Psi d_L^\Psi\frac{(Y^{\Psi})^2}{4}]\,a\,B^{\mu \nu} \tilde{B}_{\mu \nu},
\end{equation}
in which $\alpha_Y\equiv{g^{\prime 2}/4\pi}$, and the product $d_C^\Psi d_L^\Psi$ of the $\Psi_{L,R}$ representation dimensions 
of SU(3)$_C$ and SU(2)$_L$ counts the number of degrees of freedom with hypercharge ${Y^\Psi}$ 
running in the loop. For the effective coupling of  
$a(x)$ to the nonabelian  SU(2)$_L$ gauge fields $W^i_\mu(x)$ we have 
\begin{equation}
    \label{aww}
    \mathscr{L}_{aWW} = \frac{\alpha_{2}}{4\pi v_\Phi}\times[d_C^\Psi k^\Psi_L]a\,W^{i,\mu \nu} \tilde{W}^i_{\mu \nu},
\end{equation} 
in which $\alpha_2\equiv{g^2/4\pi}$ 
and $d_C^\Psi k^\Psi_L$ is the product of the SU(3)$_C$ degrees of freedom -- the  dimension 
$d_C^\Psi$ -- in the SU(2)$_L$ representation of $\Psi_{L,R}$ by the normalization 
coefficient defined with the trace Tr$[\tau^\Psi_i\tau^\Psi_j]=k^\Psi_L\delta_{ij}$ 
in the computation of the fermion loop in the diagram. Similarly,  the effective coupling 
of $a(x)$ to the nonabelian  SU(3)$_C$ gauge fields $G^a_\mu(x)$ is 
\begin{equation}
    \label{agg}
    \mathscr{L}_{aGG} = \frac{\alpha_s}{4\pi v_\Phi}\times[d_L^\Psi k^\Psi_C]a\,G^{a,\mu \nu} \tilde{G}^a_{\mu \nu}, 
\end{equation} 
in which $\alpha_s\equiv{g^2_s/4\pi}$. 
Observe that $a(x)$ does not couple to gluons if $\Psi_{L,R}$ is a color singlet 
once in that case $k^\Psi_C=0$. 

For the scalar field $S(x)$ its effective couplings to the gauge bosons 
at the lowest order are also obtained through computation of one loop diagrams 
similar to the ones shown in Figure~\ref{avv}. The effective interaction Lagrangians 
of $S(x)$ with $B_\mu(x)$, $W^i_\mu(x)$ and $G^a_\mu(x)$ are given by 
\begin{equation}
    \label{SBB}
    \mathscr{L}_{SBB} = \frac{\alpha_Y}{6\pi v_\Phi}\times[d_C^\Psi d^\Psi_L\frac{(Y^{\Psi})^2}{4}]
    S\,B^{\mu \nu} {B}_{\mu \nu}, 
\end{equation}
\begin{equation}
    \label{Sww}
    \mathscr{L}_{SWW} = \frac{\alpha_2}{6\pi v_\Phi}\times[d_C^\Psi k^\Psi_L]
    S\,W^{i,\mu \nu} {W}^i_{\mu \nu}, 
\end{equation}
\begin{equation}
    \label{Sgg}
    \mathscr{L}_{SGG} = \frac{\alpha_s}{6\pi v_\Phi}\times[d_L^\Psi k^\Psi_C]
    S\,G^{a,\mu \nu} {G}^a_{\mu \nu}. 
\end{equation}

The Eqs. (\ref{abb})--(\ref{Sgg}) furnish the expressions for the effective 
couplings of $a(x)$ and $S(x)$ to the gauge bosons assuming the generic ultraviolet 
completed theory in Eq.~(\ref{smext}). Details of the calculations of loop diagrams 
using both the linear and the polar representation of $\Phi$ for similar  
models can be found, for example, in Ref.~\cite{Quevillon:2019zrd}. 

We can now identify the coefficients of $S(x)$ and $a(x)$ couplings to the gauge bosons 
in the effective Lagrangian of Eq.~(\ref{eq:Lagrangian}) according to 
(with $\alpha_Y\approx 0.009$, $\alpha_2\approx 0.03$ and $\alpha_s\approx 0.1$):  
\begin{equation}
    \label{matchcS}
    \begin{split}
         \frac{c_{BB}}{\Lambda} & =\frac{2}{3}\frac{k_{BB}}{\Lambda}=\frac{\alpha_Y}{6\pi v_\Phi}\times[d_C^\Psi d_L^\Psi\frac{(Y^{\Psi})^2}{4}], \\
         \frac{c_{WW}}{\Lambda} & =\frac{2}{3}\frac{k_{WW}}{\Lambda}=\frac{\alpha_{2}}{6\pi v_\Phi}\times[d_C^\Psi k^\Psi_L],\\ 
         \frac{c_{GG}}{\Lambda} & =\frac{2}{3}\frac{k_{GG}}{\Lambda}=\frac{\alpha_s}{6\pi v_\Phi}\times[d_L^\Psi k^\Psi_C]\; .
    \end{split}
\end{equation}

The trilinear interaction $S\,a^2$, allowing the decay $S\rightarrow a\,a$, arises directly 
from the scalar potential. Taking into account Eq. (\ref{Vphi}) we have that the coupling 
is proportional to the vacuum expectation value $v_\Phi$, i. e.,
\begin{equation}
    \label{saa}
    \mathscr{L}_{Saa} = -(\lambda_{1\Phi}-3\lambda_{2\Phi})v_\Phi S\,a^2. 
\end{equation} 
From this we see that in Eq. (\ref{eq:Lagrangian}) the trilinear coupling is $f_S=(\lambda_{1\Phi}-3\lambda_{2\Phi})v_\Phi$. There is still a quadrilinear interaction $S^2a^2$  from the potential in Eq. (\ref{Vphi}) giving 
\begin{equation}
    \label{ssaa}
    \mathscr{L}_{SSaa} = -\frac{1}{2}(\lambda_{1\Phi}-3\lambda_{2\Phi}) S^2a^2. 
\end{equation}
This interaction leads to the decay process $S\rightarrow S^*\,a\,a\rightarrow a\,a\,\,a\,a$ which is suppressed by the mass of $S$ relatively to the decay $S\rightarrow a\,a$.

\subsection{ALP couplings to the sterile neutrinos and the seesaw mechanism}

We now discuss the light sterile neutrinos in which the ALP could decay into. 
A coupling of the ALP with sterile neutrinos might exist if there is 
a pair of electrically neutral fermionic chiral  fields, ${\cal N}_{L,R}$, 
and an interaction term $y_N\Phi\,\overline{{\cal N}}_L \,{\cal N}_R+h.c.$. 
With the vacuum expectation value $\langle\Phi\rangle$, such interaction gives rise 
to the term $-i\frac{m_{\cal N}}{v_\Phi} a\,\overline{{\cal N}}\gamma_5 {\cal N}$, where 
$m_{\cal N}=\frac{y_{\cal N}}{\sqrt{2}}{v_\Phi}$ is the mass of the sterile neutrino. 
For example, if $v_\Phi=1$ TeV then a sterile neutrino with mass in the interval 
$1\, {\rm MeV}\leq m_N\leq 200 \,{\rm MeV}$ require the Yukawa coupling $y_N$ such 
that $1.4\times 10^{-6}\leq y_N\leq 3.5\times 10^{-4} $. Although this is a fine tuning, 
we note that it is not much worst of what occurs for 
the electron and the light quarks in the context of the SM. 

It is possible to have sterile neutrinos with masses at the 
1--$10^2$ MeV scale resulting from a mass generation mechanism leading 
to sub-eV masses for the active neutrinos. In order to see this we add, 
besides the fermionic field $\Psi$, two sets of sterile neutrinos fields: 
$N_{kL}\sim (1,\,1,\,0)$ and $S_{kL}\sim (1,\,1,\,0)$, with the index $k=4,\,5,\,6$. 
Taking into account also the SM lepton multiplets, $L_{l}\sim(1,\,2,\,-1/2)$ and  
$l_{R}\sim(1,\,1,\,-1)$, with $l=e,\,\mu,\tau$, it is assumed the $Z_4$ symmetry 
presented above under which 
\begin{equation}
\begin{split}
      & N_{kL}\rightarrow -N_{kL},\hspace{1.0 cm}S_{kL}\rightarrow S_{kL}, \\
      &  L_{l}\rightarrow -L_{l},\hspace{1.5 cm}l_{R}\rightarrow -l_{R}.
\end{split}
\end{equation} 
Along with $\Phi$ and $\Psi$, we also assume that the field $N_{kL}$ is charged under the 
approximate U(1) symmetry transforming as $N_{kL}\rightarrow e^{-i\alpha}N_{kL}$. 
This symmetry is taken to be broken in the interaction terms involving the 
SM fields $L_{l}$, $H$ (the Higgs doublet) and the singlet
$N_{kL}$. Thus, the renormalizable Yukawa Lagrangian involving the new singlet fields 
and invariant under the $Z_4$ symmetry is  
\begin{equation}
    \label{lseesaw}
    \begin{split}
       - \mathscr{L}_{LNS}  = y_{kl}\overline{N_{kR}^c}{\tilde{H}}^\dagger L_{l} 
       +g_{k^\prime k}\Phi\overline{S_{k^\prime R}^c}N_{kL}
        +\frac{1}{2}{M_S}_{k^\prime k}\overline{S_{k^\prime R}^c} S_{{kL}}+H.c.,
    \end{split}
\end{equation}
where: $\tilde{H}\equiv \epsilon H^*$, with the antisymmetric matrix 
$\epsilon_{12}=-\epsilon_{12}=1$; $y_{kl}$ and $g_{k^\prime k}$ are $3\times 3$ 
complex matrices; and the $3\times 3$ matrix 
$M_{S k^\prime k}$ is taken diagonal with entries larger than $v_\Phi$. We have defined the right-handed field through the 
conjugation operation such that  $S_{kR}^c\equiv (S_{kL})^c$. It has to be pointed out that the Lagrangian in Eq.~(\ref{lseesaw}) is similar to the ones proposed years ago in extended seesaw models, allowing also sterile neutrinos at the keV--MeV scale, motivated by other purposes such as the quark-lepton complementarity \cite{Lindner:2005pk}, low scale leptogenesis and dark matter \cite{Kang:2006sn,SungCheon:2007nw}. 

Under the assumption that $M_{Sk^\prime k},\,v_\Phi > v_W\approx246$ GeV it is reasonable 
to consider a seesaw mechanism with two stages. The first stage involves $N_{kL}$ and 
$S_{kL}$, with the mass matrix for these sterile neutrinos having the typical texture of  
the canonical seesaw mechanism which is, in the basis $(N_{kL}\,\,{S}_{kL})$, 
\begin{equation}
    \label{mns}
    {\mathbf{M}}_{N\,S}=\left[
    \begin{array}{cc}
        0  & M_D^T \\
        M_D  & M_S 
    \end{array}
    \right],
\end{equation}
where ${M_D}_{k^\prime k}=\frac{g_{k^\prime k}}{\sqrt{2}}v_\Phi$. Block diagonalization leads 
to the approximate states 
\begin{equation}
    \label{nse}
    \begin{split}
        & N_{k^\prime L}\approx {\cal N}_{k^\prime L} +{B}_{k^\prime k} {\cal S}_{k L},\\ 
        & S_{k^\prime L}\approx -{B}^\dagger_{k^\prime k}{\cal N}_{k L}+ {\cal S}_{k^\prime L},
    \end{split}
\end{equation}
where ${B}=M_D^\dagger [M^{*}_S]^{-1}$, with ${\cal N}_{kL}$ and ${\cal S}_{kL}$ 
having the following mass matrices 
\begin{equation}
    \label{bdmm}
    \begin{split}
    & {\mathbf{M}}_{{\cal N}}\approx - M_D^T M_S^{-1} M_D,\\
    & {\mathbf{M}}_{{\cal S}}\approx M_S.
    \end{split}
\end{equation} 
Taking $v_\Phi\approx 10^3$ GeV and ${M_S}_{k^\prime k}=n_{k^\prime k}V/2$ we have 
\begin{equation}
    {\mathbf{M}}_{{\cal N}}\approx -g^T n^{-1} g\,{\rm MeV}\times 
    \left(\frac{10^9{\rm GeV}}{V}\right).
\end{equation}
For $g_{k^\prime k},\,n_{k^\prime k}$ of order one  we could have sterile neutrinos with masses 
1--100 MeV for $10^9\,{\rm GeV}\geq V\geq 10^7\,{\rm GeV}$. In the second stage 
the Higgs doublet gets a vacuum expectation value $\langle H\rangle=[0\,\,v_W]^T$. 
The fields ${\cal S}_{k L}$ are much heavier than ${\cal N}_{k L}$ and the active 
neutrinos so that they can be integrated out leading to the following mass 
effective Lagrangian  

\begin{equation}
    \label{nuseesaw}
    \begin{split}
       - \mathscr{L}_{\nu {\cal N}}  = {m_L}_{ll^\prime}\overline{{\nu}_{lR}^c}\nu_{l^\prime L} + {m}_{Dkl}\,\overline{{\cal N}_{kR}^c}\nu_{lL} +\frac{1}{2}{\mathbf{M}}_{{\cal N}_{k^\prime k}} \overline{{\cal N}_{k^\prime R}^c}{\cal N}_{kL}+H.c.,
    \end{split}
\end{equation}
in which the new $3\times3$ matrices are defined according to
\begin{equation}
\label{mMmD}
    {m}_{L}= -{m}_D^T {B}\,{\mathbf{M}}_{{\cal S}}^{-1} {B}^T {m}_D, \hspace{0.95 cm} 
    {m}_{D}= \frac{{y}}{\sqrt{2}}v_W. 
\end{equation}
The mass matrix involving the active, $\nu_{lL}$, and the sterile neutrinos, 
${\cal N}_{kL}$, in the basis $(\nu_{L}\,\,{\cal N}_{L})$ is
\begin{equation}
    \label{mnun}
    {\mathbf{M}}_{\nu\,{\cal N}}=\left[
    \begin{array}{cc}
        {m}_{L}  & {m}_{D}^T \\
        {m}_{D}  & {\mathbf{M}}_{{\cal N}} 
    \end{array}
    \right].
\end{equation}
Again, performing an approximate block diagonalization of ${\mathbf{M}}_{\nu {\cal N}}$ the flavor states 
of the active neutrinos $\nu_{l L}$ mix with the sterile neutrinos states ${\cal N}_{k L}$ to form the 
block diagonal states
\begin{equation}
    \label{nse2}
    \begin{split}
        & {\cal N}_{kL}^\prime\approx {\cal N}_{kL} +{b}_{kl}\,\nu_{l L},\\ 
        & \nu_{lL}^\prime\approx -{b}^\dagger_{lk}\,{\cal N}_{k L}+ \nu_{l L},
    \end{split}
\end{equation}
in which ${b}={m}_{D}^\dagger\mathbf{M}_{\cal N}^{*-1}$ is the mixing matrix; 
with ${\cal N}_{kL}^\prime$ and $\nu_{lL}^\prime$ having the following approximate mass matrices 
\begin{eqnarray}
\label{Nnumass}
  & &  {\mathbf{M}}_{{\cal N}^\prime}\approx {\mathbf{M}}_{{\cal N}},\\
    & & {\mathbf{M}}_{\nu_{L}^\prime}\approx {m}_{L}- {m}_{D}^T {\mathbf{M}}_{{\cal N}}^{-1} {m}_{D}.\label{mmnu} 
\end{eqnarray}
A discussion about the diagonalization procedure for the neutrinos mass matrices, including 
higher order corrections, involved in several seesaw mechanisms can be found  
in \cite{Lee:1977tib,Schechter:1981cv,Hettmansperger:2011bt}. 

Given that ${B}_{k^\prime k}\approx {\cal O}(v_\Phi/V)\leq 10^{-4}$ we have 
\begin{equation}
    {m}_{L} \leq 6\times 10^{-2}\,{y}^T{g}^\dagger 
    {n}^{*-1}{n}^{-1} {n}^{\dagger-1}{g}^*{y}
    \,\,{\rm eV}.
\end{equation}
On the other hand, by the fact that the term $y_{kl}\overline{N_{kR}^c}{\tilde{H}}^\dagger L_{l}$ 
in Eq. (\ref{lseesaw}) breaks the U(1) symmetry it is technically natural to have $y_{kl}\ll 1$. 
The mass scale of ${m}_{L}$ is then well below to the one required to 
be the compatible with the neutrinos mass square differences inferred from the experiments. 
Therefore, we disregard ${m}_{L}$ so that the active neutrinos states $\nu_{lL}^\prime$ 
have the mass matrix  
\begin{equation}
    {\mathbf{M}}_{\nu_{L}^\prime}\approx 3\times 10^{16}\,{y}^T{g}^{-1}{n} 
    {g}^{T-1}{y}\times
     \left(\frac{V}{10^9{\rm\,GeV}}\right)\,{\rm eV}. 
\end{equation}
Depending on the scale $V$ it would be possible to obtain active neutrinos mass eigenstates 
with mass at the sub-eV scale if $y_{kl}\leq 10^{-7.5}$, corresponding to $V\geq 10^7\,{\rm GeV}$. 
As said before, such small values would reflect the fact that the U(1) symmetry is almost exact. 

The mixing matrix of the active and sterile light neutrinos has entries such that
\begin{equation}
    {b}_{kl}\leq -1.7\times 10^{5}\,[y^\dagger g^{*-1}n^* g^{\dagger -1}]_{kl}
\times\left(\frac{V}{10^9{\rm GeV}}\right). 
\label{samix}
\end{equation}
 For the considered range of values for $V$ 
this implies we estimate the active sterile neutrinos mixing for the particular model example 
presented here to be such that $|{b}_{kl}|^2 \leq 10^{-5}$ ($V\leq 10^9\,{\rm GeV}$). 
This estimation could be compared with the limits from laboratory searches on sterile neutrinos through 
their mixing with active neutrinos. 

As far as we know, the most restrictive direct limits over sterile neutrinos are on their mixing couplings $|U_{e4}|^2$ and $|U_{\mu 4}|^2$ with  the electron neutrino and the muon neutrino, respectively. Under the assumption that there is just a fourth sterile neutrino in addition to the three active ones, Refs.~\cite{Bryman:2019ssi,Bryman:2019bjg} presented a collection of constraints on $|U_{e4}|^2$ and $|U_{\mu 4}|^2$ from various nuclear and particles decays for masses of the sterile neutrino in the MeV to GeV range. In principle, these constraints cannot be directly applied to the example model we give here once, according to the seesaw mechanism, it contains three sterile neutrinos. But we could devise a scenario in the model where the lightest sterile neutrino, which we denote as the ${\bf N}_4$ mass eigenstate of Eq. (\ref{Nnumass}), has mass in the range $1\,{\rm MeV}\leq m_{{\bf N}_4} \leq 100\, {\rm MeV}$ and the other two have masses just above 1 GeV and are not constrained by the direct limits from nuclear and particles decays. From the results of the PIENU  Collaboration~\cite{Aguilar-Arevalo:2015cdf,Aguilar-Arevalo:2017vlf}, the upper limits extracted for the mixing electron neutrino with the sterile neutrino varies from $|U_{e4}|^2\lesssim 4\times 10^{-4}$ to 
$|U_{e4}|^2\lesssim  10^{-7}-10^{-8}$, for $m_{{\bf N}_4}$ ranging from 1 MeV to 100 MeV, at 90$\%$ C.L. as shown in Refs.~\cite{Bryman:2019ssi,Bryman:2019bjg}. These upper limits could be compared 
to $|{b}^\dagger_{e4}|^2\leq 10^{-5}$ from Eq. (\ref{samix})  
assuming this is the mixing of the electron neutrino with the lightest sterile neutrino ${\bf N}_4$. For the mixing of the sterile neutrino with the muon neutrino, in the same $boldsymbol{N}_4$ mass range, the current 
upper limits are less restrictive, varying from  $|U_{\mu 4}|^2\lesssim 4\times 10^{-2}$ to 
$|U_{\mu 4}|^2\lesssim {\rm few}\times 10^{-5}$ at 90$\%$ C.L.~\cite{Bryman:2019bjg} and might be compared to the $|{b}^\dagger_{\mu 4}|^2\leq 10^{-5}$ from Eq. (\ref{samix}). 

 Thus, in principle, the model example we present here, with its mechanism for generating neutrinos mass at the eV and MeV scales, could be compatible with the direct search experiments on sterile neutrinos. Of course, a more detailed study has to be done in order to settle the precise limits from the seesaw model with three sterile neutrinos we presented here, but this is outside the scope of this work. 

There are cosmological constraints over sterile neutrinos that depend on their masses,  
mixing angle with the active neutrinos and, consequently, their lifetime as can be seen in Refs.~\cite{Asaka:2011pb,Ruchayskiy:2011aa,Ruchayskiy:2012si,Hernandez:2014fha,Vincent:2014rja,Bolton:2019pcu,Drewes:2019mhg}. There it is shown that the indirect constraints can be stronger than those from direct searches, for masses at the MeV scale and above, in different models of interactions with the active neutrinos. Within the standard cosmological theory, one main issue is that if sterile neutrinos decay during or after the Big Bang Nucleosynthesis (BBN) they could affect the observed abundance of light elements present in the universe. Many results were obtained in the context of scenarios with two sterile neutrinos, in the 1--140 MeV scale, that decay during or after the BBN, and which are also solely responsible for the type-I seesaw model leading to the observed neutrinos oscillations~\cite{Asaka:2011pb,Ruchayskiy:2011aa,Ruchayskiy:2012si}. Still in a scenario with two sterile neutrinos, the neutrinoless double beta decay can constrain masses $\lesssim 100$ MeV~\cite{Bolton:2019pcu}. An study in Ref.~\cite{Hernandez:2014fha} with three sterile neutrinos in the type-I seesaw with low mass scale, between 1 eV--100 MeV, also put severe constraints for masses sterile neutrinos below 100 MeV. But the authors of Ref.~\cite{Hernandez:2014fha} argue that it would be possible a scenario, with the lightest neutrino having mass below ${\cal O}(10^{-3}\rm eV)$, where one sterile neutrino might never thermalize so that its mass is unconstrained, considering the other two sterile neutrinos having mass above 100 MeV. It could also be  possible to easy the constraints from cosmology over sterile neutrinos taking a significant coupling with the tau neutrino, or if the three right-handed neutrinos are not the uniquely responsible for the light neutrinos masses \cite{Drewes:2019mhg}. The model example we consider falls in this last case once its seesaw mechanism contains six singlets neutrinos fields plus a complex singlet field which host the ALP.

On the other hand, if sterile neutrinos produced in the primordial universe 
have a lifetime well below $1$ second they decay before the BBN. 
In this circumstance, their decay products might have reached thermal equilibrium 
with the other particles not affecting the BBN. 
It is possible to have lifetimes well below $1\,s$, as we estimate 
below. We observe that the current constraints from cosmology may not be directly applied to the model example presented here once they consider scenarios with at most sterile neutrinos. Thus, an additional study is required to precisely 
define the extension of the cosmological constraints to a scenario like the one we take 
into account here. It still possible that the heavy sterile neutrinos decay in particles of a dark sector, evading in this way the constraints of the Big Bang Nucleosynthesis~\cite{Drewes:2019mhg}.

The second term in Eq. (\ref{lseesaw}) leads to the interactions 
\begin{equation}
    \begin{split}
    \label{lnalp}
        \mathscr{L}_{a\,N} & =  \frac{1}{2v_\Phi} \overline{{\bf N}_k}\gamma^\mu\gamma_5 {\bf N}_k\,{\partial_\mu}a +\cdots,
    \end{split}
\end{equation}
where ${\bf N}_k$ are the mass eigenstates of ${\mathbf{M}_{\cal N}}_{k^\prime k}$ 
in Eq. (\ref{Nnumass}). If the ALP can just decay into the lightest sterile neutrino 
${\bf N}_4\equiv{N}$ then we have from Eq. (\ref{eq:Lagrangian}) an interaction 
of the same form of the effective Lagragian in Eq. (\ref{eq:Lagrangian}) 
for describing the ALP-sterile neutrino interaction. From Eq. (\ref{lnalp}) we identify 
the effective coupling in Eq. (\ref{eq:Lagrangian}) according to 
\begin{equation}
    \frac{c_{aN}}{\Lambda}=\frac{1}{2v_\Phi}.
\end{equation}
Observe that typical values $c_{aN}= 50$ and $\Lambda= 100$ TeV give in fact $v_\Phi= 1$ TeV. 
Thus, with the potential in Eq. (\ref{Vphi}) and the Lagrangians in 
Eqs. (\ref{axint}) and (\ref{lseesaw}) the effective Lagrangian 
in Eq. (\ref{eq:Lagrangian}) can be generated considering simple models 
which could be part of a complete theory at the ultra-violet regime.

It is important to observe that the lightest sterile neutrino $N$ produced 
in the ALP decay, $a\rightarrow N+N $, studied here might have 
a timelife long enough to decay outside the LHC detectors. The main decay channels 
available for the sterile neutrinos with mass below that of the pion are:  
$N\rightarrow e^- + W^{*+}\rightarrow e^- +e^+ +\nu$; 
$N\rightarrow \nu + Z^{*}\rightarrow \nu+e^- +e^+$;
and $N\rightarrow \nu + Z^{*}\rightarrow \nu+\nu+\overline{\nu}$. 

{These decays occur via mixing with SM neutrinos and have the decay widths, following Ref.~\cite{Atre:2009rg}, as follows.}
For $N\rightarrow \nu_{\ell_1}+ \ell_2^-+ \ell_2^+$, we have both charged and neutral current contributions, given by 
\begin{eqnarray}
\Gamma(N\rightarrow \nu_{l_1} l_2^- l_2^+) &=& \Theta(m_N - m_{\nu_{\ell_1}} - 2 m_{\ell_2}) \frac{G_F^2}{96 \pi^3} |\boldsymbol{b}_{\ell_1 4}|^2 m_N^5 \times [ ( g^\ell_L g_R^\ell + \delta_{\ell_1 \ell_2} g_R^\ell  ) I_2(x_{\nu_{\ell_1}}, x_{\ell_2}, x_{\ell_2}) \nonumber \\ 
&+& ( (g_L^{\ell})^2 + (g_R^\ell)^2 + \delta_{\ell_1 \ell_2}(1 - 2 g_L^\ell)) I_1(x_{\nu_{\ell_1}}, x_{\ell_2}, x_{\ell_2}) ],\\
I_1(x,y,z) &=& 12 \int_{(x+y)^2}^{(1-z)^2} \frac{d\xi}{\xi} (\xi - x^2 - y^2) (1 + z^2 - \xi) \lambda^{\frac{1}{2}}(\xi,x^2,y^2) \lambda^{\frac{1}{2}}(1,\xi,z^2), \\
I_2(x,y,z) &=& 24yz \int_{(x+z)^2}^{(1-x)^2} \frac{d\xi}{\xi} (1 + x^2 - \xi) \lambda^{\frac{1}{2}}(\xi,y^2,z^2) \lambda^{\frac{1}{2}}(1,\xi,x^2),
\end{eqnarray}
where $\ell_1, \ell_2 = e, \mu, \tau$, $I_1(0,0,0) = 1$, $I_2(0,0,0) = 1$, $x_i=m_{\ell_i}/m_N$,  $g_L^\ell = -\frac{1}{2} + x_w$, $g_R^\ell = x_w$ and $x_w = sin^2 \theta_w = 0.231$, where $\theta_w$ is the Weinberg angle, and $\lambda(x,y,z)=x^2+y^2+z^2-2xy-2xz-2yz$.

For $N\rightarrow \nu+\nu+\overline{\nu}$ we have only the neutral current
\begin{eqnarray}
\sum_{\ell_2 = e}^{\tau} \Gamma(N \rightarrow \nu_{\ell_1} \nu_{\ell_2} \overline{\nu_{\ell_2}}) = \frac{G_F^2}{96 \pi^3} |\boldsymbol{b}_{\ell_1 4}|^2 m_N^5.
\end{eqnarray}
Thus, we have

\begin{eqnarray}
& &\Gamma(N\rightarrow e^- + W^{*+}\rightarrow e^- +e^+ +\nu) 
\simeq 2.28 \times 10^{-24} \; \textrm{GeV},\nonumber\\
& &\Gamma(N\rightarrow \nu + Z^{*}\rightarrow \nu+e^- +e^+) 
\simeq 2.86 \times 10^{-25} \; \textrm{GeV},\\
& &\Gamma(N\rightarrow \nu + Z^{*}\rightarrow \nu+\nu+\overline{\nu}) 
\simeq 1.71 \times 10^{-24} \; \textrm{GeV},\nonumber
\label{Nwidths}
\end{eqnarray}
resulting in a lifetime $\tau = \frac{1}{\Gamma_N} \simeq 0.15 \; s$, with $m_N = 100$ MeV, 
$|\boldsymbol{b}_{ij}|^2 = 10^{-5}$ and $\Gamma_N$ is the sum of the three partial decay widths above. Once these sterile neutrino partial widths are proportional to $m_N^{5}$, small masses lead to very large lifetimes.

We now give a set of benchmarks for this model assuming $v_\Phi= 1$ TeV and $m_S = 1.4$ TeV and with the relations of couplings 

\begin{equation}
\label{cks}
    k_{BB}=\frac{3}{2}c_{BB},\; 
    k_{WW}=\frac{3}{2}c_{WW},\; 
    k_{GG}=\frac{3}{2}c_{GG}\; . 
\end{equation}

\begin{table}[t!]
\centering
\begin{tabular}{c|c|c|c|c|c|c|c|c } 
\hline
Representation of $\Psi$ & $c_{BB}$ & $c_{WW}$ & $c_{GG}$ & $c_{aN}$ & $\Lambda$(GeV) & $m_N$(MeV) & $m_a$(MeV) & $g_{\gamma \gamma}$(GeV$^{-1}$)\\
\hline
\hline
$(3,\,1,\,2/3)$ & 0.01 & 0 & 0.05 & 9.26 & 18532 & 7.0 & 400 & $2.97 \times 10^{-6}$ \\
\hline
$(3,\,2,\,1/6)$ & 0.001 & 0.05 & 0.1 & 9.76 & 19515 & 10 & 350 & $3.6 \times 10^{-6}$ \\
\hline
$(3,\,3,\,-1/3)$ & 0.005 & 0.11 & 0.09 & 5.81 & 11630 & 8.0 & 370 & $1.5 \times 10^{-5}$ \\ 
\hline 
\hline
\end{tabular}
\caption{Benchmarks for a quark singlet, doublet and triplet of SU(2)$_L$, assuming $v_\Phi= 1$ TeV and $m_S = 1.4$ TeV.}
  \label{tab2}
\end{table}

From Eq.~(\ref{saa}) we have $f_S\approx m_S^2/2v_\Phi=1$ TeV. This model corresponds 
to the effective  theory in Eq. (\ref{eq:Lagrangian}) with $c_{ae}=c_{a\mu}=0$. 
Also, from Eqs.~\eqref{eq:gaa} and \eqref{cks} we have
\begin{eqnarray}
    k_{\gamma\gamma} = \frac{c_w^2 \alpha_Y}{\pi v_{\Phi}}\times[d_C^\Psi d^\Psi_L\frac{(Y^\Psi)^2}{4}] + \frac{s_w^2 \alpha_2}{\pi v_{\Phi}}\times[d_C^\Psi k^\Psi_L],
\end{eqnarray}
which depends only on $v_{\Phi}$ and the representation of the fermionic field $\Psi$. 
Some benchmarks are shown in Table~\eqref{tab2} for some choices of representation of $\Psi$. 
As examples we consider: a quark singlet with electric charge 2/3, 
$\Psi_{L,R}\sim(d_C^\Psi,\,d_L^\Psi,\,Y^\Psi/2)=(3,\,1,\,2/3)$; a doublet of quarks, 
$\Psi_{L,R}\sim(3,\,2,\,1/6)$; and a triplet of quarks, $\Psi_{L,R}\sim(3,\,3,\,-1/3)$.

These benchmarks are represented as dots(singlet), asterisks(doublet) and stars(triplet) in Figs.~\eqref{plot-lifetime}--\eqref{fig:gaa-params} for various slices of the parameters space, and Figs.~\eqref{fig:money-plot1} and \eqref{fig:money-plot2} in the ALP-photon coupling {\it versus} ALP mass plane. They are right 
in the region where the LHC is able to probe the EFT parametrization of the model. As a general remark, based on these examples, the triplet case might be easier to observe than doublet and singlet SU(2)$_L$ representations due the enhanced couplings to photons. This expectation is reflected in Figs.~\eqref{plot-lifetime}--\eqref{fig:gaa-params} where we can see that triplet example lie in a region of the parameters space where the chance to observe the signal at the LHC is larger. The singlet case, in special, seems to be more challenging. For example, at the right panel of Fig.~\eqref{fig:money-plot2}, we see that the singlet case is already excluded by supernova data, contrary to the triplet and doublet cases. On the other hand, the doublet and triplet representations can be confirmed by other experiments like Belle and beam dump experiments.

The signal significance of these three points are all above $5\sigma$ for 300 fb$^{-1}$, the larger significance occurs for the triplet, and the lower for the singlet case. This is a consequence of having suppressed couplings to leptons that increases the product 
$BR(a\to\gamma\gamma)\times BR(a\to NN)$. Moreover, an 1.4 TeV $S$ scalar 
has a much favorable cut efficiency compared to the backgrounds which are very efficiently 
eliminated by an 1 TeV cut in the missing transverse momentum.

It is worthwhile to observe that once the parameter $f_S$ 
scales with $m_S^2\simeq 2\lambda_{1\Phi}v_\Phi^2$, the decay width of $S\rightarrow a\,a$  
behaves as $\Gamma (S \to a a) \simeq \frac{m_S^3}{32 \pi v_\Phi^2}$ and
the branching ratio $BR(S \to a a)$ might increase with $m_S$ favoring the resonant ALP production.

\section{Conclusions}
\label{conclu}

ALPs are predicted in many extensions of the SM and their discovery could bring 
light to many problems in physics, astrophysics and cosmology. Sterile neutrinos, by their turn, might well be the first particles beyond the SM paradigm to be discovered if the signals from neutrino baseline experiments are confirmed. On the other hand, an extended scalar sector with more scalars would be plausible to give mass to an extended matter sector. These three new ingredients thus sum up to a good target to phenomenological studies and model building.

In this work, we have investigated an effective field theory where a heavy scalar and an sterile neutrino couple to an ALP. We provided prospects to probe regions of the EFT parameters space at the 13 TeV LHC after 3000 fb$^{-1}$ respecting several experimental constraints. The LHC is able to constrain and discover such models in monophoton searches with large missing transverse energy for ALP masses from 10 to 400 MeV, sterile neutrino masses from 100 keV to 100 MeV and $S$ scalars from 500 GeV to 2 TeV. Effective ALP-sterile neutrinos and ALP-photons couplings, $c_{aN}/\Lambda$ and $k_{\gamma\gamma}$, as small as $10^{-5}$ GeV$^{-1}$ and $10^{-6}$ GeV$^{-1}$, respectively, can be probed. Comparisons with existing constraints and prospects from other experiments show that the LHC in the monophoton channel can probe regions of the parameters space not accessible by those experiments establishing its complementary role in the quest for these kinds of physics beyond the SM as shown in Figs.~\eqref{fig:money-plot1} and \eqref{fig:money-plot2}.

To give a concrete example in model building, we present an ultra-violet complete model leading to the effective low energy Lagrangian studied in Section~\ref{sec-efcl}, where the ALP couples to sterile neutrinos whose masses can be well within the 1--100 MeV scale. The model contains a seesaw mechanism, which involves 
the scale of spontaneous breakdown of the approximate U(1) symmetry associated to the ALP, for generating mass for both sterile and active neutrinos. We found that singlet, doublet and triplet matter representations can be discovered after 300 fb$^{-1}$ by the proposed strategy and respecting all the constraints considered in the EFT approach if the ALP couplings to charged leptons are suppressed.

\section*{Acknowledgements}
The authors are grateful to Joerg Jaeckel for his valuable suggestions on our paper. The work of A. A. and A. G. D. is supported by the Conselho Nacional de Desenvolvimento Cient\'{\i}fico e Tecnol\'ogico (CNPq), under the grants 307265/2017-0 and 306636/2016-6, respectively.
D. D. L. acknowledges the financial support from the Coordena\c{c}\~ao de Aperfei\c{c}oamento de Pessoal de N\'{\i}vel Superior - Brasil (CAPES) - 23 Finance Code 001.

\appendix
\label{Appendix}
\section{Probability of decay of ALP pairs inside the detector}
The probability that an ALP of mass $m_a$ and proper decay length $c\tau_0=\frac{c}{\Gamma_a}$, produced at the point of interaction, decays within an annulus of inner radius $R_1$ and outer radius $R_2$, is given by~\cite{Accomando:2016rpc}
$$
P(R_1\leq r\leq R_2) = \int_{R_1}^{R_2} \frac{1}{\beta\gamma c\tau_0} \exp\left(-\frac{x}{\beta\gamma c\tau_0}\right)\; dx\; ,
$$
where $\gamma=1/\sqrt{1-\beta^2}$, with $\beta\approx \sqrt{1-4m_a^2/m_S^2}$, is the Lorentz factor for a particle that comes from the decay of the heavy scalar of mass $m_S$. In this case, $\gamma c\tau_0$ is the decay length of the ALP in the laboratory frame, $\ell_{decay}$. This is the distance that we measure the ALP to travel inside the detector before it decays.

Intuitively, a light ALP receives a large boost from a heavy particle decay, however, if its lifetime is too small, the distance it travels inside the detector before it decays might be very short.

Let us take $R_1=0$ and $R_2=R$ m, then the probability that the lighter particle decays after traveling $R$ meters inside the detector is
$$
P(r\leq R|\ell_{decay}) = 1-\exp\left(-\frac{R}{\ell_{decay}\beta}\right)\approx 1-\exp\left(-\frac{R}{\ell_{decay}}\right)
$$
for the case we are considering in this work where $m_a$ is of order of hundreds of MeV and $m_S$ is of order of hundreds of GeV, $\beta\approx 1$.

\begin{figure}[t!]
\centering
\includegraphics[scale=0.5]{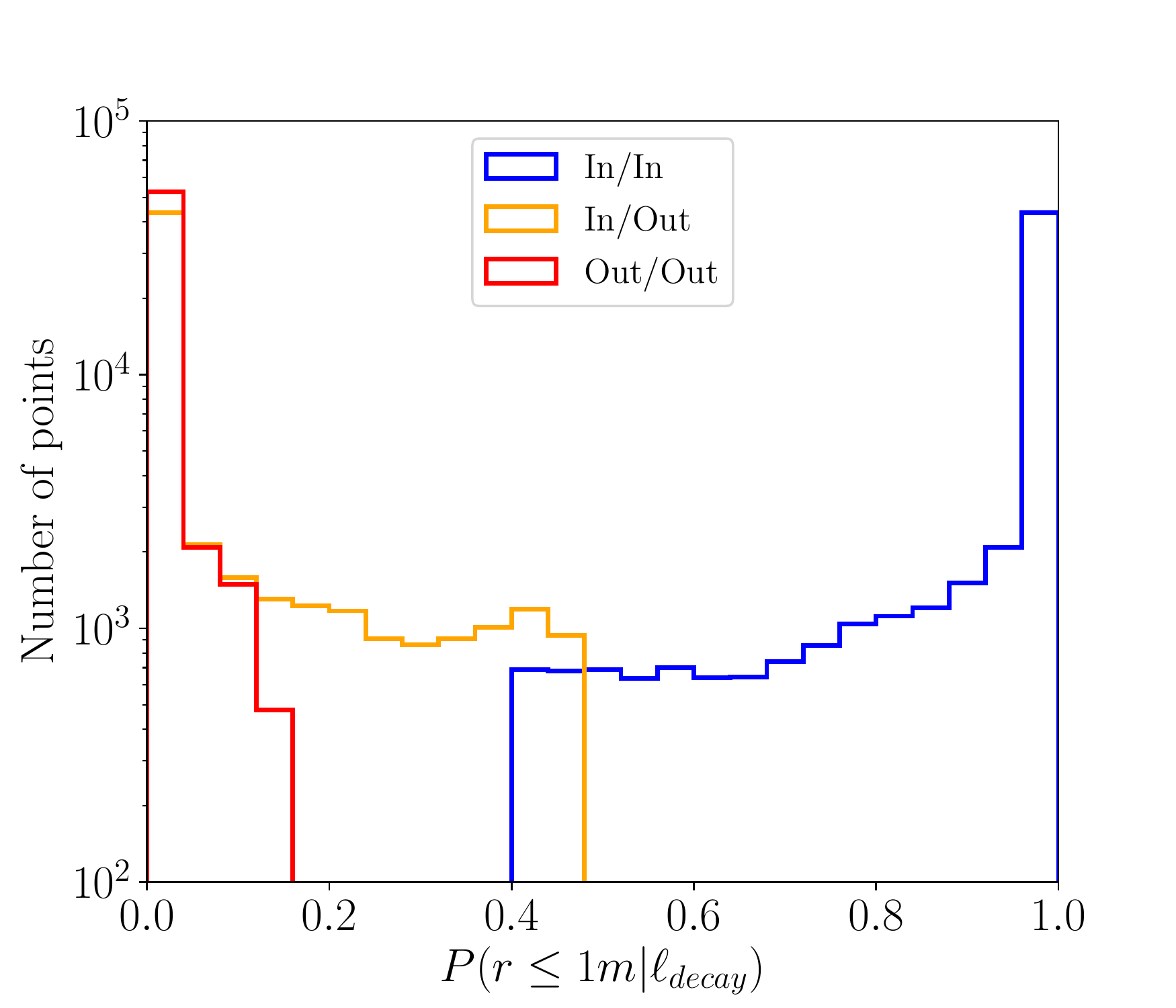}
\caption{Probability that a pair of ALPs coming from the decay of a heavy particle decays both within 1 meter inside the detector (In/In), both outside 1 meter inside the decetor (Out/Out), and one before 1 meter and the other after 1 meter inside the detector, given the decay length in the laboratory frame, $\ell_{decay}$.}
\label{fig:prob_decay}
\end{figure}

In the Fig.~\eqref{fig:prob_decay} shown right below, we show the histogram for $P(r\leq 1m|\ell_{decay})$, the probability that the ALP decays within 1 m inside the detector given the decay length of the ALP, for the cases where the two ALPs decay within 1 m (In/In), the two ALPs decay outside 1 m (Out/Out) and one of them decay inside 1 m of the detector and the other outside 1 m (In/Out). The selected sample has around 57000 points. We see that almost all these points have a probability to decay within 1 m very close to 100\%. There is a small fraction of points with a non-negligible probability for the In/Out case. These points could be taken as signal points if the ALP that decays inside 1 meter becomes a photon pair, however the large majority of points have a null probability for this case. We conclude that assuming that all ALPs decay inside 1 meter is a very good approximation.

\bibliographystyle{apsrev4-1} 
%


\end{document}